\documentclass[onecolumn,superscriptaddress,showpacs,preprintnumbers,amsmath,amssymb,amsfonts,aps,nofootinbib,floatfix]{revtex4}

\usepackage{graphicx}
\usepackage{dcolumn}
\usepackage{bm}

\begin{document}

\preprint{APS/123-QED}

\title{\large Measurement of the semi-inclusive $\pi^+$ electroproduction
off the proton}

\newcommand*{\ANL}{Argonne National Laboratory, Argonne, Illinois 60439}
\affiliation{\ANL}
\newcommand*{\ASU}{Arizona State University, Tempe, Arizona 85287-1504}
\affiliation{\ASU}
\newcommand*{\CSU}{California State University, Dominguez Hills, Carson, CA 90747}
\affiliation{\CSU}
\newcommand*{\CMU}{Carnegie Mellon University, Pittsburgh, Pennsylvania 15213}
\affiliation{\CMU}
\newcommand*{\CUA}{Catholic University of America, Washington, D.C. 20064}
\affiliation{\CUA}
\newcommand*{\SACLAY}{CEA-Saclay, Service de Physique Nucl\'eaire, 91191 Gif-sur-Yvette, France}
\affiliation{\SACLAY}
\newcommand*{\CNU}{Christopher Newport University, Newport News, Virginia 23606}
\affiliation{\CNU}
\newcommand*{\WM}{College of William and Mary, Williamsburg, Virginia 23187-8795}
\affiliation{\WM}
\newcommand*{\ECOSSEE}{Edinburgh University, Edinburgh EH9 3JZ, United Kingdom}
\affiliation{\ECOSSEE}
\newcommand*{\FU}{Fairfield University, Fairfield CT 06824}
\affiliation{\FU}
\newcommand*{\FIU}{Florida International University, Miami, Florida 33199}
\affiliation{\FIU}
\newcommand*{\FSU}{Florida State University, Tallahassee, Florida 32306}
\affiliation{\FSU}
\newcommand*{\ISU}{Idaho State University, Pocatello, Idaho 83209}
\affiliation{\ISU}
\newcommand*{\INFNPA}{INFN, Gruppo Collegato di Parma, 43100 Parma, Italy}
\affiliation{\INFNPA}
\newcommand*{\INFNFR}{INFN, Laboratori Nazionali di Frascati, 00044 Frascati, Italy}
\affiliation{\INFNFR}
\newcommand*{\INFNGE}{INFN, Sezione di Genova, 16146 Genova, Italy}
\affiliation{\INFNGE}
\newcommand*{\ORSAY}{Institut de Physique Nucleaire ORSAY, Orsay, France}
\affiliation{\ORSAY}
\newcommand*{\ITEP}{Institute of Theoretical and Experimental Physics, Moscow, 117259, Russia}
\affiliation{\ITEP}
\newcommand*{\BONN}{Institute f\"{u}r Strahlen und Kernphysik, Universit\"{a}t Bonn, Germany 53115}
\affiliation{\BONN}
\newcommand*{\JMU}{James Madison University, Harrisonburg, Virginia 22807}
\affiliation{\JMU}
\newcommand*{\KYUNGPOOK}{Kyungpook National University, Daegu 702-701, South Korea}
\affiliation{\KYUNGPOOK}
\newcommand*{\MIT}{Massachusetts Institute of Technology, Cambridge, Massachusetts  02139-4307}
\affiliation{\MIT}
\newcommand*{\MOSCOW}{Moscow State University, Skobeltsyn Institute of Nuclear Physics, 119899 Moscow, Russia}
\affiliation{\MOSCOW}
\newcommand*{\CHEP}{National Scientific and Educational Centre of Particle and High Energy Physics of the Belarusian State University, 220040  Minsk,  Belarus}
\affiliation{\CHEP}
\newcommand*{\NSU}{Norfolk State University, Norfolk, Virginia 23504}
\affiliation{\NSU}
\newcommand*{\OHIOU}{Ohio University, Athens, Ohio  45701}
\affiliation{\OHIOU}
\newcommand*{\ODU}{Old Dominion University, Norfolk, Virginia 23529}
\affiliation{\ODU}
\newcommand*{\RPI}{Rensselaer Polytechnic Institute, Troy, New York 12180-3590}
\affiliation{\RPI}
\newcommand*{\RICE}{Rice University, Houston, Texas 77005-1892}
\affiliation{\RICE}
\newcommand*{\GWU}{The George Washington University, Washington, DC 20052}
\affiliation{\GWU}
\newcommand*{\JLAB}{Thomas Jefferson National Accelerator Facility, Newport News, Virginia 23606}
\affiliation{\JLAB}
\newcommand*{\UNIONC}{Union College, Schenectady, NY 12308}
\affiliation{\UNIONC}
\newcommand*{\UNICH}{Universidad T\'{e}cnica Federico Santa Mar\'{i}a Av. Espa\~{n}a 1680 Casilla 110-V Valpara\'{i}so, Chile}
\affiliation{\UNICH}
\newcommand*{\PARMA}{Universit\`a di Parma, 43100 Parma, Italy}
\affiliation{\PARMA}
\newcommand*{\UCLA}{University of California at Los Angeles, Los Angeles, California  90095-1547}
\affiliation{\UCLA}
\newcommand*{\UCONN}{University of Connecticut, Storrs, Connecticut 06269}
\affiliation{\UCONN}
\newcommand*{\ECOSSEG}{University of Glasgow, Glasgow G12 8QQ, United Kingdom}
\affiliation{\ECOSSEG}
\newcommand*{\UMASS}{University of Massachusetts, Amherst, Massachusetts  01003}
\affiliation{\UMASS}
\newcommand*{\UNH}{University of New Hampshire, Durham, New Hampshire 03824-3568}
\affiliation{\UNH}
\newcommand*{\PITT}{University of Pittsburgh, Pittsburgh, Pennsylvania 15260}
\affiliation{\PITT}
\newcommand*{\URICH}{University of Richmond, Richmond, Virginia 23173}
\affiliation{\URICH}
\newcommand*{\SCAROLINA}{University of South Carolina, Columbia, South Carolina 29208}
\affiliation{\SCAROLINA}
\newcommand*{\VIRGINIA}{University of Virginia, Charlottesville, Virginia 22901}
\affiliation{\VIRGINIA}
\newcommand*{\VT}{Virginia Polytechnic Institute and State University, Blacksburg, Virginia   24061-0435}
\affiliation{\VT}
\newcommand*{\YEREVAN}{Yerevan Physics Institute, 375036 Yerevan, Armenia}
\affiliation{\YEREVAN}

\newcommand*{\deceased}{Deceased}

\newcommand*{\NOWMIT}{Massachusetts Institute of Technology, Cambridge, Massachusetts  02139-4307}
\newcommand*{\NOWCUA}{Catholic University of America, Washington, D.C. 20064}
\newcommand*{\NOWGEISSEN}{Physikalisches Institut der Universitaet Giessen, 35392 Giessen, Germany}
\newcommand*{\NOWOHIOU}{Ohio University, Athens, Ohio  45701}
\newcommand*{\NOWURICH}{University of Richmond, Richmond, Virginia 23173}
\newcommand*{\NOWUNH}{University of New Hampshire, Durham, New Hampshire 03824-3568}
\newcommand*{\NOWECOSSEE}{Edinburgh University, Edinburgh EH9 3JZ, United Kingdom}
\newcommand*{\NOWUMASS}{University of Massachusetts, Amherst, Massachusetts  01003}

\author {M.~Osipenko} 
\affiliation{\INFNGE}
\affiliation{\MOSCOW}
\author {M.~Ripani} 
\affiliation{\INFNGE}
\author {G.~Ricco} 
\affiliation{\INFNGE}
\author {H.~Avakian} 
\affiliation{\JLAB}
\author {R.~De~Vita} 
\affiliation{\INFNGE}
\author {G.~Adams}
\affiliation{\RPI}
\author {M.J.~Amaryan}
\affiliation{\ODU}
\author {P.~Ambrozewicz} 
\affiliation{\FIU}
\author {M.~Anghinolfi} 
\affiliation{\INFNGE}
\author {G.~Asryan} 
\affiliation{\YEREVAN}
% ...
\collaboration{The CLAS Collaboration}
     \noaffiliation
%
 
%
%

\date{\today}

\begin{abstract}
Semi-inclusive $\pi^+$ electroproduction on protons has been measured with the
CLAS detector at Jefferson Lab. The measurement was performed on a liquid-hydrogen
target using a 5.75 GeV electron beam. The complete five-fold differential
cross sections were measured over a wide kinematic range
including the complete range of azimuthal angles between hadronic and leptonic planes, $\phi$,
enabling us to separate the $\phi$-dependent terms.
Our measurements of $\phi$-independent term of the cross section at low Bjorken $x$
were found to be in fairly good agreement with
pQCD calculations.
Indeed, the conventional current fragmentation calculation can account for almost all of
the observed cross section, even at small $\pi^+$ momentum.
The measured center-of-momentum spectra are in qualitative agreement with high energy data, which
suggests a surprising numerical similarity between the spectator diquark
fragmentation in the present reaction and the anti-quark fragmentation measured in $e^+ e^-$ collisions.
We have observed that the two $\phi$-dependent terms of the cross section are small.
Within our precision the $\cos 2\phi$ term is compatible with zero, except for low-$z$ region,
and the measured $\cos\phi$ term is much smaller in magnitude
than the sum of the Cahn and Berger effects.
\end{abstract}



\maketitle

\section{\label{sec:intro} Introduction}
\begin{figure}[htb]
\includegraphics[bb=2cm 11cm 20cm 20cm, scale=0.25,angle=270]{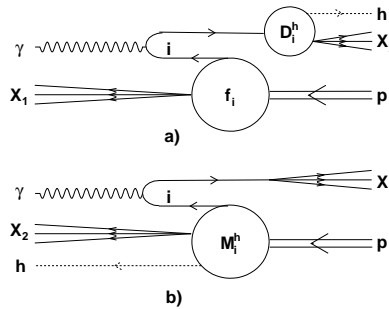}
\caption{\label{fig:scheme}Schematic representation of current (a)
and target (b) fragmentation processes in the virtual photon-proton center
of momentum frame neglecting transverse momenta of particles.
Blobs represent non-perturbative functions:
the parton density function $f_i(x)$, parton fragmentation function $D_i^h(z)$
and fracture function $M_i^h(x,z)$ as given in Eq.~\ref{eq:lo_pqcd}.
$X_1$ and $X_2$ indicate two components of the undetected final state
hadronic system $X$.}
\end{figure}

The semi-inclusive leptoproduction of hadrons off the nucleon, $eN\rightarrow e^\prime h X$,
is an important tool allowing to study simultaneously the internal structure of
the target nucleon and hadron creation mechanism.
In Deep Inelastic Scattering (DIS) regime the semi-inclusive leptoproduction of hadrons
can be described by perturbative quantum chromodynamics (pQCD)
combining non-perturbative distribution/fragmentation functions.
Semi-Inclusive leptoproduction of hadrons in DIS (SIDIS) can occur through 
current or target fragmentation~\cite{Trentadue} (see Fig.~\ref{fig:scheme}).
Current fragmentation is the hadronization of the struck quark,
while target fragmentation is hadronization of the spectator.
Both non-perturbative, soft fragmentation mechanisms factorize
from the hard virtual-photon/parton scattering amplitude
in pQCD (see Ref.~\cite{current_factorization} for the current fragmentation and
Ref.~\cite{Grazzini,Collins} for the target fragmentation).

Inclusive lepton scattering off the nucleon and hadron production
in $e^+e^-$ collisions allow one to study separately the fractional momentum dependence of
the parton distribution functions for the nucleon
and the parton fragmentation functions, respectively.
The leptoproduction of hadrons in the current fragmentation region
combines these two and provides additional information about 
hadronization and nucleon structure.  
In fact, for DIS, semi-inclusive measurements provide 
new information about the Transverse Momentum Distribution (TMD) of partons, which
is important for understanding the role of orbital angular momenta of quarks
and gluons~\cite{TMD}.
Furthermore, the detection of a hadron in SIDIS introduces
a flavor selectivity for the observed parton distributions.
In contrast, target fragmentation is described by fracture functions,
present only in the semi-inclusive reactions.

The finite transverse momentum of partons in the initial state
leads to an azimuthal variation in the cross section, as does 
the transverse spin of partons in the unpolarized nucleon.

In order to achieve SIDIS regime sufficiently high beam energy is mandatory.
With decreasing the beam energy higher order (in pQCD) and higher twist effects
appear, spoiling the agreement between the experimental data and
theoretical pQCD expectations. Therefore, only a comparison between
the actual data and theoretical calculations can reveal
the dominance of SIDIS dynamics in the experiment.
Though, a good agreement between data and theory in one
observable does not necessarily guarantee SIDIS dominance
in others.

Previous measurements~\cite{emc_h2,E665_h2,H1_h2,ZEUS_h2,HERMES_h2} have 
verified these factorizations experimentally and have tested pQCD predictions.
Measurements of unpolarized semi-inclusive lepton-nucleon scattering have been performed
at several facilities such as CERN (EMC~\cite{emc_h2}), Fermilab (E655~\cite{E665_h2}),
DESY (H1~\cite{H1_h2}, ZEUS~\cite{ZEUS_h2}, HERMES~\cite{HERMES_h2}), SLAC~\cite{Dakin},
Cornell~\cite{Bebek_prd15,Bebek_prd16,Browman_prl37} and Jefferson Lab (Hall C)~\cite{hallc,hallc_new}.
The last two experiments covered a kinematical region similar to the present measurement.
Despite all of these measurements, open questions remain
about  the target fragmentation mechanism and the
physics behind the azimuthal distributions. The measurements at
high beam energies (EMC, E655, H1 and ZEUS) covered a broad kinematic
range, but lacked particle identification and the statistics to look at differential cross sections
in more than two kinematic variables (the latter applies also to HERMES).
Experiments at lower energies (SLAC, Cornell and Hall C of Jefferson Lab) using 
classical spectrometers measured cross sections only at a few kinematic points.
To improve the current knowledge of semi-inclusive lepton-nucleon scattering
one has to combine the broad coverage of high energy experiments with high luminosity
and particle identification in order to measure the fully differential cross section
for a specified hadron.

\begin{figure}[htb]
\includegraphics[bb=6cm 11cm 20cm 18cm, scale=0.75]{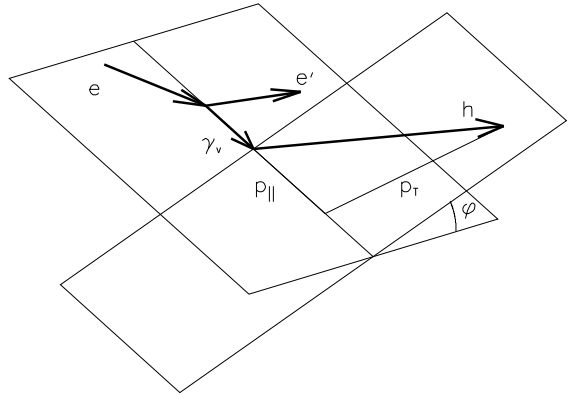}
\caption{\label{fig:define}Definition of the azimuthal angle
between the leptonic and hadronic planes $\phi$ and hadron
momentum components $p_T$ and $p_\parallel$.}
\end{figure}

Semi-inclusive hadron electroproduction, $\gamma_v(q) + p(P)\rightarrow h(p_h) + X$, 
is completely described by a set of five kinematic variables. The variables $q$, $P$ and
$p_h$ in parentheses denote four-momenta of the virtual photon $\gamma_v$, the proton $p$ and the observed hadron $h$.
The letter $X$ denotes the unobserved particles in the reaction. In this article
we have chosen a commonly used set of independent variables:
the virtual photon four-momentum transfer squared $Q^2=-q^2\stackrel{\rm Lab}{=}4E_0E^\prime \sin^2{\frac{\theta}{2}}$,
the Bjorken scaling variable $x=-\frac{q^2}{2P\cdot q}\stackrel{\rm Lab}{=}\frac{Q^2}{2M\nu}$,
the virtual photon energy fraction carried by the hadron 
$z=\frac{P\cdot p_h}{P\cdot q}\stackrel{\rm Lab}{=}\frac{E_h}{\nu}$,
the squared hadron spatial transverse momentum with respect to the virtual photon direction $p_T^2$
and the angle between the leptonic and hadronic planes $\phi$~\cite{Trento_convention} (see Fig.~\ref{fig:define}).
Here $E_0$ is the beam energy, $E^\prime$ and $\theta$ are the scattered electron energy and angle,
$\nu=E_0-E^\prime$ is the virtual photon energy in the lab frame,
and $M$ is the proton mass. We will also use momentum transfer $t=(q-p_h)^2$,
Feynman $x_F=2 p_\parallel^{CM}/W$
and the projection of the hadron momentum onto the photon direction $p_\parallel$
as alternative variables when they help with the physical interpretation.
Here $W=\sqrt{(q+P)^2}$ is the invariant mass of the final hadronic system
and the $CM$ label denotes the center-of-momentum frame.

\begin{figure*}
\includegraphics[bb=2cm 4cm 22cm 24cm, scale=0.4]{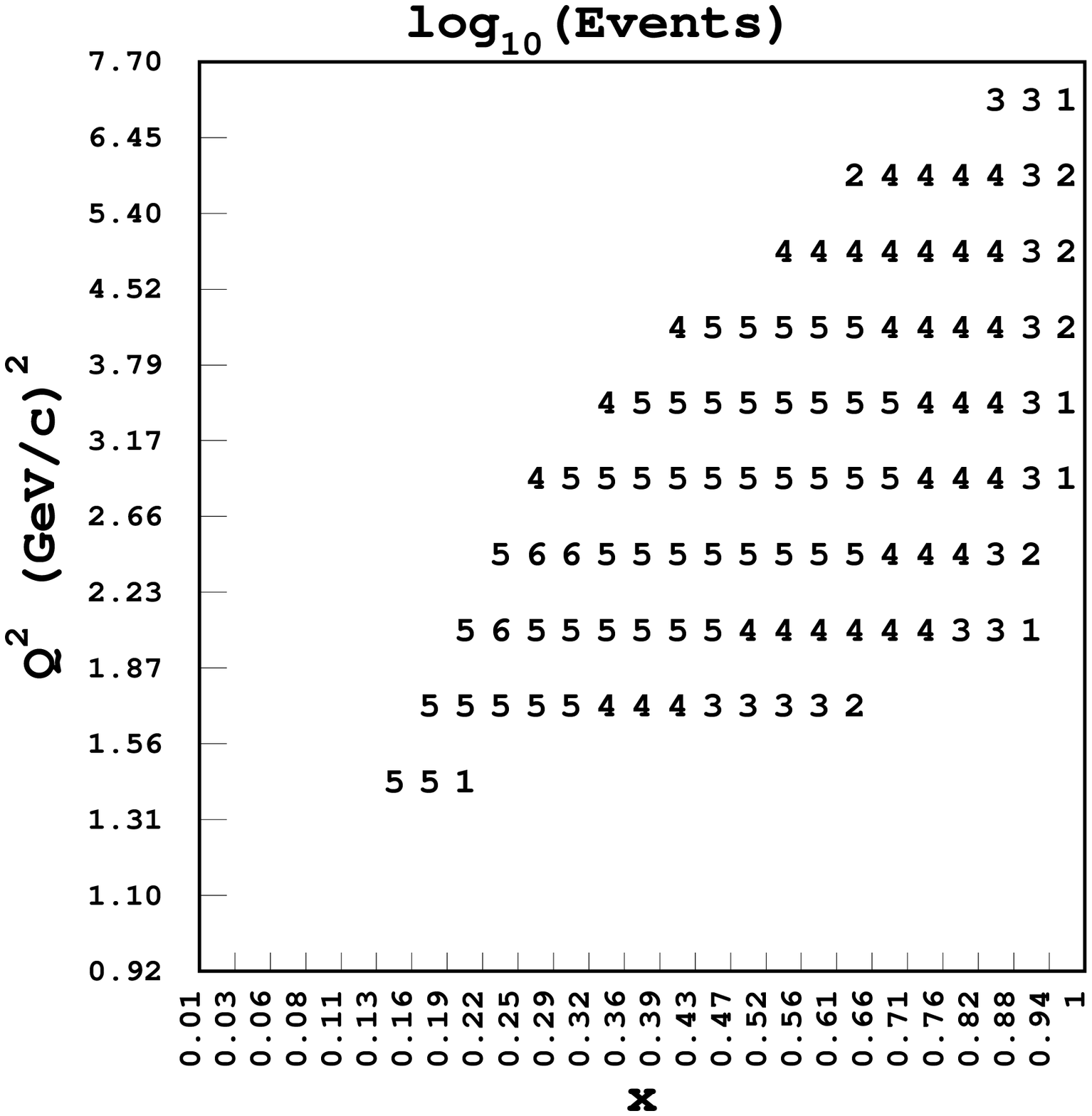}~%
\includegraphics[bb=2cm 4cm 22cm 24cm, scale=0.4]{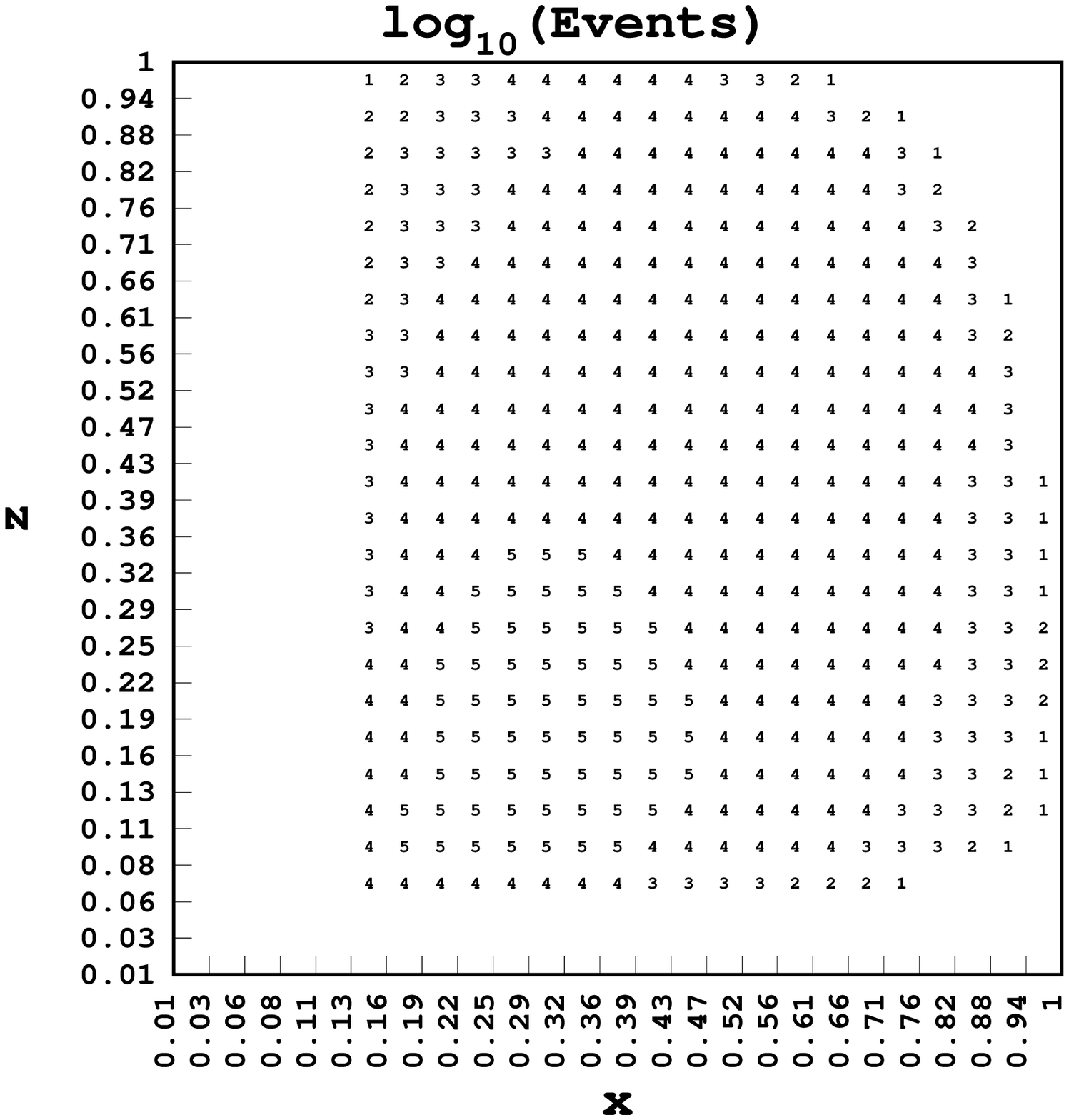}
\includegraphics[bb=2cm 4cm 22cm 24cm, scale=0.4]{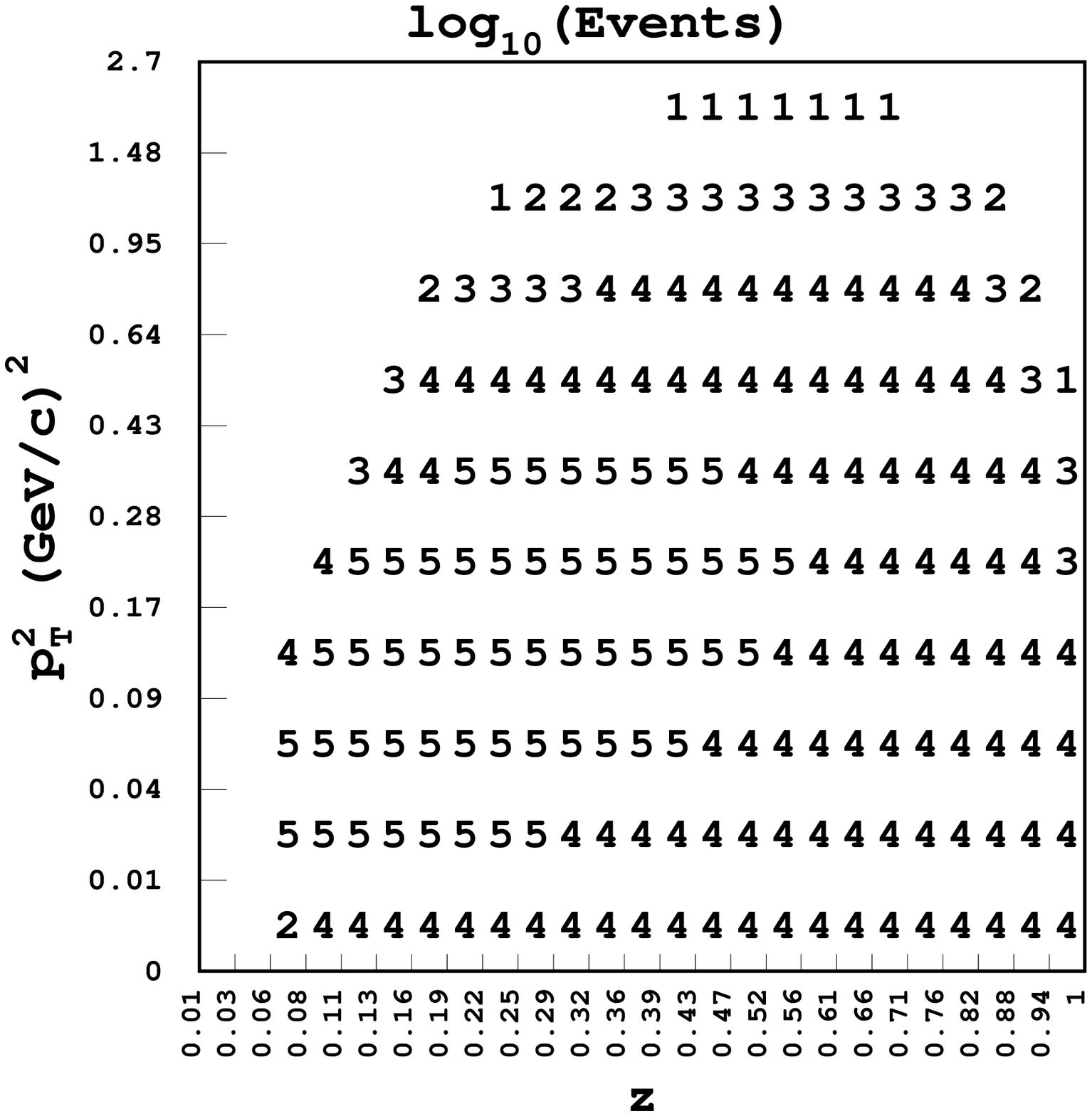}~%
\includegraphics[bb=2cm 4cm 22cm 24cm, scale=0.4]{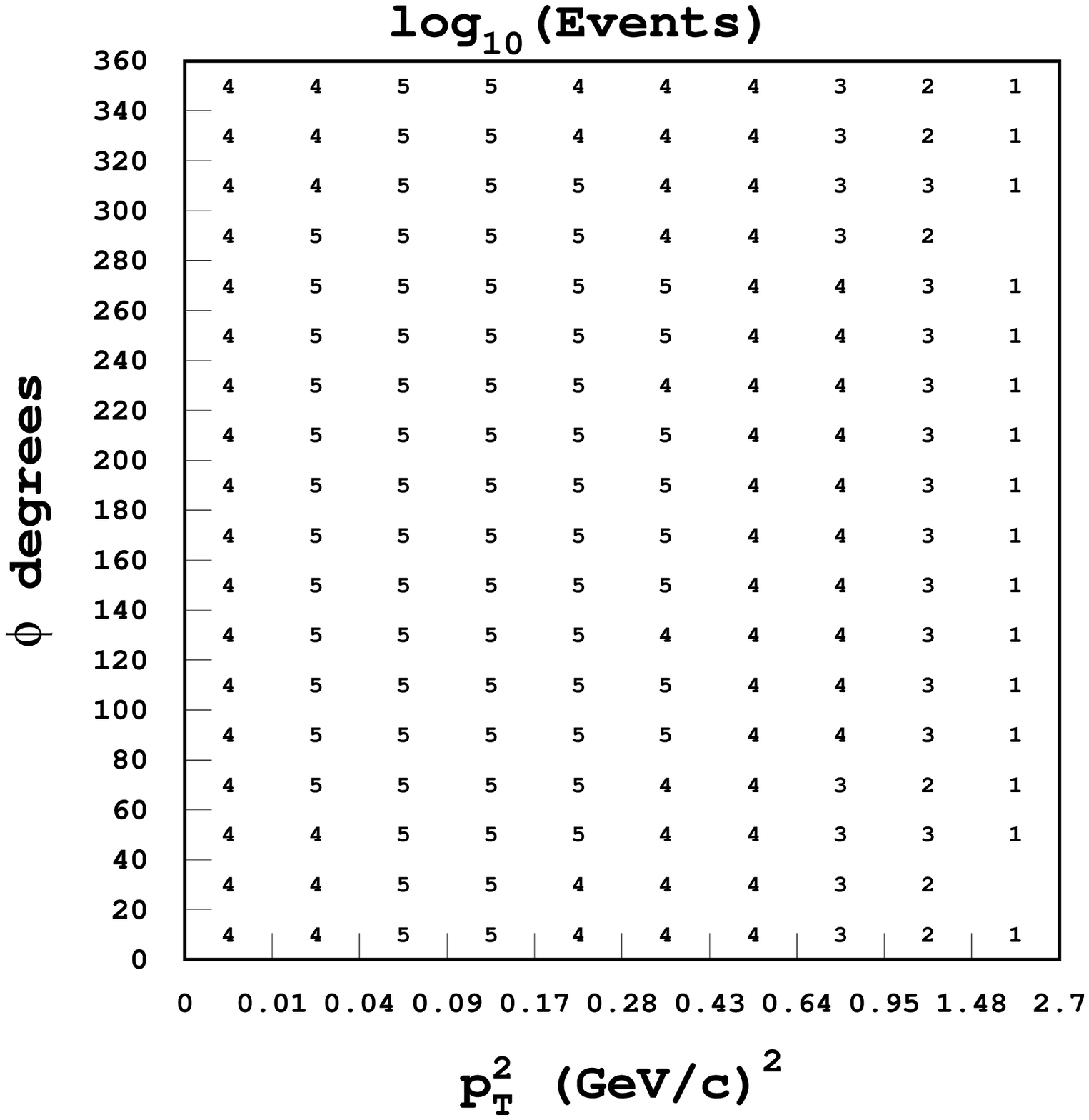}
\caption{\label{fig:kinreg} Kinematical regions covered by the present experiment
in different independent variables. Numbers on the plots give base-10 logarithm
of the number of events.}
\end{figure*}

The CEBAF Large Acceptance Spectrometer (CLAS) in Hall B at Jefferson Lab 
allows us to study the five-fold differential
semi-inclusive cross section over a large range of
four-momentum transfer $Q^2$ from 1.4 to 7 (GeV/c)$^2$
and Bjorken $x$ from 0.15 to 1 (see Fig.~\ref{fig:kinreg}).
CLAS enables us to measure distributions of the outgoing meson
($z$ from 0.07 to 1 and $p_T^2$ from 0.005 to 1.5 (GeV/c)$^2$),
in particular, full coverage in the azimuthal distributions that is very important.
However, the covered kinematical interval is not rectangular
in all five dimensions leading to a shrinkage of the four-dimensional acceptance
when one of variables reaches its limits.
The CLAS detector also has good particle identification
capabilities, resulting in a clean selection of pions for this analysis.

\section{\label{sec:kinem} Formalism and Theoretical Expectations}
The unpolarized semi-inclusive cross section can be written in terms
of four independent Lorentz-invariant structure functions\cite{Mulders}:
\begin{eqnarray}\label{eq:cs}
&& \frac{d^5 \sigma}{dx dQ^2 dz dp_T^2 d\phi} =
\frac{2\pi\alpha^2}{x Q^4}\frac{E_h}{|p_\parallel|} \zeta
\Biggl [
\epsilon \mathcal{H}_1 + \mathcal{H}_2 ~~~~~~~~~\\ \nonumber
&& + 2 (2-y)\sqrt{\frac{\kappa}{\zeta}} \cos{\phi} \mathcal{H}_3
+ 2 \kappa \cos{2\phi} \mathcal{H}_4
\Biggr ] ~,
\end{eqnarray}
\noindent where inelasticity $y=\nu/E_0$, $\gamma=\frac{2Mx}{\sqrt{Q^2}}$, $\zeta=1-y-\frac{1}{4}\gamma^2 y^2$,
$\epsilon=\frac{x y^2}{\zeta}$, $\kappa=\frac{1}{1+\gamma^2}$
and $\mathcal{H}_i=\mathcal{H}_i(x,z,Q^2,p_T^2)$. 
In contrast to Ref.~\cite{Mulders}, we absorbed the $\sqrt{p_T^2/Q^2}$ and $p_T^2/Q^2$ coefficients
in front of the structure functions $\mathcal{H}_3$ and $\mathcal{H}_4$, respectively,
into the structure function definition to let $\mathcal{H}_4$ reflect the recently identified
leading twist contribution by Boer and Mulders~\cite{Boer}. Both structure functions
include also an additional factor ${1\over 2}$ to simplify relation with the azimuthal
moments.

In order to disentangle all four structure functions $\mathcal{H}_i$, one has to
measure the complete five-fold differential semi-inclusive cross section at a few different
beam energies (as proposed in Ref.~\cite{RLT_proposal}).
In the present work we have data for only a single beam energy,  and we relied upon
the separation between the longitudinal ($\sigma_L \sim \mathcal{H}_2/\kappa-2x\mathcal{H}_1$)
and transverse ($\sigma_T \sim 2x\mathcal{H}_1$) cross sections performed in Ref.~\cite{Bebek-R}
and found to be compatible with $R=0.12 \pm 0.06$ (the weighted average over proton and deuteron data):
\begin{equation}\label{eq:rlt_h}
\frac{\mathcal{H}_1}{\mathcal{H}_2}
=\frac{1}{2\kappa x} \frac{\sigma_T}{\sigma_T+\sigma_L}
=\frac{1}{2\kappa x} \frac{1}{1+R} ~~,
\end{equation}
\noindent where $R=\sigma_L/\sigma_T$ is the longitudinal to transverse cross section ratio.

The azimuthal moments can be expressed in
terms of the structure functions $\mathcal{H}_i$ as follows:
\begin{eqnarray}\label{eq:asym}
&& \langle\cos{\phi}\rangle= (2-y)\sqrt{\frac{\kappa}{\zeta}} \frac{\mathcal{H}_3}{\mathcal{H}_2 + \epsilon\mathcal{H}_1} ~, ~~~~~~\\ \nonumber
&& \langle\cos{2\phi}\rangle= \kappa  \frac{\mathcal{H}_4}{\mathcal{H}_2 + \epsilon\mathcal{H}_1} ~~.
\end{eqnarray}
\noindent These relations allow us to extract azimuthal moments from the data in order to
determine $\mathcal{H}_3$ and $\mathcal{H}_4$.

The SIDIS cross section integrated over $\phi$ and $p_T^2$ is given by~\cite{Chay,Mulders}:
\begin{eqnarray}\label{eq:cs_dis}
&& \frac{d^3 \sigma}{dx dQ^2 dz} = 
\frac{4\pi\alpha^2}{x Q^4}
\Biggl [
x y^2 H_1(x,z,Q^2) + \\ \nonumber
&& + \Bigl(1-y-\frac{Mxy}{2E_0}\Bigr) H_2(x,z,Q^2)
\Biggr ] ~~~,
\end{eqnarray}
\noindent where roman $H_i$ are defined as calligraphic $\mathcal{H}_i$ structure functions
integrated over $p_T^2$.
In the parton model, the initial momentum of the struck quark is given by the proton momentum
multiplied by the light-cone fraction $x$. If we instead consider the momentum carried
by the struck quark after absorption of the virtual photon, then $z$ represents
the light-cone fraction of the momentum taken away by the produced hadron.
In the region of forward going hadron (frame dependent) this cross section can be evaluated
as the convolution of the parton density function $f(x,Q^2)$ obtained in inclusive processes
and the parton fragmentation function $D^h(z,Q^2)$ measured in $e^+e^-$ collisions:
\begin{equation}\label{eq:cur_fra}
H_2(x,z,Q^2)=\sum_i e_i^2 x f_i(x,Q^2) \otimes D_i^h(z,Q^2) ~,
\end{equation}
\noindent where the sum runs over quark flavors $i$ and $e_i$ is the charge of $i$th flavor quark.
Instead, in the region of backward going hadron (frame dependent) the cross section is
proportional to the fracture function~\cite{Trentadue} $M^h(x,z,Q^2)$ uniquely defined
in the semi-inclusive processes:
\begin{equation}\label{eq:tar_fra}
H_2(x,z,Q^2)=\sum_i e_i^2 x (1-x) M_i^h(x,z,Q^2) ~.
\end{equation}
\noindent The separation between the two processes is frame dependent
and can only be studied phenomenologically.

Values of the parton density function $f(x,Q^2)$ and the parton fragmentation function $D^h(z,Q^2)$
can be obtained in pQCD inspired world data fits e.g. in Refs.~\cite{CTEQ,GRV,MRST} and \cite{Kretzer,KKP,AKK}, respectively,
but only Ref.\cite{Kretzer} allows for hadron charge separation.
The fracture function is only studied for proton and neutron production~\cite{diff_proton,diff_neutron} and $\pi^+$
fracture function is completely unknown.

In practice the measured cross section also depends on the transverse
momentum $p_T$ of the hadron. The intrinsic motion of partons in the proton (Cahn effect~\cite{Cahn})
leads to an exponential $p_T^2$-behavior of the structure function $\mathcal{H}_2$:
\begin{equation}\label{eq:exp_pt2}
\mathcal{H}_2(x,z,Q^2,p_T^2)=
\frac{H_2(x,Q^2,z)}{\pi\langle p_T^2 \rangle}\exp{\Bigl[-p_T^2/\langle p_T^2 \rangle\Bigr]}\,.
\end{equation}
\noindent The mean squared transverse momentum in the naive parton model
is given by the sum of two terms~\cite{Ans1,Ans2,Barone2}:
\begin{equation}\label{eq:mean_pt2}
\langle p_T^2 \rangle = b^2 + a^2 z^2 ~,
\end{equation}
where $a^2$  is the mean squared intrinsic transverse momentum of the partons,
$a^2z^2$ is the mean squared parton transverse momentum transferred to the hadron, 
and $b^2$ is the mean squared transverse momentum acquired during fragmentation.

The $\phi$-dependent terms in Eq.~\ref{eq:cs} also appear in NLO pQCD
because the radiation of hard gluons leads to an azimuthal variation~\cite{Chay}.
However, this effect is expected to be important at energies higher
than that of the present experiment~\cite{Barone2}.
This is because the transverse momentum generated by the hard gluon
is generally larger than that accessible in our experiment.
In our energy range, the main contributions
to the $\phi$-dependence of the cross sections are expected to be 
the Cahn and Berger~\cite{Berger} effects for
$\mathcal{H}_3$ and the Boer-Mulders function~\cite{Boer} for $\mathcal{H}_4$
(see also Refs.~\cite{Zhang,Oganesian,Gamberg2,Gamberg1}).
The Cahn effect arises from the simple kinematics of partons with transverse momentum
and can be calculated explicitly in the limits $Q^2 \to \infty$ and $z \to 1$
(see Refs.~\cite{Ans1,Ans2,Barone2,DiSalvo}).

The Berger effect is the exclusive production of a single pion from a free, struck
quark that radiates a gluon, produces a $q\bar q$ pair, and recombines with the $\bar q$.
The formation of this pion through one-gluon exchange yields a $\cos\phi$ dependence
proportional to the hadron wave function.
Since such production mechanism does not require any initial transverse momentum
of struck parton it is completely orthogonal to the Cahn effect.

Explicitly neglecting intrinsic parton transverse momentum one has~\cite{Brandenbug}:
\begin{equation}
\frac{\mathcal{H}_3}{\mathcal{H}_2+\epsilon\mathcal{H}_1}=
\frac{zI_1(I_2-\frac{p_T^2}{Q^2}I_1)}{\eta I_2^2 + (4z^2+\eta\frac{p_T^2}{Q^2})\frac{p_T^2}{Q^2}I_1^2}
\sqrt{\frac{p_T^2}{Q^2}}
\end{equation}
\noindent and
\begin{equation}
\frac{\mathcal{H}_4}{\mathcal{H}_2+\epsilon\mathcal{H}_1}=
-\frac{I_1 I_2}{\eta I_2^2 + (4z^2+\eta\frac{p_T^2}{Q^2})\frac{p_T^2}{Q^2}I_1^2}
\frac{p_T^2}{Q^2} ~~~.
\end{equation}
\noindent Here we defined
\begin{equation}
I_1=z\int_0^1 d\xi \frac{\psi(\xi)}{z-\xi(z^2-\frac{p_T^2}{Q^2})}
\end{equation}
\noindent and
\begin{equation}
I_2=\int_0^1 d\xi\frac{\psi(\xi)}{1-\xi}-z^2 I_1~~~,
\end{equation}
\noindent with $\psi(\xi)$ being the pion wave function and $\eta=1+\epsilon/2x$.

The contribution of the Boer-Mulders function gives the probability to find a transversely
polarized quark in the unpolarized proton.
Explicitly in LO pQCD and $p_T^2/Q^2 \to 0$ one has~\cite{Boer}:
\begin{equation}
\frac{\mathcal{H}_4}{\mathcal{H}_2+\epsilon\mathcal{H}_1}=
\frac{1-y}{1+(1-y)^2} \frac{p_T^2}{M m_h}
\frac{8}{\kappa}\frac{\sum_i e_i^2 x h_i^\perp(x) H_i^{\perp h}(z)}{\sum_i e_i^2 x f_i(x) D_i^h(z)} ~~~,
\end{equation}
\noindent where $m_h$ is the mass of the detected hadron, $h_i^\perp(x)$ is
the momenum distribution of transversely polarized quarks in the unpolarized proton (Boer-Mulders function)
and $H_i^{\perp h}(z)$ is the Collins fragmentation function~\cite{Collins_hperp} describing
fragmentation of a transversely polarized quark into a polarized hadron.
The Collins fragmentation function was parametrized using $e^+e^-$ data in Ref.~\cite{Anselmino_collins}.

These three main effects predict different kinematic dependencies.
For example, the contribution of the Boer-Mulders function in $\mathcal{H}_4$ is of leading order,
and therefore should scale with $Q^2$.  On the other hand, both the Cahn and Berger
effects have a non-perturbative origin and should decrease with rising $Q^2$.
The Cahn and Berger effects have opposite signs, but both increase in magnitude with $z$.
The Berger effect should also increase in magnitude with $x$
as the exclusive limit is approached, whereas the Cahn effect does not have any $x$ dependence.
To distinguish among these physical effects, one needs to perform a complete
study of all kinematic dependencies in the data.

\section{\label{sec:analysis} Data Analysis}
The data were collected at Jefferson Lab in Hall B with CLAS~\cite{CLAS}
using a $0.354$~g/cm$^2$ liquid-hydrogen target and a 5.75-GeV electron beam during
the period October 2001 to January 2002.
The average luminosity was 10$^{34}$ cm$^{-2}$s$^{-1}$.
CLAS is based on a six-sector
torus magnet with its field pointing  azimuthally around the beam
direction. 
The torus polarity was set to bend negatively charged particles toward the beam line.
The sectors delimited by the magnet coils are individually
instrumented to form six independent magnetic spectrometers. The particle
detection system includes drift chambers (DC) for track
reconstruction~\cite{dc},
scintillation counters (SC) for time-of-flight measurements~\cite{sc},
Cherenkov counters (CC) for electron identification~\cite{cc}, and
electromagnetic calorimeters (EC) for
electron-pion separation~\cite{ec}.  
The CLAS can detect and identify charged particles with momenta
down to 0.2 GeV/c for polar angles between 8$^\circ$ and 142$^\circ$,
while the electron-pion separation is limited up to about 50$^\circ$ by the CC acceptance.
The total angular acceptance for electrons is about 1.5 sr.
The CLAS superconducting coils limit the acceptance 
for charged hadrons to  about $80$\% at $\theta=90^{\circ}$
and about $50$\% at $\theta=20^{\circ}$ (forward angles).

The electron momentum resolution is a function of the scattered electron angle
and varies from 0.5\% for $\theta \leq 30^{\circ}$ up to
1-2\% for $\theta > 30^{\circ}$. The angular resolution
is approximately constant, approaching 1~mrad for polar and 4~mrad
for azimuthal angles.  Therefore, the momentum transfer resolution
ranges from 0.2 to 0.5 \%. For the present experiment
the invariant mass of the struck proton ($W = \sqrt{(P+q)^2}$) has an
estimated resolution of 2.5 MeV for beam energies less than 3 GeV and
about 7 MeV for larger energies.
In order to study all possible multi-particle states, we set the data acquisition trigger
to require at least one electron candidate in any of
the sectors, where an electron candidate was defined as
the coincidence of a signal in the EC and Cherenkov modules
for any one of the sectors.


\subsection{\label{sec:gen} Generic procedures}
Both the $e^-$ and $\pi^+$ were detected within
the volume defined by fiducial cuts. These geometrical cuts
selected regions of uniform high efficiency
by removing areas near the detector boundaries and regions
corresponding to problematic SC counters or DC readout.
For electrons the fiducial volume limitations are mostly due to the Cherenkov counter, which is
necessary for electron identification, and the electromagnetic calorimeter, which is used in the trigger.
The CLAS Cherenkov counter's optics reduce significantly its azimuthal acceptance,
in particular in the region of small polar scattering angles, where the 
light collection mirrors are small.
Moreover, the Cherenkov counter extends only up to $\sim$50$^\circ$
in the polar scattering angle
of an inbending charged particle. The trigger threshold for the electromagnetic
calorimeter limits the lowest
electron momentum, which in our case was about 0.64 GeV/c.

CLAS achieves its best charged-particle acceptance for $\pi^+$, since
complete identification requires only information from the drift chambers and
the scintillation counters, which are limited in coverage only by the CLAS torus magnet's coils.
For the standard torus configuration, $\pi^+$ particles bend outward toward larger angles,
where the useful detector area between the coils is greater.

Small corrections to the momenta of the $e^-$ and $\pi^+$ were necessary because of
distortions in the drift chambers and magnetic fields not accounted for in the tracking routines.
Correction parameters were determined by minimizing the difference in the missing mass for 
$ep\rightarrow e p$ and $ep\rightarrow e \pi^+ n$  from known values (see Ref.~\cite{Kijun}).
The magnitude of these kinematic corrections was well below the CLAS resolution
leading to sub-percent changes in the measured cross section.

Events were selected by a coincidence of an electron and a $\pi^+$
whose identification criteria are described in the next section.
The trigger gate time in CLAS was 150 ns, but, due to the limited range of
particle momenta, the effective time window for a coincidence
was much smaller. This, and the relatively low beam current
in CLAS (about 7 nA),  ensured that accidental coincidences were negligible.

\subsection{\label{sec:pid} Particle Identification and Backgrounds}
The electron identification is based on combined information
from the CC, EC, DC and SC.
The fastest (as measured by the SC)
negatively charged (as determined from DC tracking) particle having
EC and CC hits is assumed to be an electron. However, the large rate of
negatively charged pions contaminates the sample of reconstructed electrons, in particular
in the region of low momenta and large polar scattering angles.
Moreover at lowest accessible polar scattering angles CC efficiency
is reduced due to geometrical constraints.
This contamination can be eliminated by using SC and DC information
to better correlate the particle track and the time of the SC hit with the 
CC signal~\cite{Osipenko_f2d}.
We estimated the electron efficiency after this process to be greater than 97\%
and the corresponding inefficiency was propagated to $e^-$ identification
systematic uncertainty. The inefficiency is maximum at the lowest $Q^2$.

\begin{figure}[htb]
\includegraphics[bb=5cm 5cm 16cm 24cm, scale=0.4]{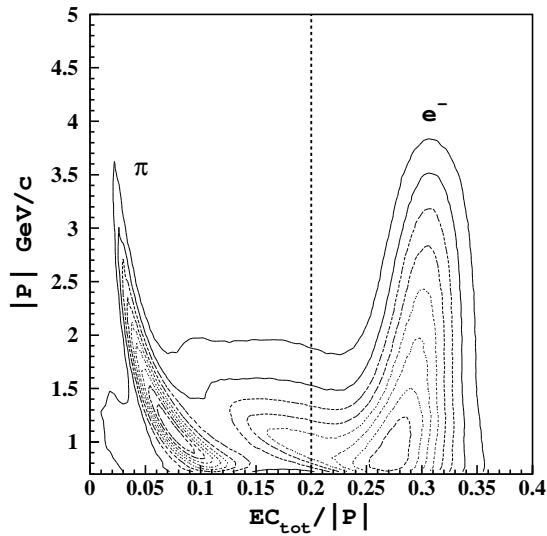}
\caption{\label{fig:el_pid}Contour plot of particle momentum $\textbf{p}$ from tracking versus particle
energy deposited in the calorimeter $EC_{\rm tot}$ normalized by $|\textbf{p}|$. Events on the left
correspond to pions and those on the right to electrons.
Only fiducial cuts were applied. The Cherenkov detector, providing the basic electron identification,
allows to identify clearly the electrons up to $|\textbf{p}| \approx 2.7$ GeV/c.
The dashed line shows the cut applied to the data to remove remaining pion contamination
at $|\textbf{p}| > 2.7$ GeV/c.}
\end{figure}

The CC becomes less efficient at distinguishing electrons from pions for 
momenta above the Cherenkov light threshold for pions
($|\textbf{p}| \approx 2.7$ GeV/c). However, in this kinematic region the EC signal can
be used to remove the remaining pion contamination. The minimum-ionizing
pion releases a nearly constant energy in the EC, independent of its momentum,
whereas an electron releases an almost constant energy fraction of about 30\%
in the EC.
Fig.~\ref{fig:el_pid} shows a contour plot of events with momentum $\textbf{p}$ determined from
the DC and total energy in the EC normalized by $|\textbf{p}|$.
Pion-electron separation, in this case, increases with particle momentum.

Pion identification is based on time of flight as measured with the SC and
momentum as measured with the DC. Since the distance between the target
and SC is independent of the scattering angle, the efficiency of pion identification
depends only on the pion momentum and therefore on $z$.
The time-of-flight resolution decreases with pion momentum leading to
larger pion identification inefficiency. A contribution proportional to
this inefficiency was added to $\pi^+$ identification systematic uncertainty.
Furthermore, the time-of-flight interval between different hadron species
decreases with hadron momentum resulting in larger contamination.

\begin{figure}[htb]
\includegraphics[bb=5cm 5cm 16cm 24cm, scale=0.4]{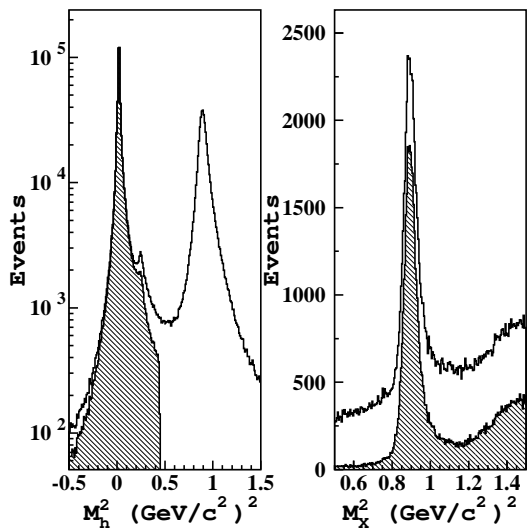}
\caption{\label{fig:pid}Measured squared mass of positive hadrons (left) and
squared missing mass for the $e^-h^+$-system
in the region of the $\pi^+n$ exclusive peak (right).
The shaded area indicates hadrons identified as pions.
The exclusive $\pi^+n$ was removed from the analysis by
a cut $M_X^2 > 1.08$ (GeV/c$^2$)$^2$.}
\end{figure}

Fig.~\ref{fig:pid} shows how effectively this procedure
removes the background
under the exclusive $\pi^+ n$ peak without a significant
loss of good events. The example of the exclusive $\pi^+ n$ peak is important because
these pions have large momenta,
which makes their separation by time of flight more difficult
than in the semi-inclusive case,
where the pions have slightly lower momenta.

%

A positively charged particle identified as a pion may in some cases be a positron from
$e^+e^-$ pair production. This background becomes important at
low momenta and at $\phi\approx 0$ or $\phi\approx 180^\circ$.
To remove this contamination we applied the cut
\begin{equation}
M^2(e^-h^+) > 0.012 \exp{\Bigl[-M_{TOF}^4/(2 \sigma_{TOF}^2)\Bigr]} ~,
\label{eq:d_pp1}
\end{equation}
\noindent where $M_{TOF}$ is the mass of the positive particle as measured by the TOF and
$M(e^-h^+)$ is the invariant mass of the measured system of two particles in GeV/c$^2$
(assuming them to be $e^+$ and $e^-$ and $\sigma_{TOF}=0.01$~(GeV/c$^2$)$^2$).
The remaining contribution from $e^+e^-$ pairs is negligible over the entire kinematic range
after the cut.

The electron, detected in coincidence with the pion, may be a secondary electron,
whereas the scattered electron is not observed. To remove this background
we measured lepton charge symmetric $e^+\pi^+$ coincidence cross section from the same data
and removed its contribution from our data. This contamination
is limited to a few lowest-$x$ points where it achieves 5\% at most.

%
%

Another source of contamination comes from $K^+$ production at high hadron
momenta. At low hadron momenta the TOF system is able to distinguish pions
from kaons, but above $|\textbf{p}_h| \approx 1.2$ GeV/c the peaks
of the two particles begin to mix. However,
large hadron momenta make two-kaon production less likely due to the correspondingly
high energy threshold, and therefore most of the background comes from single kaons
associated with $\Lambda$ and $\Sigma^0$ production.
In order to suppress the kaon contamination we applied two cuts:
a kinematical cut that removes $\Lambda$, $\Sigma^0$ and $\Lambda^*(1520)$,
and a TOF cut $M_h^2 < m_\pi^2 + 2\sigma_{M^2(TOF)}$ that suppresses low-momentum kaons.
The mass resolution of the TOF system was determined by fitting
the width of the pion peak, which yielded 
$\sigma_{M^2(TOF)}=0.022 |\textbf{p}_h| \exp{(0.6\sqrt{|\textbf{p}_h|})}$,
where $\textbf{p}_h$ is given in GeV/c and $\sigma_{M^2(TOF)}$ in (GeV/c$^2$)$^2$.
Corrections for the remaining kaons from semi-inclusive production
above the two kaon threshold were made using
the ratio of $K^+$ to $\pi^+$ semi-inclusive cross sections
obtained
from a pQCD-based Monte Carlo (MC) event generator (see the following section),
weighted with the kaon/pion rejection factor obtained from the simulation itself.
Kaons from the MC were propagated through the entire
chain of the reconstruction procedure exactly in the same way
as was done for pions, and the fraction $f(K^+)$ of kaons reconstructed
as pions was obtained. This number was normalized to the fraction $f(\pi^+)$
of simulated pions reconstructed by the procedure. This
kaon/pion rejection factor was
parameterized as a function of the hadron momentum.
The contribution from the $K^+$ background varied from 0 to 20\% with an average of 1\%,
and our procedure reduced the kaons by a factor of two at  2.3 GeV/c  with an increasing
reduction factor at lower hadron momenta.

\subsection{\label{sec:empty} Empty target contribution}
Empty target runs were analyzed in exactly in the same
way as the full target runs and subtracted from the data to eliminate
scattering from the target endcaps.
The total charge collected on the empty target is an order of magnitude
smaller than the one for the full target data.
In order to increase the statistics of the empty target distributions,
we made the assumption that the ratio of full to empty target event rates
factorizes as a function of all variables.
Thus one can obtain the ratio of empty to full target rates
(ranging from 0 to 18\% with an average value of 4.7\%)
for the five-fold differential cross section
as a product of one-fold differential ratios.
The contribution of empty target is typically smaller
than the total systematic uncertainty, reaching its maximum
at the two pion threshold.

\subsection{\label{sec:mc} Monte Carlo Simulations}
Detector efficiencies and acceptances were studied
with a standard CLAS simulation package GSIM~\cite{GSIM}.
The simulated data obtained from GSIM
can then be analyzed using the event reconstruction routine exactly in
the same way as the measured data. This allows  a complete determination
of the detector efficiency plus acceptance.

\begin{figure*}[ht]
\includegraphics[bb=2cm 6cm 22cm 24cm, scale=0.4]{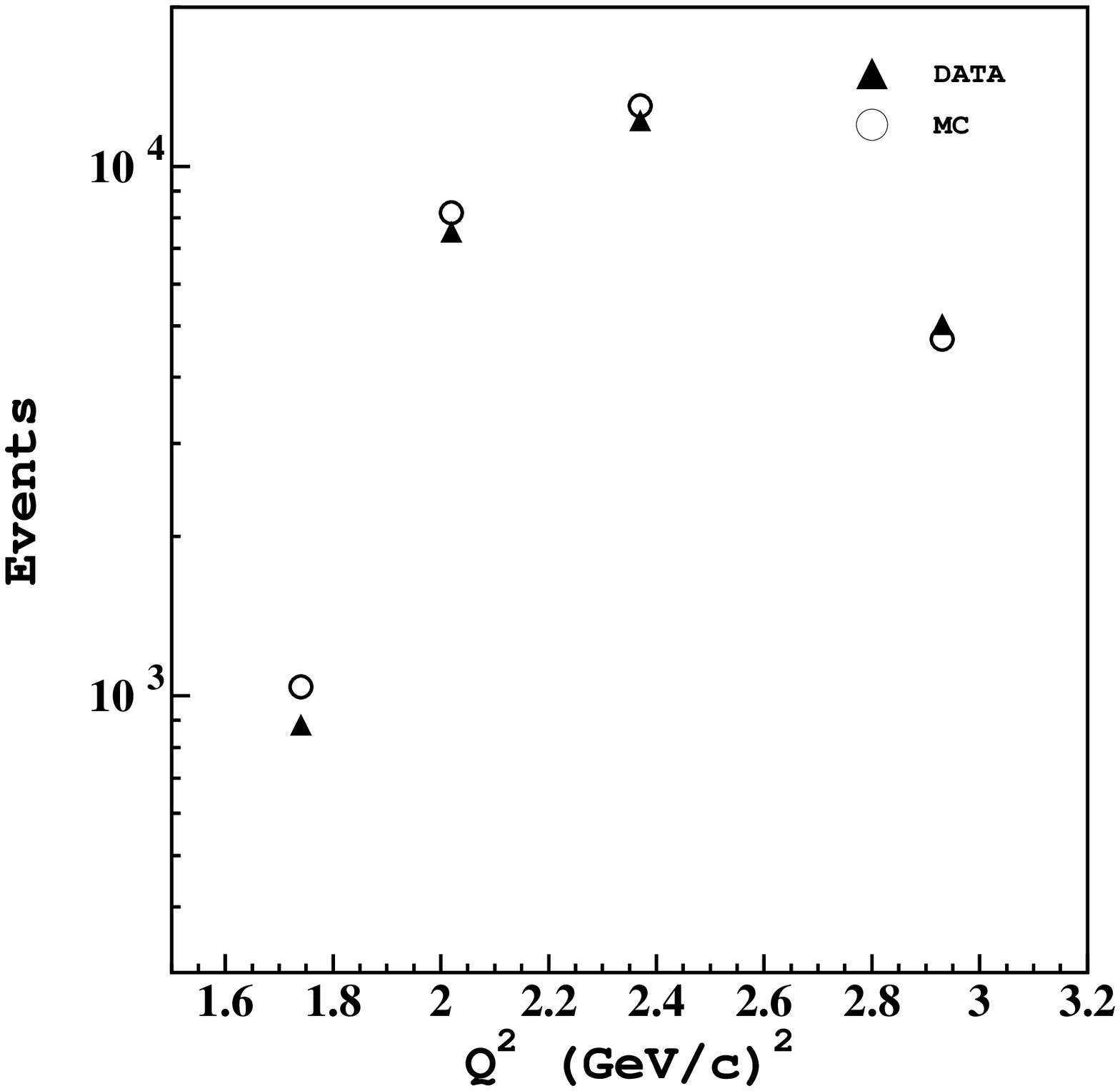}~%
\includegraphics[bb=2cm 6cm 22cm 24cm, scale=0.4]{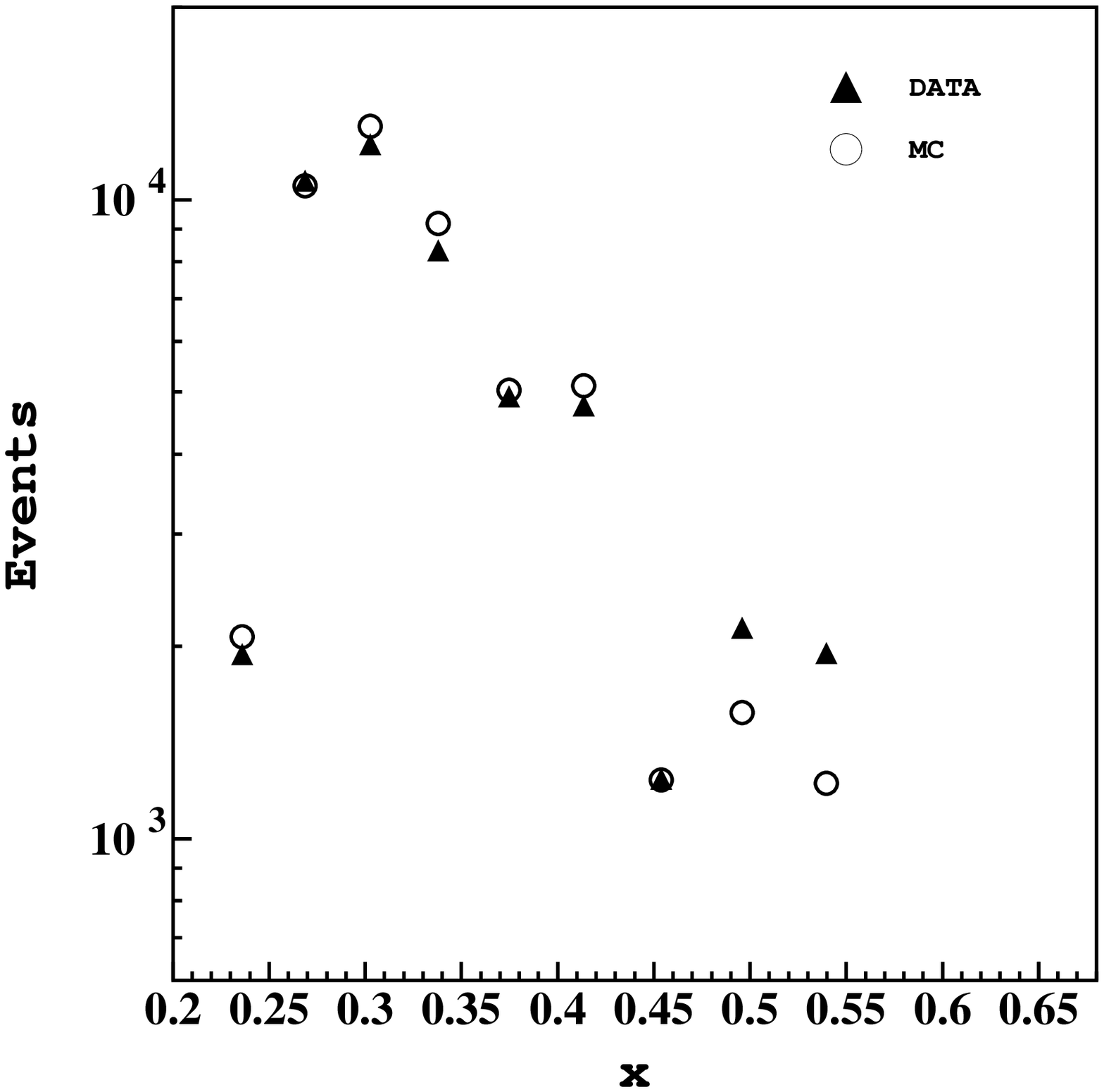}
\includegraphics[bb=2cm 6cm 22cm 24cm, scale=0.4]{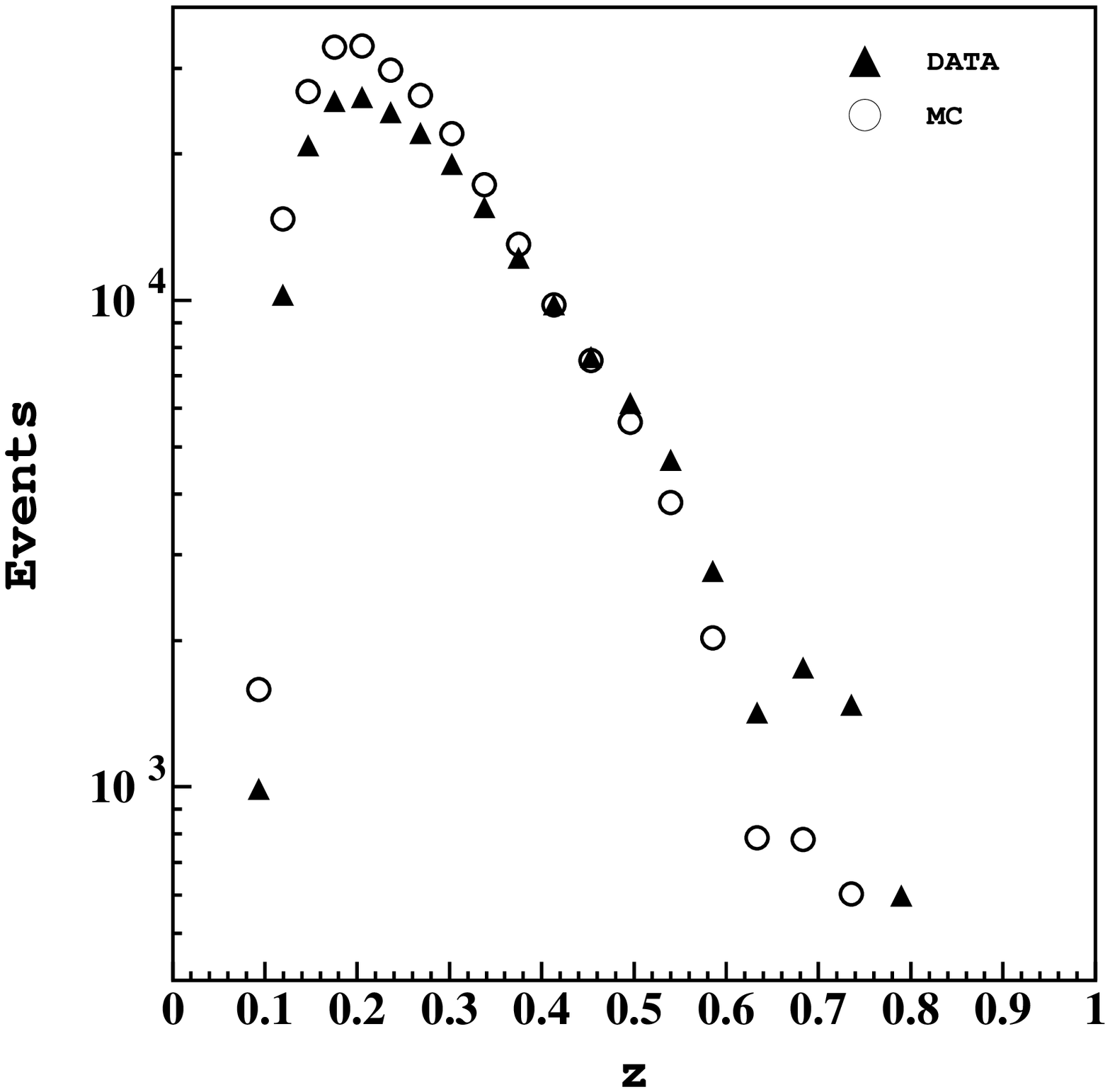}~%
\includegraphics[bb=2cm 6cm 22cm 24cm, scale=0.4]{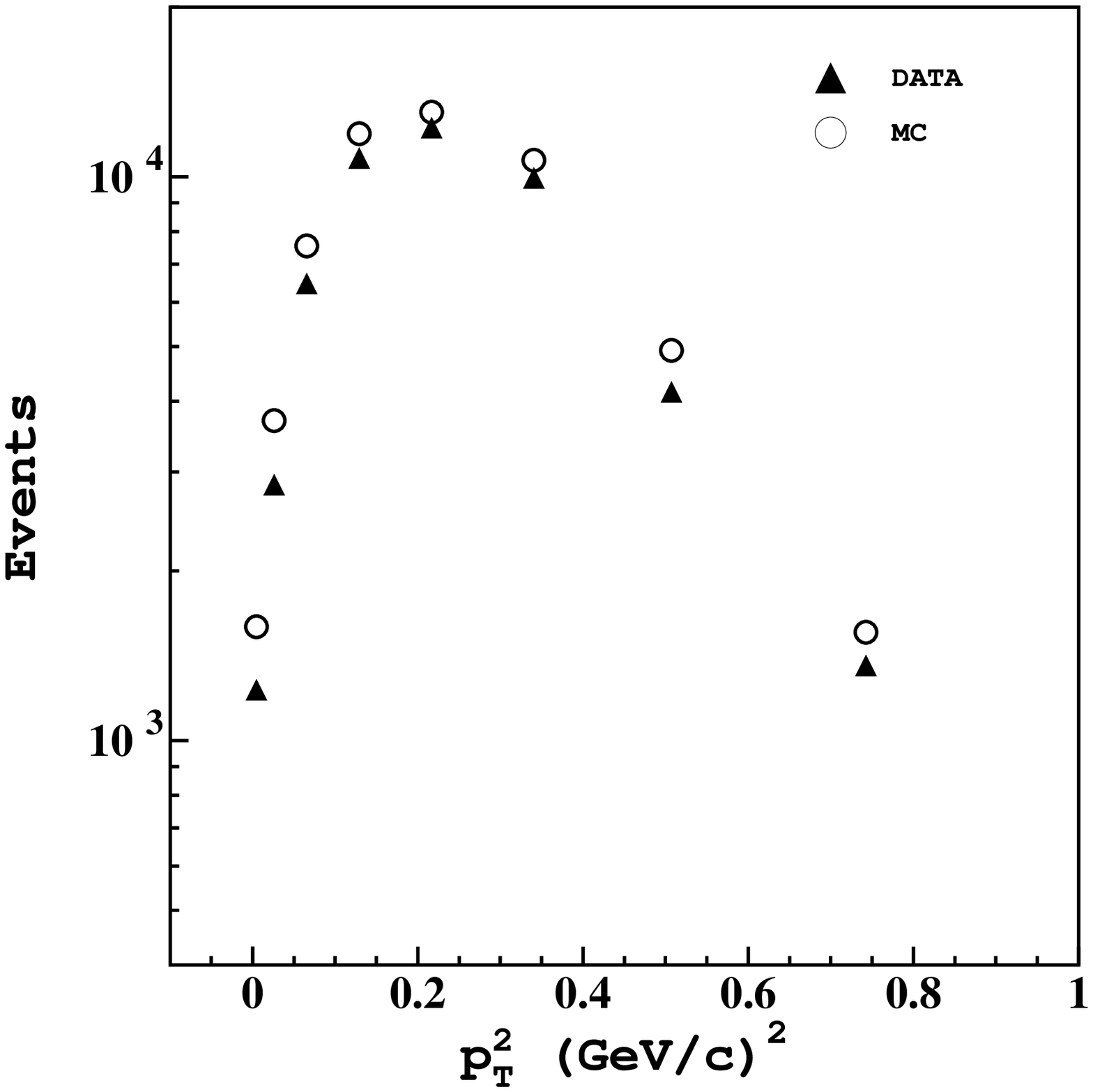}
\caption{\label{fig:mc_dat}Comparison of $e^-\pi^+$ coincidence data (full triangles)
and MC raw yields (open circles) as a function of one of the kinematic variables.
The other variables were kept fixed
at $Q^2=2.4$ (GeV/c)$^2$, $x=0.30$, $z=0.37$, $p_T^2=0.22$ (GeV/c)$^2$.
The MC simulation yield was normalized to the integrated luminosity of the
experiment. Error bars are statistical only and they are smaller than
the symbol size.}
\end{figure*}

The first step of the simulation is to generate $e^-\pi^+$ coincidence
events based on a pQCD-like SIDIS parameterization~\cite{plb2007}
at leading order for the semi-inclusive contribution and on the MAID2003 model~\cite{MAID}
extrapolated to the $W>2$ GeV/c$^2$ region with the parameterization from Ref.~\cite{Browman_prl35}
for the exclusive $\pi^+n$ reaction.
Distributions of counts from the experimental data and  GSIM simulations
are shown in Fig.~\ref{fig:mc_dat}. The same cuts are applied to both data and 
MC as described in the previous section.

The MC yield reproduces the shape of the experimental data fairly well.
In order to keep systematic uncertainties on the acceptance plus efficiency  
small (we estimated them to be 10\%)
we had to extract fully-differential cross sections in narrow kinematic bins.
Bins with combined acceptance and efficiency $< 0.1$ \%,
corresponding to 25\% filling per each dimension, were discarded.
The average value of combined acceptance and efficiency was about 25\%.

\begin{figure}[ht]
\includegraphics[bb=2cm 6cm 22cm 24cm, scale=0.4]{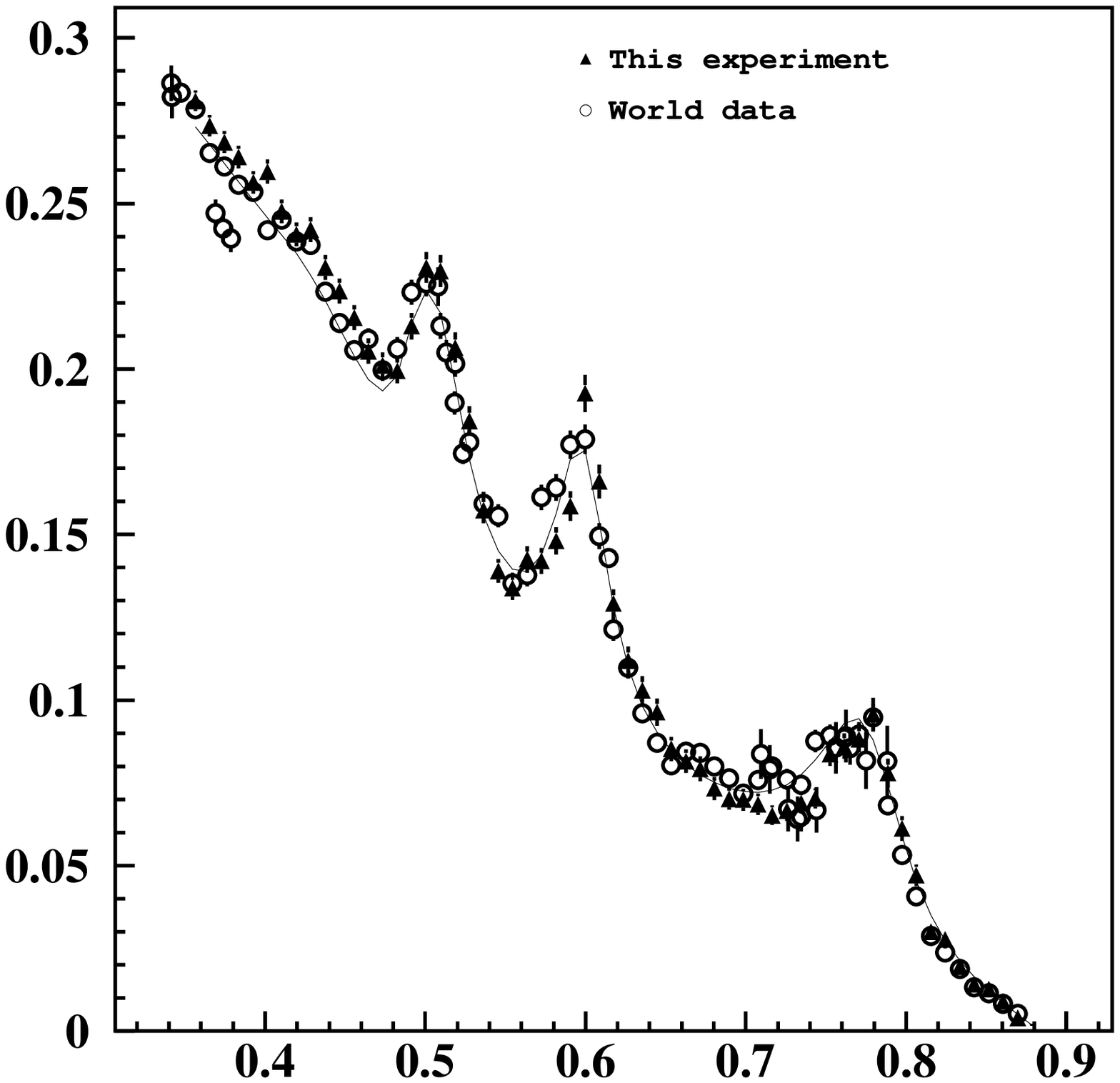}
\caption{\label{fig:f2p}Inclusive structure function $F_2$ at $Q^2=2$ (GeV/c)$^2$
extracted from the present experiment (full triangles) in comparison with
previous world data (open circles) from Ref.~\cite{Osipenko_f2p} and references therein.
The curve is a parameterization from Ref.~\cite{Osipenko_f2p}.
The error bars are statistical only and an overall
systematic uncertainty for the present experiment of the order or 5\% is estimated.}
\end{figure}

To test Monte Carlo simulations of electrons we extracted the inclusive structure function $F_2$
and compared it to the world data in our kinematic range. An example
of this comparison at $Q^2=2$ (GeV/c)$^2$ is shown in Fig.~\ref{fig:f2p}.

\begin{figure}[ht]
\includegraphics[bb=2cm 6cm 22cm 24cm, scale=0.4]{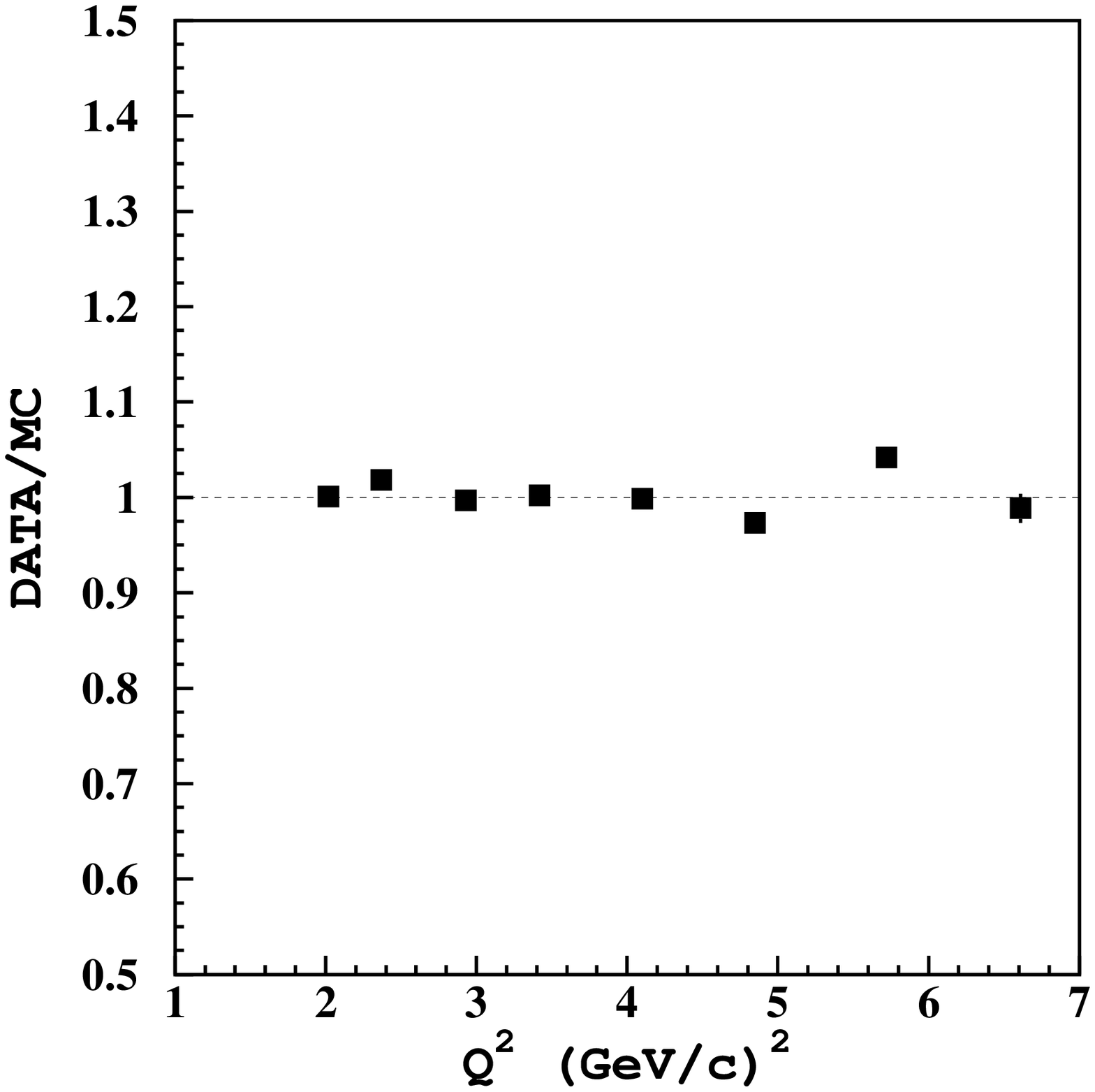}
\caption{\label{fig:el_ratio}Ratio of $e^-p$ coincidence events from
the data and GSIM Monte Carlo simulations. The simulation event generator was based
on proton form-factors from Ref.~\cite{AMT}.
The coincidences were selected by the following set of cuts:
$|W^2-M^2|<0.2$ (GeV/c$^2$)$^2$, $|M_X^2|<0.01$ (GeV/c$^2$)$^2$ and $||\phi_h-\phi_e|-\pi|<4$ degrees.
The error bars are statistical only and an overall
systematic uncertainty of the order or 10\% is estimated.
Error bars are smaller than the symbol size.}
\end{figure}

For positively charged hadrons we tested Monte Carlo simulations by extracting
the elastic scattering cross section measured in electron-proton coincidences.
The normalized event yield was compared to normalized GSIM simulation yield,
based on form-factors from Ref.~\cite{AMT}.
The obtained ratio, shown in Fig.~\ref{fig:el_ratio}, is in good agreement
with unity in the central region of $Q^2$, but rises at large $Q^2$ due to
unresolved inelastic contamination.

Furthermore, the efficiency of $\pi^+$ production reconstruction
in the present data set was tested in the measurement
of the exclusive pion production published in Ref.~\cite{Kijun}.

\subsection{\label{sec:bins} Binning}
The data were divided into kinematic bins as follows:
\begin{itemize}
\item $Q^2$ - 10 logarithmic bins (with centers at)
1.31-1.56 (1.49), 1.56-1.87 (1.74), 1.87-2.23 (2.02), 2.23-2.66 (2.37), 2.66-3.17 (2.93),
3.17-3.79 (3.42), 3.79-4.52 (4.1), 4.52-5.4 (4.85), 5.4-6.45 (5.72), 6.45-7.7 (6.61) (GeV/c)$^2$;
\item $x$ - 25 logarithmic bins in the interval from 0.01 to 1;
\item $z$ - 25 logarithmic bins in the interval from 0.01 to 1;
\item $p_T$ - 10 logarithmic bins (with centers at):
0-0.1 (0.07), 0.1-0.2 (0.16), 0.2-0.3 (0.26), 0.3-0.41 (0.36), 0.41-0.53 (0.47),
0.53-0.65 (0.58), 0.65-0.8 (0.71), 0.8-0.97 (0.86), 0.97-1.22 (1.04), 1.22-1.64 (1.25) GeV/c;
\item $\phi$ - 18 linear bins in the interval from 0 to 360$^\circ$.
\end{itemize}
The bin sizes were chosen large enough to accommodate CLAS angular and momentum
resolutions reducing bin migrations, but they were small enough to avoid reconstruction
efficiency model dependence. Indeed the average bin migration is about 30\%,
consistent with expectation for bin size close to $1\sigma$ of detector resolution.
The centers of the $Q^2$ and $p_T$ bins coincide with the mean values
of these variables in the raw data after all cuts but before acceptance corrections.

The general rectangular grid described above was used to sort the data.
But the measured kinematic volume is not rectangular. Hence a fraction
of bins in one dimension can result empty depending on the values of the other four variables.
This is in particular the case when one of variables is close to
the kinematic limits of the accessible region.

\subsection{\label{sec:cs} Five-fold differential Cross Section}
The five-fold differential semi-inclusive cross section was extracted
for each kinematic bin from the number of measured events $N_{dat}$ according to
the relation:
\begin{eqnarray}\label{eq:d_cs1}
&& \frac{d^5\sigma}{dxdQ^2dzdp_T^2d\phi}=
\frac{G_{dat}}{\Delta x\Delta Q^2\Delta z\Delta p_T^2 \Delta\phi} \\ \nonumber
&& \frac{N_{dat}(x,Q^2,z,p_T^2,\phi)}{F_{eff/acc}(x,Q^2,z,p_T^2,\phi)} ~,
\end{eqnarray}
\noindent with data inverse luminosity given by
\begin{equation}
G_{dat}=\frac{1}{\mathcal{L}}=\frac{1}{\rho \frac{N_A}{M_A} L Q_{FC}} ~,
\end{equation}
\noindent $\rho=0.0708$ g/cm$^3$ is the liquid-hydrogen target density, 
$L=5$ cm is the target length
and $Q_{FC}$ is the total charge collected in the Faraday Cup (FC), corrected for dead
time.
$F_{eff/acc}(x,Q^2,z,p_T^2,\phi)$ is the acceptance/efficiency correction
obtained with Monte Carlo simulations:
\begin{eqnarray}\label{eq:d_e_s2}
&& F_{eff/acc}(x,Q^2,z,p_T^2,\phi)=
\frac{G_{sim}}{\Delta x\Delta Q^2\Delta z\Delta p_T^2 \Delta\phi} \\ \nonumber
&& \frac{N_{rec}(x,Q^2,z,p_T^2,\phi)}{\sigma_M(x,Q^2,z,p_T^2,\phi)} ~,
\end{eqnarray}
\noindent with its normalization factor given by
\begin{equation}
G_{sim}=\frac{\int_{\tau} d\sigma_M}{N_{tot}} ~.
\end{equation}
\noindent Here $N_{rec}(x,Q^2,z,p_T^2,\phi)$ is the number of Monte Carlo
events reconstructed in the current bin, $\sigma_M(x,Q^2,z,p_T^2,\phi)$ is
the cross section model used in the event generator, $\tau$ is the complete
phase space volume of event generation and
$N_{tot}$ is the total number of generated events.

The final cross sections were corrected for radiative effects
using the analytic calculations described in Refs.~\cite{plb2007,Akushevich}
implemented in the Monte Carlo generator.
It includes both radiative corrections
to the SIDIS spectrum and the radiative tail from exclusive $\pi^+n$ production.
The average contribution of radiative corrections is about 6\% with
largest contribution close to the two pion threshold. The magnitude
of radiative corrections increases with $z$ and $p_T^2$.

\subsection{\label{sec:asym} Azimuthal dependence}
A separation of the constant, $\cos\phi$ and $\cos 2\phi$ terms 
in Eq.~\ref{eq:cs} has been performed  using
two methods,
either a fit to the $\phi$-distributions or
an event-by-event determination of azimuthal moments.
Both methods should give compatible results.
By studying the two methods in detail we concluded that both
give unreliable results if the $\phi$-distribution contains regions of
poor detector acceptance.
Therefore we excluded kinematic points where the $\phi$-acceptance was inadequate.
This reduced significantly the kinematic range of the extracted moments.
Nevertheless, the kinematic bins
with incomplete $\phi$-coverage can still be
used in a multidimensional fit exploiting continuity
in the other variables.

In the first method we fit the $\phi$-distribution 
to the function $\sigma_0(1+2B\cos{\phi}+2C\cos{2\phi})$ using MINUIT~\cite{cernlib}
and extracted the coefficients $\sigma_0$, $B$ and $C$ and  their statistical uncertainties. 
These coefficients give the $\phi$-integrated cross section,
$\langle\cos{\phi}\rangle$ and $\langle\cos{2\phi}\rangle$,
respectively.

\begin{figure}[htb]
\begin{center}
\includegraphics[bb=2cm 4cm 22cm 24cm, scale=0.4]{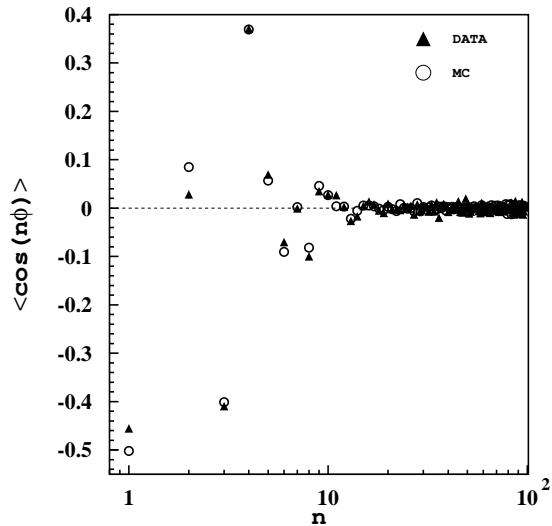}
\caption{\label{fig:fourier_spectra} Extracted Fourier components
of the raw data (full triangles) and the Monte Carlo (open circles) yields.
Error bars are statistical only and they are smaller than the symbol size.}
\end{center}
\end{figure}

The second method of moments was used in a previous CLAS paper~\cite{Harut},
but due to the strong effect of acceptance on even moments, we
developed the necessary corrections described below.
The fully differential cross section can be written as:
\begin{equation}\label{eq:fourier_sig}
\sigma=V_0+V_1\cos{\phi}+V_2\cos{2\phi} ~.
\end{equation}
\noindent The acceptance/efficiency correction can be expanded in a Fourier
series in $\phi$ as:
\begin{equation}\label{eq:fourier_acc}
F_{eff/acc}=\frac{A_0}{2}+\sum_{n=1}^{\infty}A_n\cos{n\phi}+
\sum_{n=1}^{\infty}B_n\sin{n\phi} ~.
\end{equation}
\noindent The coefficients $B_n$ are fairly small in CLAS.
Fourier spectrum of the raw data and MC yields
is shown in Fig.~\ref{fig:fourier_spectra}.

Let us define the Fourier coefficients of raw yield
(before acceptance/efficiency correction)
\begin{equation}\label{eq:fourier_int}
Y_n=\frac{1}{\pi}\int_0^{2\pi} \tilde{\sigma}(\phi)\cos{n\phi}d\phi ~,
\end{equation}
\noindent where $\tilde{\sigma}=F_{eff/acc}\sigma$ is 
the cross section distorted by
the acceptance/efficiency correction $F_{eff/acc}$.
Then combining Eqs.~\ref{eq:fourier_sig}, \ref{eq:fourier_acc} and \ref{eq:fourier_int}
we obtain an infinite series of linear equations relating
Fourier coefficients of the raw yield and the physical cross section:
\begin{equation}\label{eq:recurrence}
Y_n=A_n V_0+\frac{A_{n-1}+A_{n+1}}{2} V_1+\frac{A_{n-2}+A_{n+2}}{2} V_2 ~,
\end{equation}
\noindent where $A_n=A_{-n}$ and $n=0,1,2,\dots$.
The magnitudes of $A_n$ and $Y_n$
decrease rapidly with $n$ and are consistent with zero for  $n>10$
(see Fig.~\ref{fig:fourier_spectra}).
As one can see in the Figure the Monte Carlo simulations describe
fairly well Fourier spectrum of the data, in particular for large $n$.
Therefore, we can cut the infinite set of equations for MC $Y_n$ 
at some arbitrary $n=N$ and solve the resulting system of $N$ linear equations
to obtain $A_n$ coefficients for $n=0,1,2,\dots N$.
Assuming that GSIM reproduces the CLAS acceptance and efficiency (within systematic
uncertainties treated later), coefficients  $A_n$ should be the same as in the data.
We used these
efficiency/acceptance Fourier coefficients $A_n$ in the expression for data $Y_n$ 
to extract the measured cross section $\phi$-terms: $V_0$, $V_1$ and $V_2$.
We fit the overdetermined system of $N$ linear equations with these 3 unknowns
using the weighted linear least squares fitting routine TLS 
in the CERNLIB library~\cite{cernlib}.

\begin{figure}[htb]
\begin{center}
\includegraphics[bb=2cm 4cm 22cm 24cm, scale=0.4]{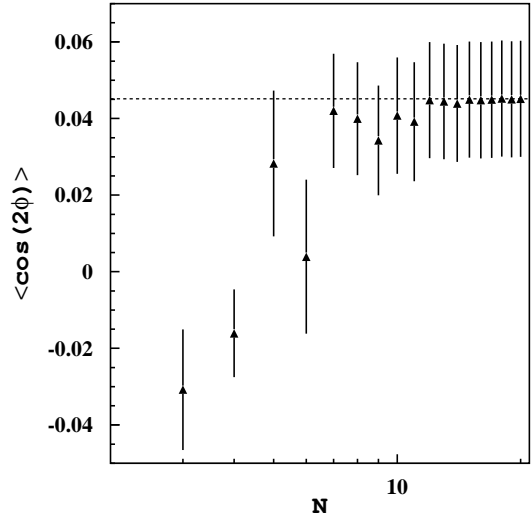}
\caption{\label{fig:fourier_stability} Stability of the $V_2$ cross section term
as a function of the number of moments $N$ taken into account in the
extraction procedure. Error bars are statistical only.}
\end{center}
\end{figure}

The stability of the solution
as a function of $N$ is shown in Fig.~\ref{fig:fourier_stability}.
From this plot we concluded that $N=7$ is the minimum number of moments
necessary to extract sensible $\langle\cos{\phi}\rangle$ and $\langle\cos{2\phi}\rangle$ for the present kinematics.
In the following we made the more conservative choice of $N=20$.

\begin{figure}[htb]
\begin{center}
\includegraphics[bb=2cm 4cm 22cm 24cm, scale=0.4]{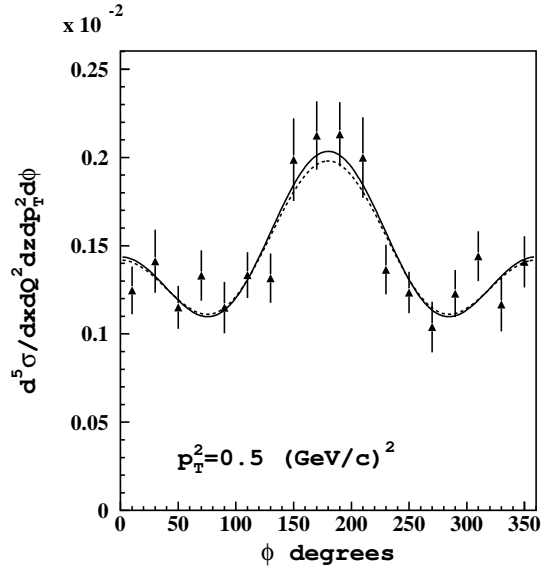}
\caption{\label{fig:moms_fits} 
The $\phi$-dependence of the data taken at
$Q^2=2$ (GeV/c)$^2$, $x=0.24$, $z=0.18$ and $p_T^2=0.5$ (GeV/c)$^2$ (full triangles)
together with the results of the azimuthal moment (solid lines) and fitting (dashed line) methods.
Error bars are statistical only.}
\end{center}
\end{figure}

A typical acceptance-corrected $\phi$-distribution is shown in Fig.~\ref{fig:moms_fits}
together with the two methods of extracting moments, which are in good agreement.
The systematic uncertainties on the $\phi$-dependent terms are larger
than the difference between the two methods (see the following section).

\begin{figure}[htb]
\begin{center}
\includegraphics[bb=2cm 4cm 22cm 24cm, scale=0.4]{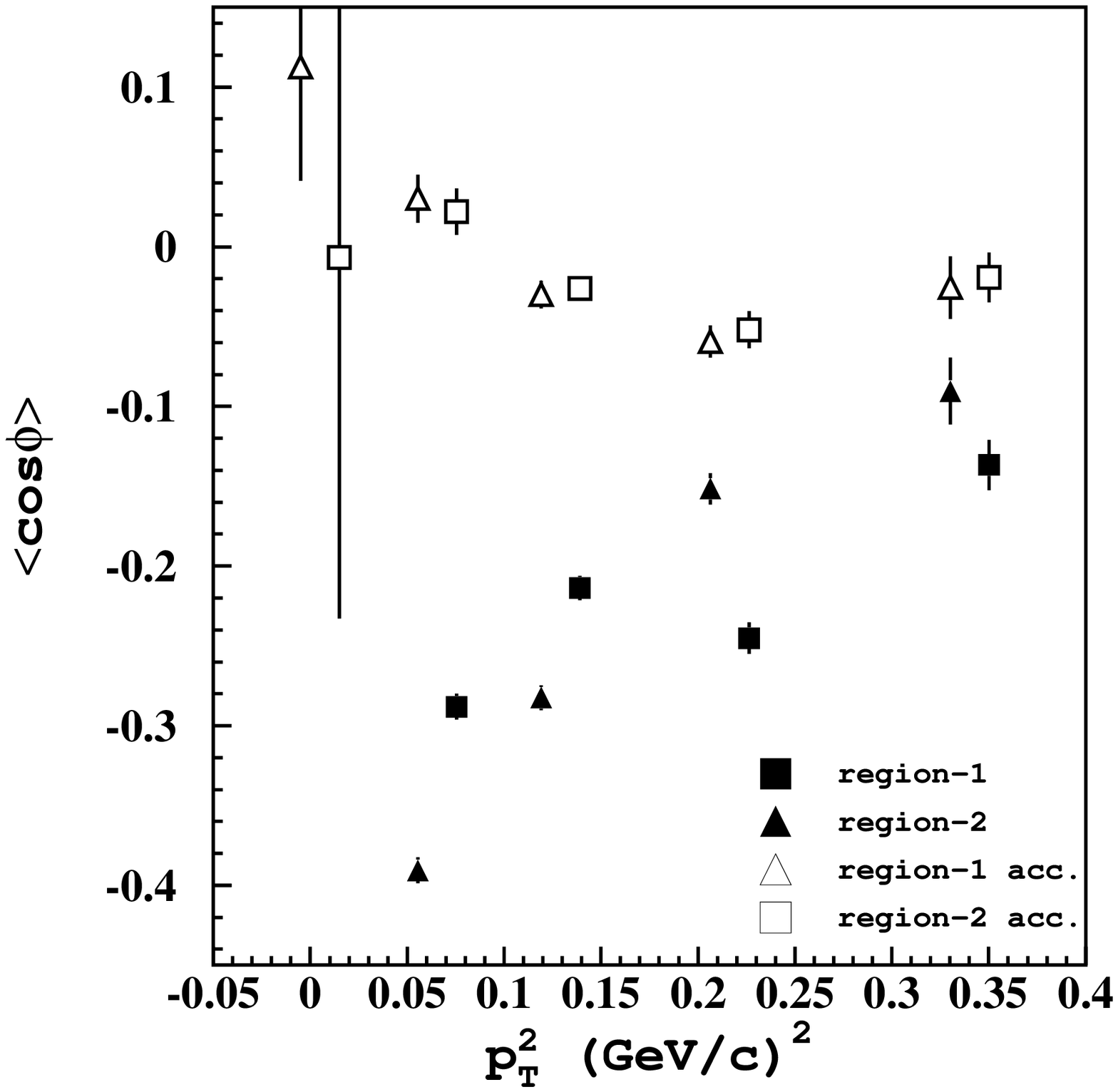}
\caption{\label{fig:acc_reg} 
The $p_T^2$-dependence of the $\langle\cos{\phi}\rangle$ taken at
$Q^2=2$ (GeV/c)$^2$, $x=0.30$ and $z=0.21$
obtained from two different detector acceptance regions (triangles and squares).
Full markers correspond to the data before the acceptance correction
and the empty markers show acceptance corrected $\langle\cos{\phi}\rangle$.
The two data sets are shifted equally along the $x$-axis in opposite directions
from their central values for visibility.
Error bars are statistical only.}
\end{center}
\end{figure}

\begin{figure}[htb]
\begin{center}
\includegraphics[bb=2cm 4cm 22cm 24cm, scale=0.4]{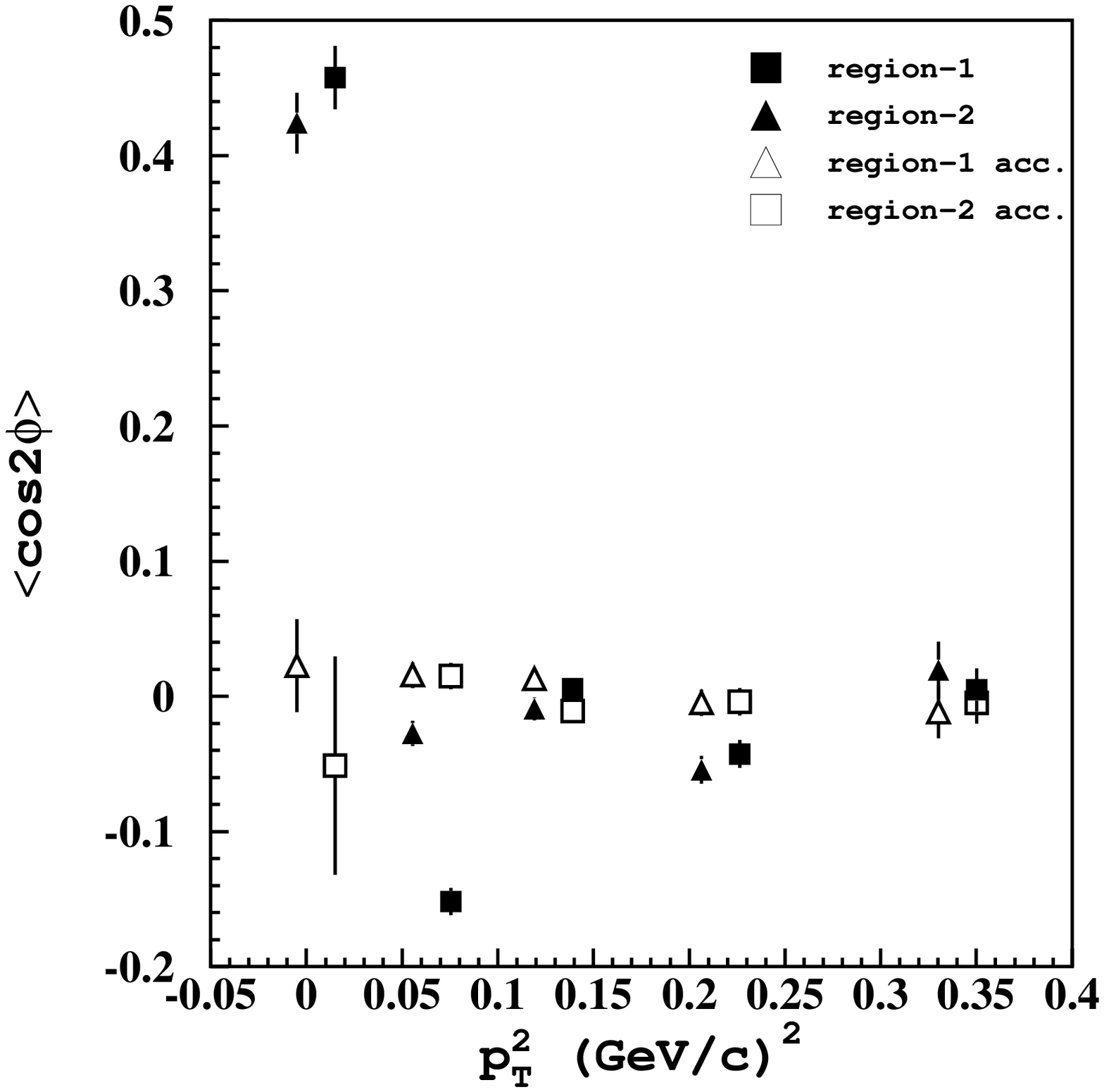}
\caption{\label{fig:acc_reg2} 
Same as Fig.~\ref{fig:acc_reg} except for $\langle\cos{2\phi}\rangle$.}
\end{center}
\end{figure}

The acceptance correction was also tested by dividing each bin
in two parts by cutting the corresponding scattered electron energy range
in two equal intervals and comparing the extracted
$\langle\cos{\phi}\rangle$ and $\langle\cos{2\phi}\rangle$ terms
in these two acceptance regions. An example of this test is shown
in Figs.~\ref{fig:acc_reg} and \ref{fig:acc_reg2}. It can be see that
the acceptance corrections are significant and sometimes differ
for the two separated regions, however the final reconstructed
$\langle\cos{\phi}\rangle$ and $\langle\cos{2\phi}\rangle$
comes out to be consistent within statistical and systematic uncertainties.


\subsection{\label{sec:sys} Systematic Uncertainties}
The systematic uncertainties for the measured absolute cross sections are considerably
different from those for the azimuthal moments, because many quantities drop out in the
ratios measured by moments. Most of systematic uncertainties are point-to-point correlated and
evaluated on a bin-by-bin basis with the exception of the overall normalization, efficiency,
and radiative and bin-centering corrections, for which a uniform relative uncertainty was assumed.
The following sections discuss these uncertainties.

\subsubsection{\label{sec:sys_cs}Cross Section}
The total systematic uncertainties on the five-fold differential cross section
vary from 11 to 44\% with a mean value of 16\%.  
Apart from systematics due to the efficiency corrections
discussed in Section~\ref{sec:mc}, the major contributions come from
detector acceptance and electron identification.
The relative value of most uncertainties is amplified at the
two pion threshold, where the cross section vanishes.

The estimate of the systematic uncertainty from efficiency modeling comes from a comparison
of cross section extractions using two different event generators:
one uses a LO pQCD model, while the other is based on the sum over several exclusive channels.

The systematic uncertainties on the acceptance 
were estimated from the variation in the absolute cross sections obtained
using each of six CLAS sectors separately to detect the electron (pion) and then integrating
over the pion (electron) wherever else it appeared. This uncertainty was estimated bin-by-bin
and reflects the ability of Monte Carlo to describe the detector non-uniformities.
The uncertainty increases at low polar scattering angle, and therefore low-$Q^2$
for electrons and low-$p_T^2$ for pions,
where the azimuthal acceptance of CLAS is reduced.

Systematic uncertainties arising from electron identification
were estimated by comparing
two different methods (as in Ref.~\cite{Osipenko_f2d}) of pion rejection,
one based on Poisson shapes of Cherenkov counter spectra
and another on the geometrical and temporal matching between the measured track
and Cherenkov signal. This uncertainty appear mostly at low-$Q^2$,
where CC is less efficient.

The systematic uncertainty arising from $\pi^+$ identification has two contributions.
One was estimated from the
difference between the ratios of events in the missing neutron peak before
and after pion identification as calculated for data and GSIM simulations.
The second part comes from our treatment of kaon contamination
which was assumed to be 20\%. The two errors were  added in quadrature.

Radiative corrections are model-dependent.
To estimate this systematic uncertainty we changed the model
used in the radiative correction code by 15\% and took the resulting difference as an
estimate of the uncertainty.

There is an additional overall systematic uncertainty of $1\%$ due to
uncertainties in the target length and density. The target length was $5\pm 0.05$ cm
and the liquid-hydrogen density
was $\rho=0.0708\pm 0.0003$ g/cm$^3$ giving approximately a $1$\% uncertainty.

The systematic uncertainty on the bin centering correction was estimated
in the same way as for the radiative corrections. The model was changed
as described above and the difference
between the two centering corrections was taken as the uncertainty.

The empty target subtraction introduces a
small systematic uncertainty due to the assumption of cross-section-ratio (empty to full target)
factorization in the individual kinematic variables.
This uncertainty was estimated by comparing the factorized
and direct bin-by-bin subtraction methods.

These main contributions are listed in Table~\ref{table:d_se}.
All systematic uncertainties shown in the table were combined in quadrature.

\begin{table}[ht]
\begin{center}
\caption{Systematic uncertainties of the semi-inclusive cross section.}
\label{table:d_se}
\vspace{2mm}
\begin{tabular}{|c|c|c|c|} \hline
Source                        & Variation range  & Mean value  \\
                              &         \%       &      \%     \\ \hline
Overall normalization         &         1        &      1      \\ \hline
$e^-$ identification          &      1.8-13      &    3.5      \\ \hline
$\pi^+$ identification        &      0.9-6.7     &    2.1      \\ \hline
$e^-$ acceptance              &      0-19        &    5.3      \\ \hline
$\pi^+$ acceptance            &        0-52      &      4      \\ \hline
Efficiency                    &         10       &     10      \\ \hline
Radiative corrections         &         2        &      2      \\ \hline
Empty target subtraction      &      0-0.7       &     0.2     \\ \hline
Bin centering correction      &        0.7       &     0.7     \\ \hline
Total                         &       11-54      &     14      \\ \hline
\end{tabular}
\end{center}
\end{table}

\subsubsection{\label{sec:sys_asym}Azimuthal Moments}
Azimuthal moments (see Eq.~\ref{eq:asym}) have the advantage of smaller systematic
uncertainties since many of them cancel in the ratio. In particular,
systematic uncertainties of overall normalization, kinematic corrections,
particle identification, efficiency, empty target subtraction and bin centering
cancel. The remaining systematic uncertainties are due to
non-uniformities in the CLAS acceptance and radiative corrections.
The uncertainties due to CLAS acceptance were estimated as the spread
between the central values of the azimuthal moments obtained using
each single CLAS sector to detect the electron or pion and then integrating over
the second particle (pion and electron respectively).
This way we obtained the influence of the electron and pion acceptances separately.
Similar conclusions about the acceptance influence on the azimuthal moment extraction
were made in Ref.~\cite{Mac}.

To estimate the systematic uncertainties of the radiative corrections,
we made a few calculations in randomly chosen kinematic points comparing
correction factors obtained with our model, changing by 15\%
the exclusive $\pi^+n$ contribution or modifying by 30\% the 
$\mathcal{H}_3$ and $\mathcal{H}_4$ structure functions. The difference
in the correction factor was taken as the estimate of this systematic uncertainty.
The variation range and averaged value of these systematic uncertainties
are given in Tables~\ref{table:d_am1} and \ref{table:d_am2} for
the $\langle\cos{\phi}\rangle$ and $\langle\cos{2\phi}\rangle$ moments, respectively.

\begin{table}[ht]
\begin{center}
\caption{Systematic uncertainties of $\langle\cos{\phi}\rangle$.}
\label{table:d_am1}
\vspace{2mm}
\begin{tabular}{|c|c|c|c|} \hline
Source                        & Variation range  & Mean value \\
                              &                  &            \\ \hline
$e^-$ acceptance              &     0-0.06       &  0.016     \\ \hline
$\pi^+$ acceptance            &     0-0.13       &  0.016     \\ \hline
Radiative corrections         &     0.005        &  0.005     \\ \hline
Total                         &   0.005-0.13     &  0.026     \\ \hline
\end{tabular}
\end{center}
\end{table}
\begin{table}[ht]
\begin{center}
\caption{Systematic uncertainties of $\langle\cos{2\phi}\rangle$.}
\label{table:d_am2}
\vspace{2mm}
\begin{tabular}{|c|c|c|c|} \hline
Source                        & Variation range  & Mean value \\
                              &                  &            \\ \hline
$e^-$ acceptance              &     0-0.08       &   0.015    \\ \hline
$\pi^+$ acceptance            &     0-0.12       &   0.011    \\ \hline
Radiative corrections         &     0.003        &   0.003    \\ \hline
Total                         &   0.003-0.12     &   0.021    \\ \hline
\end{tabular}
\end{center}
\end{table}

\subsubsection{\label{sec:sys_sf}Structure Functions}
One additional systematic uncertainty appears in the extraction of the
structure function $\mathcal{H}_2$
from the measured combination $\mathcal{H}_2+\epsilon \mathcal{H}_1$.
In this case some transverse to
longitudinal cross section ratio $R$ should be assumed.
In our results on the structure function $\mathcal{H}_2$
we included a 50\% systematic uncertainty on $R$.
This does not affect strongly the extracted structure function $\mathcal{H}_2$ 
(see Eq.~\ref{eq:rlt_h}),
in the same way as the inclusive structure function $F_2$ is weakly sensitive
to the ratio $R$ for forward-angle scattering. The assumed 50\% precision leads to the systematic
uncertainty shown in the Table~\ref{table:d_h2}.

\begin{table}[ht]
\begin{center}
\caption{Additional systematic uncertainty on $\mathcal{H}_2$.}
\label{table:d_h2}
\vspace{2mm}
\begin{tabular}{|c|c|c|c|} \hline
Source                 & Variation range  & Mean value \\
                       &       \%         &    \%     \\ \hline
$R$ ratio              &     0.6-1.9      &   1.5    \\ \hline
\end{tabular}
\end{center}
\end{table}

\section{\label{sec:results} Results}
The obtained data allow us to perform studies in four different areas:
hadron transverse momentum distributions,
comparison of the $\phi$-independent term with pQCD calculations,
search for the target fragmentation contribution
and study of azimuthal moments. We present these analyses
in the following sections.

\subsection{\label{sec:tmd} Transverse Momentum Distributions}
\begin{figure}[htb]
\includegraphics[bb=5cm 9cm 16cm 28cm, scale=0.4]{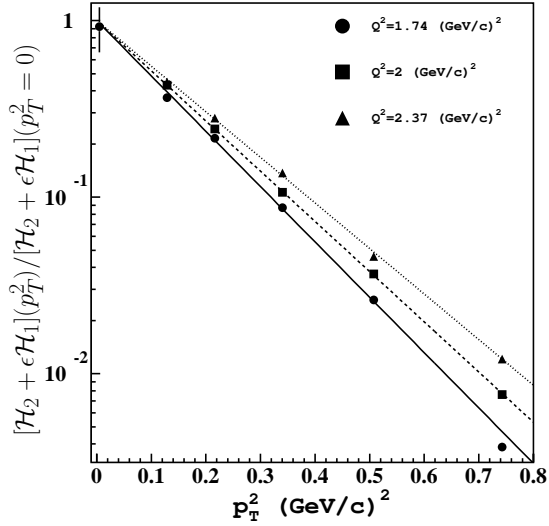}
\caption{\label{fig:pt} The $p_T^2$-dependence of the $\phi$-independent term
$\mathcal{H}_2+\epsilon \mathcal{H}_1$
at $x=0.24$ and $z=0.30$. The lines represent exponential fits to the data for
$Q^2=1.74$ (GeV/c)$^2$ (full circles and solid line),
$Q^2=2$ (GeV/c)$^2$ (full squares and dashed line), and
$Q^2=2.37$ (GeV/c)$^2$ (triangles and dotted line).
The errors bars are statistical only.}
\end{figure}

The $\phi$-independent part of the cross section falls off exponentially in $p_T^2$,
as shown in Fig.~\ref{fig:pt}. This has been predicted in Ref.~\cite{Cahn} to arise from
the intrinsic transverse momentum of partons.
We observe no deviation from this exponential behavior over the entire kinematic domain of our data.

\begin{figure}[htb]
\includegraphics[bb=5cm 5cm 16cm 24cm, scale=0.4]{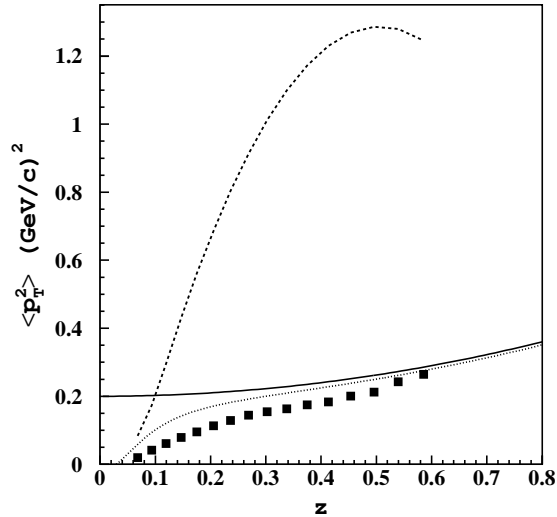}
\caption{\label{fig:ptmean_z} The $z$-dependence of $\langle p_T^2 \rangle$ at
$Q^2=2.37$ (GeV/c)$^2$ and $x=0.27$.
The points are the data from the present analysis.
The curves show the maximum allowed $p_T^2=(p_T^2)^{max}$ (dashed),
the parameterization of high energy data from Eq.~\ref{eq:mean_pt2} (solid),
and the low-$z$ modification from Eq.~\ref{eq:mean_pt2_le}.
The error bars are statistical only and they are smaller than the symbol size.}
\end{figure}

By studying the $p_T^2$-dependence in our data at various values of $z$, we have extracted the
$z$-dependence of the mean transverse momentum $\langle p_T^2 \rangle$, defined
within the Gaussian model, in Eq.~\ref{eq:exp_pt2}, and obtained by fitting $p_T^2$-distributions
in each $(x,Q^2,z)$ bin.
Fig.~\ref{fig:ptmean_z} shows
a clear rise of  $\langle p_T^2 \rangle$ with $z$.
We compared
this with the distribution given in Eq.~\ref{eq:mean_pt2}
with $a^2=0.25$ and  $b^2=0.20$ (GeV/c)$^2$ based on previous data~\cite{Ans1,Ans2,Barone2}.
Significant deviations from this behavior were found at
low-$z$, which can be explained as a threshold kinematic effect.
The maximum achievable transverse momentum $p_T^{max}\simeq z \nu$
becomes smaller at low $z$, because $\nu$ is
limited by the 5.75-GeV beam energy, and
$p_T^{max}$ is smaller than the intrinsic transverse momentum of partons
which is at first order independent of beam energy.
This leads to a cut on the $p_T^2$-distribution,
which is not present in high energy experiments. To account for this low-energy
effect we modified the parameterization as:
\begin{equation}\label{eq:mean_pt2_le}
\langle \tilde{p}_T^2 \rangle = \frac{\langle p_T^2 \rangle}{1+\langle p_T^2 \rangle/(p_T^2)^{max}} ~.
\end{equation}
The dotted curve in Fig.~\ref{fig:ptmean_z} shows that this new
parameterization follows the data points, but the absolute normalization
given by the parameters $a$ and $b$ is still too high. This modification breaks
the factorization between $x$, $Q^2$ and $p_T$ in the low-$z$ region
because the $p_T^2$-distribution now depends also on $x$ and $Q^2$.

\begin{figure}[htb]
\includegraphics[bb=5cm 5cm 16cm 24cm, scale=0.4]{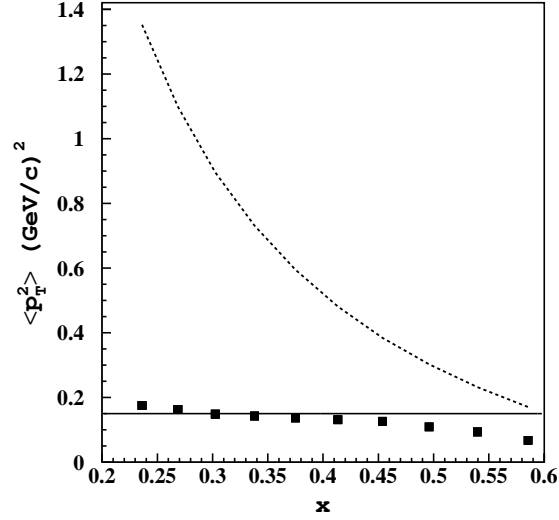}
\caption{\label{fig:ptmean_x} The $x$-dependence of $\langle p_T^2 \rangle$ at
$Q^2=2.37$ (GeV/c)$^2$ and $z=0.34$.
The points are from the present analysis. The curves show
$p_T^2=(p_T^2)^{max}$ (dashed) and
a constant fit to the data (solid).
The error bars are statistical only.}
\end{figure}

At large $z$, $p_T^{max}$ is also large. Therefore, we can check
the factorization of $p_T^2$ from $x$ and $Q^2$.
Fig.~\ref{fig:ptmean_x} shows no appreciable dependence of
the mean transverse momentum $\langle p_T^2  \rangle$
for $x < 0.5$ corresponding to the missing mass $M_X^2<1.6$ (GeV/c$^2$)$^2$,
i.e. the $\Delta$ resonance region.

The transverse momentum distribution exhibits
a small variation with $Q^2$ over the covered kinematic interval
as seen in the different slopes in Fig.~\ref{fig:pt}.
However, the $Q^2$ coverage is
insufficient to observe the logarithmic pQCD evolution of $\langle p_T^2 \rangle$
with $Q^2$ discussed in Ref.~\cite{Ceccopieri_pt_evol}.

\subsection{\label{sec:fact} Comparison with pQCD}
In order to compare the $\phi$-independent term with pQCD predictions,
we assumed a constant
longitudinal to transverse cross section ratio $R=0.12$~\cite{Bebek-R}.

Since there is no TMD-based approach to which we could directly compare
our data, we integrated the measured structure functions $\mathcal{H}_2$ in $p_T^2$
in order to compare $\mathcal{H}_2$
measured in this experiment with $H_2$ from pQCD calculations.
We integrated Eq.~\ref{eq:cs} in $\phi$ and $p_T^2$ and compared with Eq.~\ref{eq:cs_dis}
obtaining
\begin{equation}\label{eq:pt_int}
H_2(x,Q^2,z)=\pi E_h \int_0^{(p_T^2)^{max}}dp_T^2
\frac{\mathcal{H}_2(x,z,Q^2,p_T)}{\sqrt{E_h^2-m_h^2-p_T^2}} ~~~,
\end{equation}
\noindent where the upper limit of integration is given by the smaller of the quantities
$(p_T^2)^{max}=(z\nu)^2-m_h^2$ and the value defined
by the pion threshold, which limits the longitudinal hadron momentum
in the lab frame to
\begin{eqnarray}\label{eq:pl_th}
&& p_\parallel>p_\parallel^{min}=\frac{1}{2|\textbf{q}|}\Bigl\{(M_n^2-M^2)+Q^2- \\ \nonumber
&& 2M\nu(1-z)-m_\pi^2+2z\nu^2-2M_n m_\pi\Bigr\} ~.
\end{eqnarray}
\noindent This limits $p_T^2<|\textbf{p}_h|^2-(p_\parallel^2)^{min}$.
If we exploit the exponential behavior of the measured structure function $\mathcal{H}_2$
in $p_T^2$ (see Eq.~\ref{eq:exp_pt2}), the integration can be performed analytically leading to
\begin{eqnarray}
&& H_2(x,Q^2,z)=V(x,Q^2,z) E_h e^{-\frac{|\textbf{p}_h|^2}{\langle p_T^2 \rangle}} \sqrt{\frac{\pi}{\langle p_T^2 \rangle}} \\ \nonumber
&& \Biggl [
\mbox{Erfi}\Biggl(\sqrt{\frac{|\textbf{p}_h|^2}{\langle p_T^2 \rangle}} \Biggr)-
\mbox{Erfi}\Biggl(\sqrt{\frac{|\textbf{p}_h|^2-(p_T^2)^{max}}{\langle p_T^2 \rangle}}\Biggr)
\Biggr ] ~~~,
\end{eqnarray}
\noindent where $V(x,Q^2,z)$ is the $p_T$-independent part of the structure function
and \mbox{Erfi} is the imaginary error function.
By neglecting the factor $E_h/|p_\parallel|$ in Eq.~\ref{eq:cs} and
by extending the integral to infinity (as typically done in SIDIS analyses, see Eq.~\ref{eq:exp_pt2}),
we find
\begin{equation}
H_2(x,Q^2,z)=V(x,Q^2,z) ~~~.
\end{equation}

\begin{figure*}[htb]
\includegraphics[bb=1cm 2cm 22cm 24cm, scale=0.9]{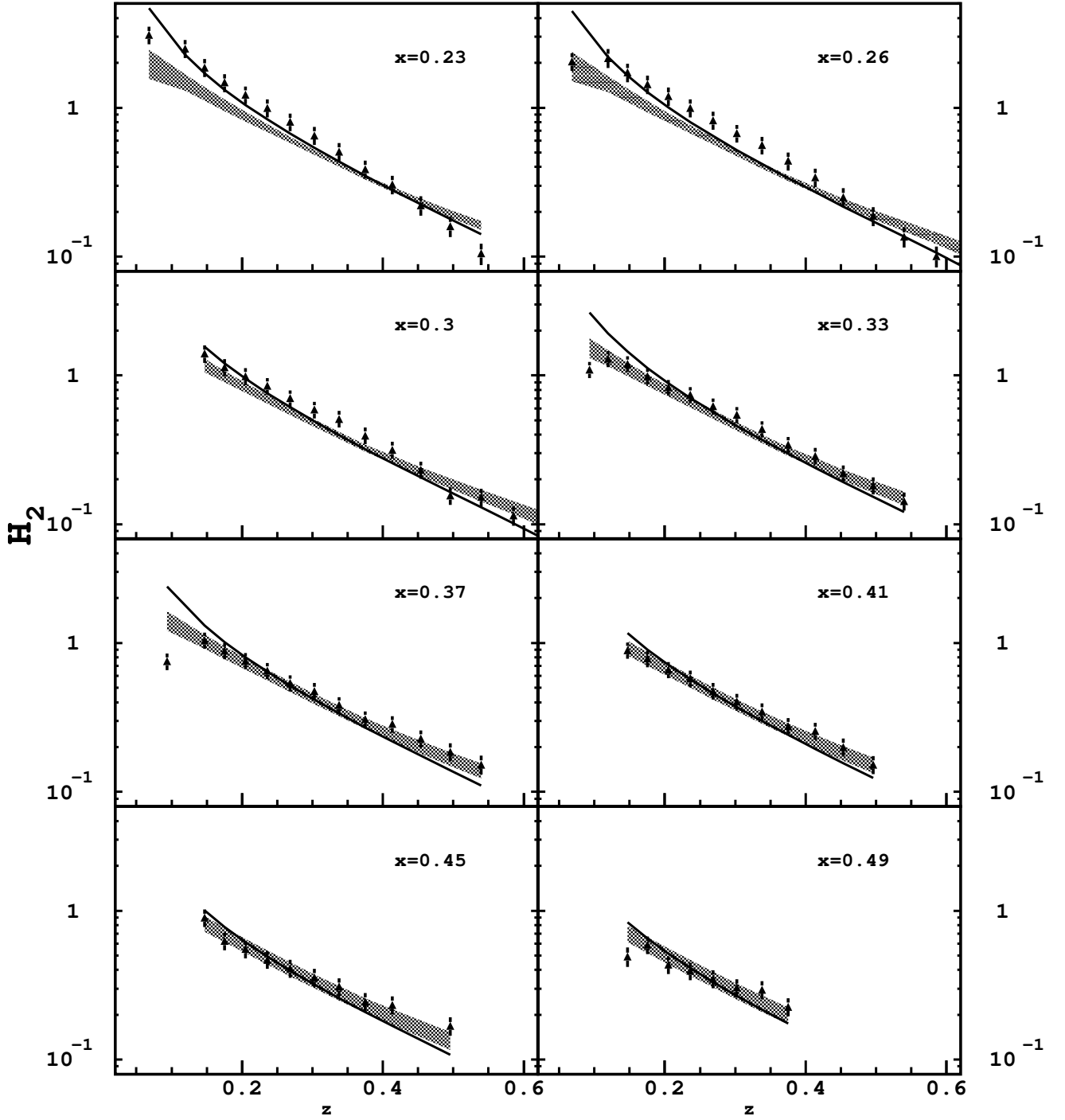}
\caption{\label{fig:pqcd} The $z$-dependence of $H_2$ at
$Q^2=2.37$ (GeV/c)$^2$. The data are shown by full triangles.
The error bars give statistical and systematic uncertainties combined in quadrature.
The solid line shows LO pQCD calculations using the prescription from Ref.~\cite{Furmanski},
CTEQ 5 parton distribution functions~\cite{CTEQ}, and the Kretzer fragmentation functions~\cite{Kretzer}.
NLO calculations performed within the same framework (using CTEQ 5M PDFs)
are shown by the shaded area, for which the width indicates
systematic uncertainties due to factorization and renormalization scale variations~\cite{Ridolfi}.}
\end{figure*}

\begin{figure*}[htb]
\includegraphics[bb=1cm 2cm 22cm 24cm, scale=0.9]{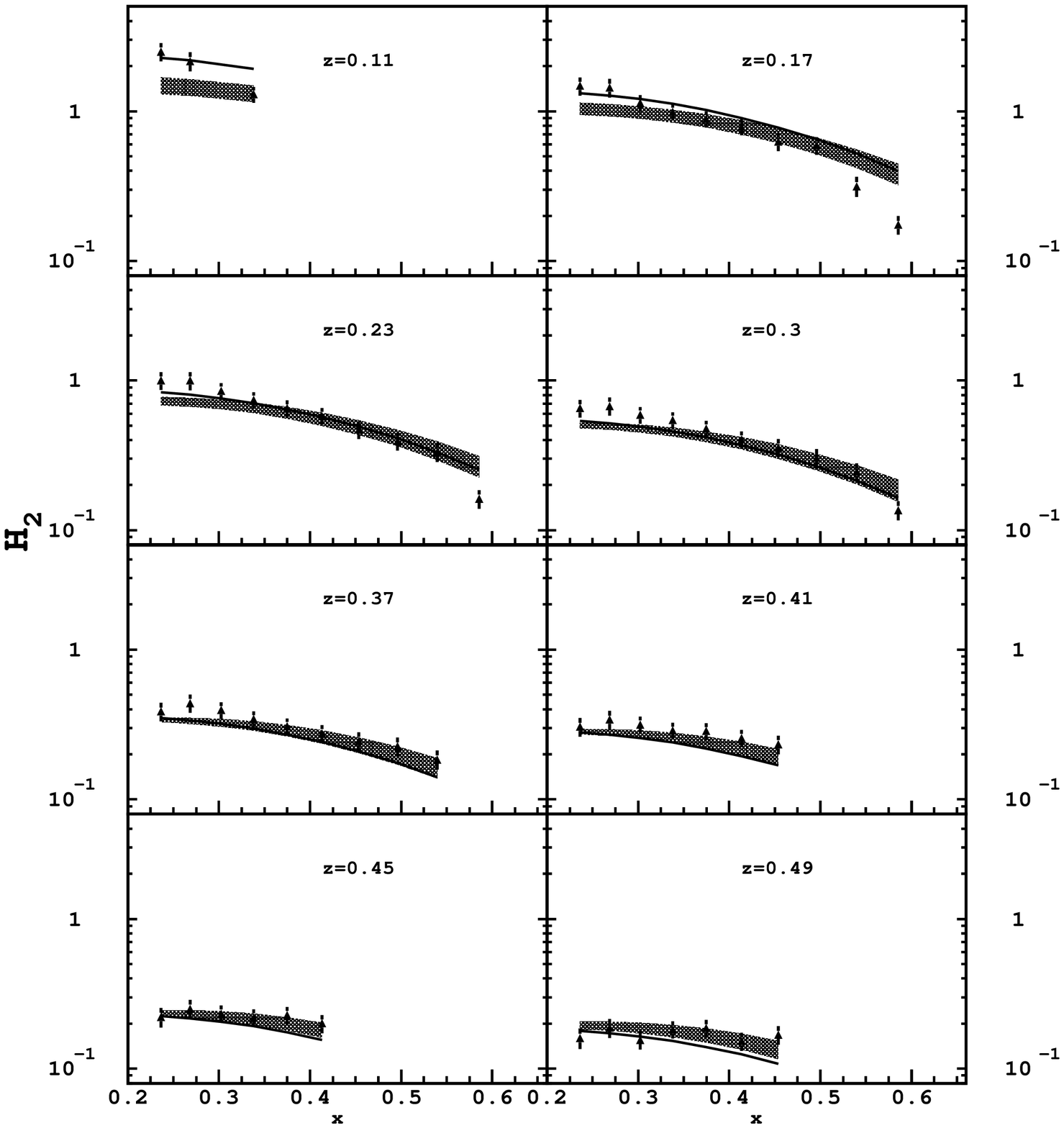}
\caption{\label{fig:pqcd_x} Same as Fig.~\ref{fig:pqcd} except with $H_2$ plotted
as a function of $x$ rather than $z$.}
\end{figure*}

\begin{figure*}[htb]
\includegraphics[bb=1cm 2cm 22cm 24cm, scale=0.9]{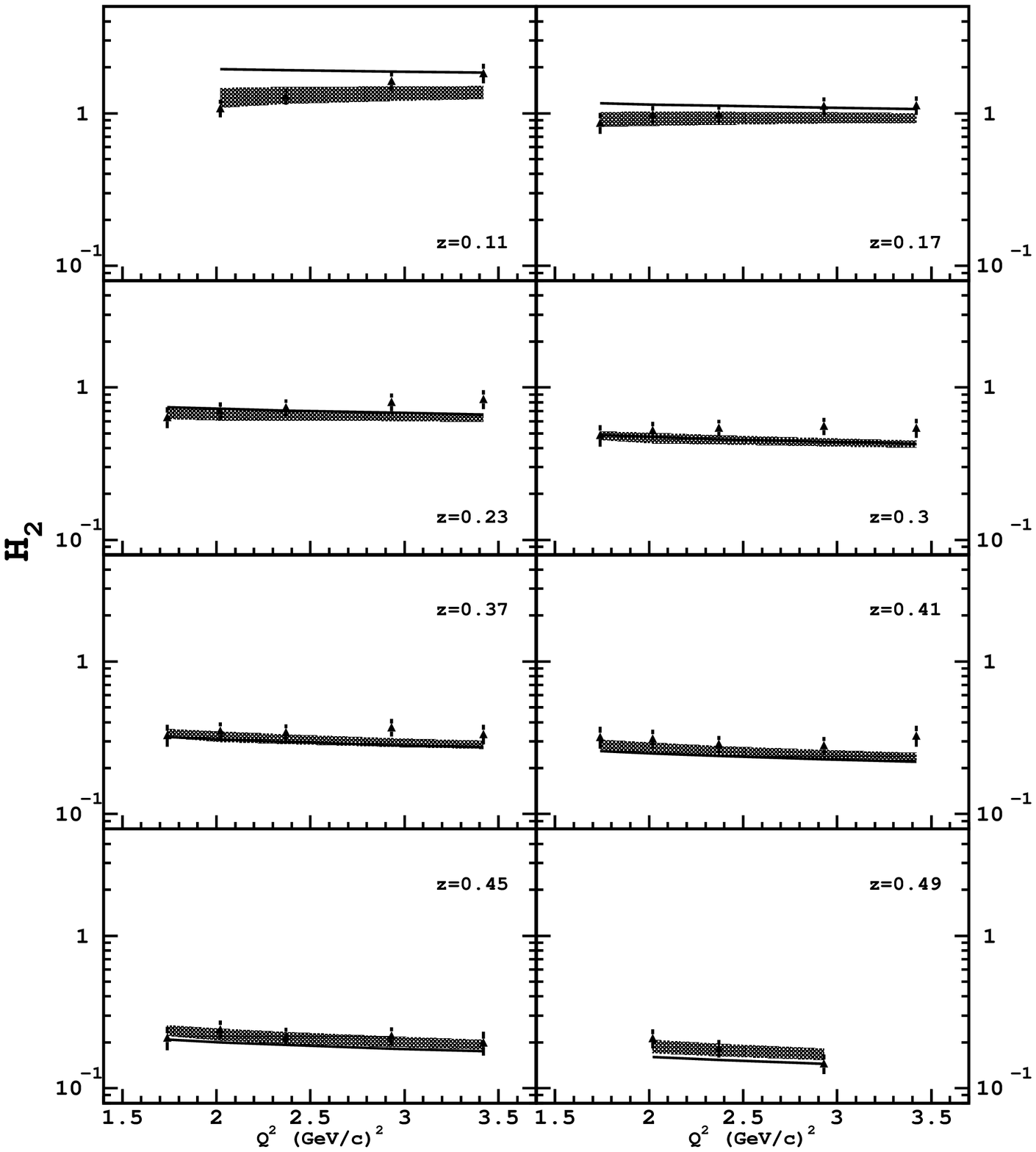}
\caption{\label{fig:pqcd_q} Same as Fig.~\ref{fig:pqcd} except with $H_2$ plotted
as a function of $Q^2$ rather than $z$ at $x=0.33$.}
\end{figure*}

\begin{figure*}[htb]
\includegraphics[bb=1cm 2cm 22cm 24cm, scale=0.9]{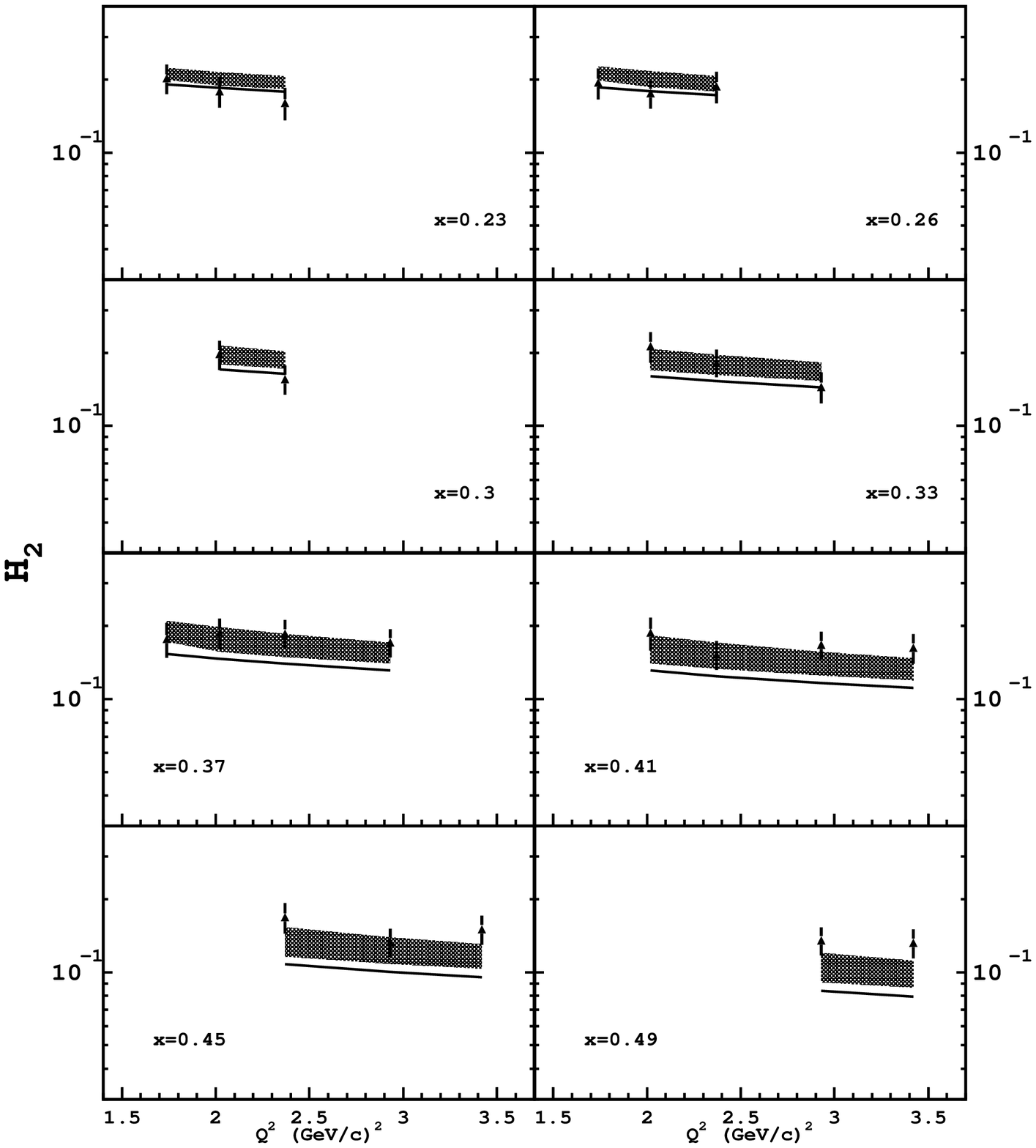}
\caption{\label{fig:pqcd_q_x} Same as Fig.~\ref{fig:pqcd} except with $H_2$ plotted
as a function of $Q^2$ rather than $z$ at $z=0.5$.}
\end{figure*}

\begin{figure*}[htb]
\includegraphics[bb=1cm 2cm 22cm 24cm, scale=0.9]{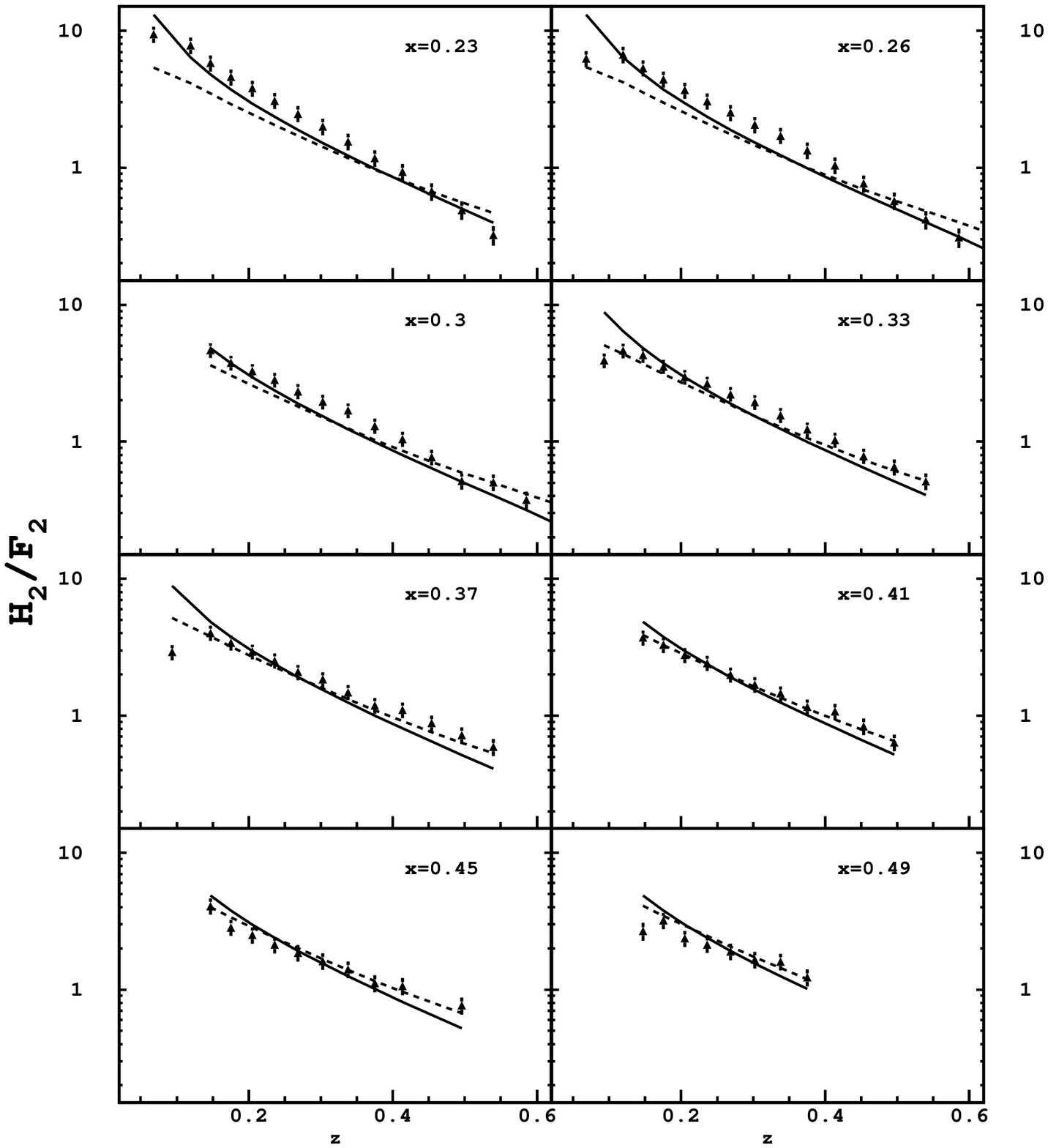}
\caption{\label{fig:pqcd_m} Same as Fig.~\ref{fig:pqcd} except with $H_2/F_2$,
where $F_2$ is the inclusive structure function obtained in the same experiment.
And NLO calculations are shown by the dashed line,
without systematic uncertainties due to factorization and renormalization scales.}
\end{figure*}

In Figs.~\ref{fig:pqcd}, \ref{fig:pqcd_x}, \ref{fig:pqcd_q} and \ref{fig:pqcd_q_x}
our integrated structure function $H_2$ is compared to pQCD calculations given by:
\begin{eqnarray}
&& H_2(x,Q^2,z)= \int_x^1 \frac{d\xi}{\xi}\int_z^1\frac{d\zeta}{\zeta} \\ \nonumber
&& \sum_{ij} \sigma_{hard}^{ij}\Bigl(\xi,\zeta,\frac{Q^2}{\mu^2},\alpha_s(\mu^2)\Bigr) \frac{x}{\xi}
f_i\Bigl(\frac{x}{\xi},\mu^2\Bigr) \frac{z}{\zeta} D_j^{\pi^+}\Bigl(\frac{z}{\zeta},\mu^2\Bigr) ~~~,
\end{eqnarray}
\noindent where $\sigma_{hard}^{ij}$ is the hard scattering cross section for incoming parton $i$
and outgoing parton $j$ given in Ref.~\cite{Furmanski},
$f_i$ is the parton distribution function for parton $i$ taken from Ref.~\cite{CTEQ},
$D_j^{\pi^+}$ is the fragmentation function for parton $j$ and hadron $\pi^+$ taken from Ref.~\cite{Kretzer},
and $\mu$ is the factorization/renormalization scale.
These next-to-leading order (NLO) calculations include a systematic
uncertainty due to arbitrary factorization/renormalization scale variations~\cite{Ridolfi},
indicating the size of possible higher order effects.
This was evaluated by variation of each scale by a factor of two in both directions and
the obtained differences for all scales were summed in quadrature.
NLO calculations within their uncertainty lie closer to the data in the low-$z$ region than leading order (LO) ones.
The difference between the data and NLO pQCD is at most about 20\%.
At low $x$ and $z<0.4$ the data are higher than NLO calculations,
while at largest $x$ both the LO and NLO calculations lie above the data.
The multiplicity ratio $H_2/F_2$ shown in Fig.~\ref{fig:pqcd_m}
demonstrates the same level of agreement between data and pQCD calculations as $H_2$ alone.
This suggests that the differences between the data and theory
do not cancel in the ratio.

The widening systematic uncertainty band in the NLO calculations at high $x$ suggests that
the discrepancy with the data here might be due to a
significant contribution
from multiple soft gluon emission, which can be resummed to
all orders in $\alpha_s$ as in Refs.~\cite{Catani,Nadolsky}.
Similar results were obtained in Ref.~\cite{DeFlorian2} from the comparison
between HERMES $\pi^+$ SIDIS data and NLO calculations.

The difference between the data and calculations
in some kinematic regions leaves room for an additional
contribution from target fragmentation of $<$20\%.
However, the possible presence of
higher twists at our relatively small $Q^2$ values
casts doubt on the attribution of data/pQCD differences to target fragmentation.
In order to better explore target fragmentation, we studied the $t$ and $x_F$-dependencies
of $\mathcal{H}_2$ as described in the following section.

The pQCD calculations are significantly biased by
the assumption of favored fragmentation~\cite{Kretzer}. In fact, using
unseparated $h^+ + h^-$ fragmentation functions as directly measured in $e^+ e^-$
collisions, one obtains curves that are systematically higher by about 20\%.


\subsection{\label{sec:target} Target Fragmentation}
In Leading Order (LO) pQCD, the structure function $H_2$ is given by:
\begin{eqnarray}\label{eq:lo_pqcd}
&& H_2(x,z,Q^2)=\sum_i e_i^2 x [ f_i(x,Q^2) D_i^h(z,Q^2) A(\theta_{\gamma h}=0) \\ \nonumber
&& + (1-x) M_i^h(x,z,Q^2) A(\theta_{\gamma h}=\pi)]~~~,
\end{eqnarray}
\noindent where
$D_i^h(z)$ is the fragmentation function,
$M_i^h(x,z)$ is the fracture function~\cite{Trentadue}
and $A(\theta_{\gamma h})$ is the angular distribution of the observed hadron~\cite{Graudenz}.
The fracture function $M_i^h(x,z)$ gives
the combined probability of striking a parton of flavor $i$ at $x$ and producing a hadron $h$ at $z$
from the proton remnant.
This function obeys the pQCD evolution equations~\cite{Trentadue,Graudenz}
similar to those for $f_i(x)$ and $D_i^h(z)$. The factorization of
the hard photon-parton scattering
and a soft part described by
$M_i^h(x,z)$ has been proved
in Refs.~\cite{Grazzini,Collins}.

Because the agreement between pQCD calculations and our data, shown in Fig.~\ref{fig:pqcd},
was rather poor we could explore only qualitative behavior of the structure functions
to search for the target fragmentation contribution.

To estimate target fragmentation we used two alternative sets of variables:
1) $z$ and $t$, where the squared 4-momentum transfer
$t$ provides added information on the direction of $p_\parallel$; and
2) $x_F$ and $p_T^2$, which included the sign of the longitudinal
hadron momentum in the center-of-momentum (CM) frame through Feynman $x_F$.

\begin{figure}[htb]
\includegraphics[bb=2cm 5cm 16cm 20cm, scale=0.25,angle=270]{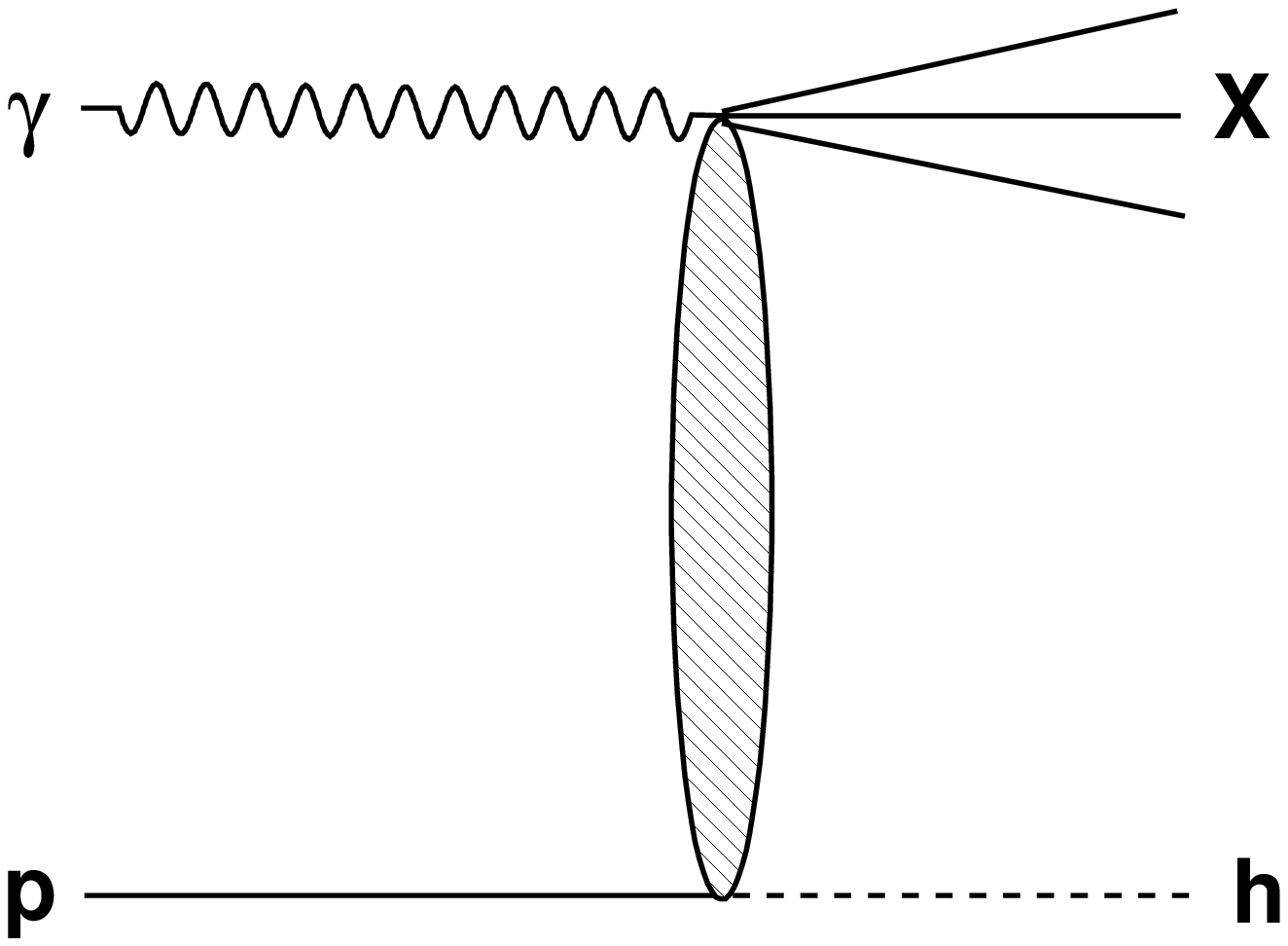}
\caption{\label{fig:lead_trg} Schematic representation of the
leading particle target fragmentation mechanism.
The hatched blob represents the structure function of a particle
or a set of particles exchanged. In case of $\pi^+$ production
the first particle in the blob is the neutron.}
\end{figure}

\begin{figure}[htb]
\includegraphics[bb=5cm 9cm 16cm 28cm, scale=0.4]{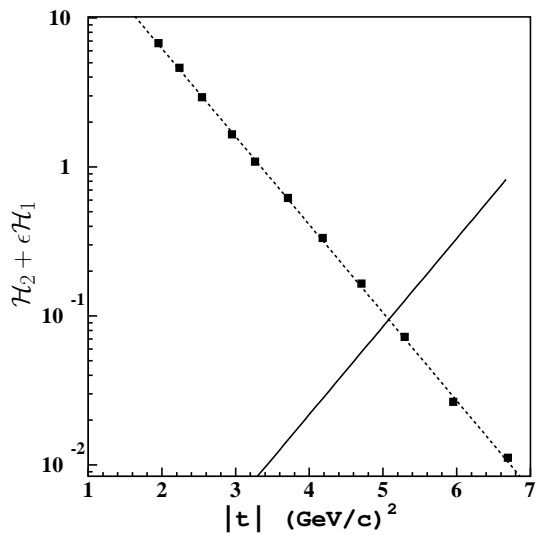}
\caption{\label{fig:t} The $t$-dependence of the $\phi$-independent term $\mathcal{H}_2+\epsilon\mathcal{H}_1$
at $Q^2=2$ (GeV/c)$^2$, $x=0.24$ and $z=0.18$.
The data are shown as solid squares.  The curves represent the exponential fit to the data (dashes) and
the expected behavior of leading-particle target fragmentation (solid),
assuming it to be 1\% of the $t$-channel exchange term.
The error bars are statistical only and they are smaller than symbol size.}
\end{figure}

Target fragmentation is expected to appear at small $z$,
where hadrons are kinematically allowed in the direction opposite
to that of the virtual photon.
In analogy with vector meson photoproduction measurements,
a contribution from target fragmentation may come from $u$-channel
exchange~\cite{Laget} by a particle or a set of particles (see Fig.~\ref{fig:lead_trg}).
In this case the cross section would be proportional to the structure function
of the exchanged particle (e.g. a neutron) or set of particles~\cite{Ingelman}.
In this case one would expect a peak at $|t|=|t|_{max}$, in addition to the dominant peak at
$|t|=|t|_{min}$ due to Regge exchange in the $t$-channel.
This $u$-channel production can be called the ``leading particle'' contribution
in the target fragmentation region because the produced hadron carries
almost all of the spectator momentum.
However, the measured $t$-distribution
shown in Fig.~\ref{fig:t} displays the exponential
behavior expected in Regge theory but
does not show any evidence of the second peak at $|t|=|t|_{max}$. In Fig.~\ref{fig:t}
the solid line shows an expected $u$-channel exchange contribution
assumed to be 1\% of the $t$-channel term. As one can see
this assumption is not supported by the data.
This observation is in agreement with a known phenomenological rule that
a particle not present in the initial state cannot be the leading particle
in a target jet~\cite{Basile}.

Another contribution may come from soft fragmentation of the spectator diquark.
One can naively define all hadrons
produced in the direction of the struck quark to be in the current fragmentation region,
whereas those produced in the direction of the spectator diquark to be in the target fragmentation region.
Since this definition is clearly frame-dependent,
in the following we will use the CM frame.

\begin{figure}[htb]
\includegraphics[bb=5cm 9cm 16cm 28cm, scale=0.4]{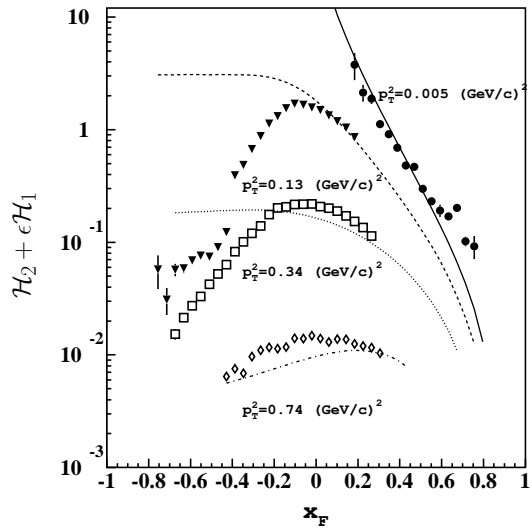}
\caption{\label{fig:x_f} The $x_F$-dependence of the $\phi$-independent term $\mathcal{H}_2+\epsilon\mathcal{H}_1$
at $Q^2=2$ (GeV/c)$^2$ and $x=0.26$.  The data are compared to LO pQCD calculations
combined with a Gaussian $p_T$-dependence from Eq.~\ref{eq:exp_pt2}
for $p_T^2=0.005$ (GeV/c)$^2$ (full circles and solid curve),
$p_T^2=0.13$ (GeV/c)$^2$ (full triangles and dashed curve),
$p_T^2=0.34$ (GeV/c)$^2$ (open squares and dotted curve), and
$p_T^2=0.74$ (GeV/c)$^2$ (open diamonds and dot-dashed curve).
The coverage in $x_F$ is limited for the data by detector acceptance.
The error bars are statistical only.}
\end{figure}

Fig.~\ref{fig:x_f} shows the data for four $p_T$ bins as a function of
$x_F$.  They exhibit a
wide distribution centered at $x_F \simeq 0$, which
corresponds to the center of momentum.
Such behavior is in good agreement with that observed in semi-inclusive
$\pi^+$ production by a muon beam at much higher energies~\cite{emc_x_f}.
According to our definition, all hadrons at $x_F>0$ come from current fragmentation,
while those at $x_F<0$ come from target fragmentation.

\begin{figure}[htb]
\includegraphics[bb=5cm 9cm 16cm 28cm, scale=0.4]{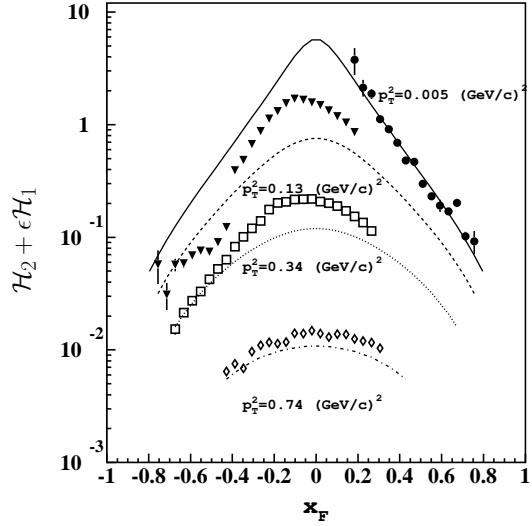}
\caption{\label{fig:x_f_zg} Same data as in Fig.~\ref{fig:x_f}.
The curves are the same as in Fig.~\ref{fig:x_f} except for the fragmentation
functions, which are evaluated at $z_G$ rather than $z$.}
\end{figure}

In the CM frame, $z$ mixes backward-angle production with the production
of low-momentum forward-going hadrons~\cite{Graudenz}. In Fig.~\ref{fig:x_f} the standard LO
pQCD calculations are combined with a Gaussian $p_T$-distribution (Eq.~\ref{eq:exp_pt2}),
plotted versus $x_F$, and compared with the data.
The theory describes approximately the $x_F>0$ behavior
beginning from the $x_F \sim 0$ peak. At negative $x_F$ values the theoretical curve is almost constant
and deviates strongly from the data. This is because at $x_F<0$,
$z$ is close to zero and varies slowly, making $D(z)$ nearly constant.
In order to distinguish target and current fragmentation,
one can use a different variable~\cite{Graudenz}
\begin{equation}
z_G=\frac{2E_h^{CM}}{W} ~~~,
\end{equation}
in which $E_h^{CM}$ is hadron energy in the CM frame. This
can still be interpreted as the parton momentum fraction carried by the measured hadron,
similar to that in $e^+ e^-$ collisions.
By simply using the fragmentation function $D(z_G)$ in Eq.~\ref{eq:lo_pqcd} for both forward and backward regions,
one obtains a qualitative agreement between theoretical and experimental
$x_F$ distributions (see Fig.~\ref{fig:x_f_zg}).
Hence the target fragmentation term in Eq.~\ref{eq:lo_pqcd} is equal to
the standard ``current fragmentation'' contribution $(1-x) M= f(x)\otimes D(z_G)$
We speculate, therefore, that the fragmentation of the spectator diquark system
may be quantitatively similar to the anti-quark fragmentation (see Ref.~\cite{LUND}) for $\pi^+$ production.
The latter mechanism is implicitly included in the fragmentation functions $D(z)$
measured in $e^+ e^-$ collisions. It is also related to the dominance of the favored $u$-quark
fragmentation in $\pi^+$, since the two proton's valence $u$-quarks are likely
to be evenly distributed between current and target fragments.
This intriguing similarity
allows us to describe qualitatively the semi-inclusive cross section by the standard
current fragmentation $f(x)\otimes D(z_G)$ term only even in the region of backward-going $\pi^+$s.

\begin{table*}[ht]
\begin{center}
\caption{$H_2$ structure function
with its statistical and systematic uncertainties.}
\label{table:h2_sidis}
\vspace{2mm}
\begin{tabular}{|c|c|c|c|c|c|c|c|c|c|c|c|c|c|c|c|} \hline
$z$ & $H_2$ & stat.err. & sys.err. & $z$ & $H_2$ & stat.err. & sys.err. & $z$ & $H_2$ & stat.err. & sys.err. & $z$ & $H_2$ & stat.err. & sys.err. \\ \hline
\multicolumn{4}{|c|}{$Q^2$=1.49 (GeV/c)$^2$, x=0.15} &  		      0.37 & 0.418 & 0.003 & 0.058 &			    0.41 & 0.321 & 0.015 & 0.051 &			  0.50 & 0.179 & 0.003 & 0.026   \\ \hline
                        0.07 & 3.746 & 0.038 & 0.548 &  		      0.41 & 0.324 & 0.003 & 0.046 &			    0.45 & 0.215 & 0.014 & 0.034 &			  0.54 & 0.134 & 0.004 & 0.020   \\ \hline
                        0.09 & 3.897 & 0.022 & 0.605 &  		      0.45 & 0.232 & 0.003 & 0.033 &\multicolumn{4}{|c|}{$Q^2$=1.74 (GeV/c)$^2$, x=0.37} &			  0.59 & 0.087 & 0.003 & 0.013   \\ \hline
                        0.12 & 2.957 & 0.018 & 0.451 &  		      0.50 & 0.176 & 0.003 & 0.026 &			    0.18 & 0.724 & 0.025 & 0.109 &			  0.68 & 0.032 & 0.003 & 0.006   \\ \hline
                        0.15 & 2.195 & 0.015 & 0.328 &  		      0.54 & 0.134 & 0.004 & 0.021 &			    0.21 & 0.631 & 0.048 & 0.096 &\multicolumn{4}{|c|}{$Q^2$=2.02 (GeV/c)$^2$, x=0.27}   \\ \hline
                        0.18 & 1.735 & 0.012 & 0.256 &  		      0.59 & 0.099 & 0.003 & 0.016 &			    0.24 & 0.490 & 0.021 & 0.074 &			  0.09 & 1.544 & 0.011 & 0.209   \\ \hline
                        0.21 & 1.359 & 0.011 & 0.200 &\multicolumn{4}{|c|}{$Q^2$=1.74 (GeV/c)$^2$, x=0.24} &			    0.27 & 0.404 & 0.018 & 0.061 &			  0.12 & 1.633 & 0.006 & 0.221   \\ \hline
                        0.24 & 1.084 & 0.010 & 0.159 &  		      0.09 & 1.527 & 0.012 & 0.209 &			    0.30 & 0.431 & 0.026 & 0.067 &			  0.15 & 1.413 & 0.005 & 0.191   \\ \hline
                        0.27 & 0.861 & 0.009 & 0.126 &  		      0.12 & 1.660 & 0.008 & 0.228 &			    0.34 & 0.296 & 0.016 & 0.046 &			  0.18 & 1.163 & 0.004 & 0.154   \\ \hline
                        0.30 & 0.661 & 0.008 & 0.097 &  		      0.15 & 1.444 & 0.007 & 0.197 &			    0.37 & 0.280 & 0.014 & 0.044 &			  0.21 & 1.009 & 0.004 & 0.133   \\ \hline
                        0.34 & 0.499 & 0.008 & 0.073 &  		      0.18 & 1.215 & 0.006 & 0.163 &			    0.41 & 0.269 & 0.013 & 0.042 &			  0.24 & 0.846 & 0.004 & 0.111   \\ \hline
                        0.37 & 0.412 & 0.007 & 0.061 &  		      0.21 & 1.049 & 0.005 & 0.140 &			    0.45 & 0.205 & 0.010 & 0.032 &			  0.27 & 0.716 & 0.003 & 0.093   \\ \hline
                        0.41 & 0.302 & 0.007 & 0.046 &  		      0.24 & 0.878 & 0.005 & 0.116 &			    0.50 & 0.177 & 0.009 & 0.028 &			  0.30 & 0.612 & 0.003 & 0.079   \\ \hline
                        0.45 & 0.238 & 0.007 & 0.037 &  		      0.27 & 0.749 & 0.005 & 0.098 &\multicolumn{4}{|c|}{$Q^2$=1.74 (GeV/c)$^2$, x=0.41} &			  0.34 & 0.503 & 0.004 & 0.065   \\ \hline
                        0.50 & 0.182 & 0.006 & 0.029 &  		      0.30 & 0.621 & 0.005 & 0.081 &			    0.21 & 0.414 & 0.031 & 0.069 &			  0.37 & 0.412 & 0.004 & 0.053   \\ \hline
                        0.63 & 0.099 & 0.008 & 0.018 &  		      0.34 & 0.535 & 0.005 & 0.070 &			    0.24 & 0.342 & 0.021 & 0.057 &			  0.41 & 0.323 & 0.004 & 0.042   \\ \hline
\multicolumn{4}{|c|}{$Q^2$=1.49 (GeV/c)$^2$, x=0.18} &  		      0.37 & 0.435 & 0.005 & 0.057 &			    0.27 & 0.354 & 0.025 & 0.060 &			  0.45 & 0.250 & 0.004 & 0.033   \\ \hline
                        0.12 & 2.374 & 0.013 & 0.334 &  		      0.41 & 0.339 & 0.006 & 0.045 &			    0.30 & 0.391 & 0.023 & 0.066 &			  0.50 & 0.176 & 0.005 & 0.024   \\ \hline
                        0.15 & 1.896 & 0.012 & 0.261 &  		      0.45 & 0.252 & 0.005 & 0.033 &			    0.34 & 0.310 & 0.017 & 0.053 &			  0.54 & 0.174 & 0.005 & 0.024   \\ \hline
                        0.18 & 1.574 & 0.010 & 0.213 &  		      0.50 & 0.203 & 0.005 & 0.028 &			    0.37 & 0.210 & 0.013 & 0.036 &			  0.59 & 0.121 & 0.003 & 0.017   \\ \hline
                        0.21 & 1.279 & 0.009 & 0.172 &  		      0.54 & 0.131 & 0.006 & 0.019 &			    0.41 & 0.216 & 0.014 & 0.037 &\multicolumn{4}{|c|}{$Q^2$=2.02 (GeV/c)$^2$, x=0.30}   \\ \hline
                        0.24 & 1.074 & 0.009 & 0.144 &  		      0.59 & 0.132 & 0.006 & 0.019 &\multicolumn{4}{|c|}{$Q^2$=1.74 (GeV/c)$^2$, x=0.45} &			  0.09 & 1.062 & 0.015 & 0.136   \\ \hline
                        0.27 & 0.847 & 0.008 & 0.113 &\multicolumn{4}{|c|}{$Q^2$=1.74 (GeV/c)$^2$, x=0.27} &			    0.27 & 0.294 & 0.028 & 0.044 &			  0.12 & 1.389 & 0.007 & 0.178   \\ \hline
                        0.30 & 0.724 & 0.008 & 0.096 &  		      0.09 & 1.133 & 0.024 & 0.153 &			    0.30 & 0.301 & 0.040 & 0.045 &			  0.15 & 1.255 & 0.006 & 0.162   \\ \hline
                        0.34 & 0.569 & 0.007 & 0.076 &  		      0.12 & 1.394 & 0.011 & 0.188 &			    0.34 & 0.232 & 0.021 & 0.035 &			  0.18 & 1.057 & 0.006 & 0.136   \\ \hline
                        0.37 & 0.437 & 0.007 & 0.058 &  		      0.18 & 1.126 & 0.010 & 0.153 &			    0.41 & 0.172 & 0.019 & 0.026 &			  0.21 & 0.899 & 0.005 & 0.116   \\ \hline
                        0.41 & 0.383 & 0.008 & 0.052 &  		      0.21 & 0.959 & 0.008 & 0.130 &\multicolumn{4}{|c|}{$Q^2$=2.02 (GeV/c)$^2$, x=0.21} &			  0.24 & 0.798 & 0.005 & 0.103   \\ \hline
                        0.45 & 0.290 & 0.008 & 0.040 &  		      0.24 & 0.836 & 0.007 & 0.114 &			    0.09 & 3.226 & 0.011 & 0.467 &			  0.27 & 0.673 & 0.004 & 0.087   \\ \hline
                        0.50 & 0.204 & 0.006 & 0.029 &  		      0.27 & 0.731 & 0.007 & 0.099 &			    0.12 & 2.519 & 0.009 & 0.361 &			  0.30 & 0.578 & 0.004 & 0.075   \\ \hline
                        0.54 & 0.128 & 0.008 & 0.019 &  		      0.30 & 0.613 & 0.007 & 0.083 &			    0.15 & 1.901 & 0.008 & 0.266 &			  0.34 & 0.454 & 0.004 & 0.059   \\ \hline
                        0.59 & 0.143 & 0.010 & 0.022 &  		      0.34 & 0.490 & 0.007 & 0.067 &			    0.18 & 1.513 & 0.006 & 0.209 &			  0.37 & 0.377 & 0.004 & 0.050   \\ \hline
                        0.07 & 3.353 & 0.019 & 0.457 &  		      0.37 & 0.397 & 0.006 & 0.055 &			    0.21 & 1.232 & 0.006 & 0.170 &			  0.41 & 0.311 & 0.004 & 0.041   \\ \hline
                        0.12 & 2.714 & 0.008 & 0.387 &  		      0.41 & 0.324 & 0.007 & 0.045 &			    0.24 & 0.997 & 0.005 & 0.137 &			  0.45 & 0.238 & 0.004 & 0.032   \\ \hline
                        0.15 & 2.013 & 0.007 & 0.281 &  		      0.45 & 0.240 & 0.008 & 0.034 &			    0.27 & 0.790 & 0.005 & 0.108 &			  0.50 & 0.198 & 0.004 & 0.027   \\ \hline
                        0.18 & 1.587 & 0.006 & 0.219 &  		      0.50 & 0.194 & 0.007 & 0.028 &			    0.30 & 0.654 & 0.005 & 0.090 &			  0.54 & 0.164 & 0.004 & 0.023   \\ \hline
                        0.21 & 1.294 & 0.005 & 0.178 &  		      0.54 & 0.168 & 0.006 & 0.025 &			    0.34 & 0.513 & 0.005 & 0.070 &			  0.59 & 0.142 & 0.005 & 0.020   \\ \hline
                        0.24 & 1.019 & 0.005 & 0.140 &  		      0.59 & 0.134 & 0.007 & 0.020 &			    0.37 & 0.401 & 0.004 & 0.056 &			  0.63 & 0.119 & 0.005 & 0.018   \\ \hline
                        0.27 & 0.812 & 0.004 & 0.111 &\multicolumn{4}{|c|}{$Q^2$=1.74 (GeV/c)$^2$, x=0.30} &			    0.41 & 0.297 & 0.004 & 0.042 &\multicolumn{4}{|c|}{$Q^2$=2.02 (GeV/c)$^2$, x=0.34}   \\ \hline
                        0.30 & 0.644 & 0.004 & 0.088 &  		      0.15 & 1.123 & 0.015 & 0.164 &			    0.45 & 0.228 & 0.004 & 0.033 &			  0.12 & 1.084 & 0.018 & 0.143   \\ \hline
                        0.34 & 0.503 & 0.004 & 0.069 &  		      0.18 & 1.011 & 0.013 & 0.147 &			    0.50 & 0.171 & 0.004 & 0.026 &			  0.15 & 1.128 & 0.009 & 0.150   \\ \hline
                        0.37 & 0.389 & 0.004 & 0.054 &  		      0.21 & 0.878 & 0.014 & 0.128 &			    0.54 & 0.122 & 0.005 & 0.019 &			  0.18 & 0.980 & 0.007 & 0.130   \\ \hline
                        0.41 & 0.302 & 0.003 & 0.043 &  		      0.24 & 0.726 & 0.011 & 0.106 &			    0.59 & 0.097 & 0.005 & 0.015 &			  0.21 & 0.897 & 0.007 & 0.119   \\ \hline
                        0.45 & 0.220 & 0.003 & 0.032 &  		      0.27 & 0.608 & 0.009 & 0.089 &			    0.63 & 0.071 & 0.004 & 0.013 &			  0.24 & 0.702 & 0.006 & 0.093   \\ \hline
                        0.50 & 0.176 & 0.003 & 0.027 &  		      0.30 & 0.535 & 0.009 & 0.078 &\multicolumn{4}{|c|}{$Q^2$=2.02 (GeV/c)$^2$, x=0.24} &			  0.27 & 0.595 & 0.005 & 0.079   \\ \hline
                        0.54 & 0.139 & 0.004 & 0.022 &  		      0.34 & 0.434 & 0.009 & 0.064 &			    0.07 & 1.683 & 0.010 & 0.224 &			  0.30 & 0.524 & 0.005 & 0.070   \\ \hline
                        0.59 & 0.127 & 0.005 & 0.020 &  		      0.37 & 0.356 & 0.008 & 0.053 &			    0.12 & 2.002 & 0.005 & 0.281 &			  0.34 & 0.412 & 0.004 & 0.055   \\ \hline
                        0.63 & 0.083 & 0.004 & 0.015 &  		      0.41 & 0.308 & 0.008 & 0.046 &			    0.15 & 1.617 & 0.004 & 0.223 &			  0.37 & 0.352 & 0.004 & 0.048   \\ \hline
\multicolumn{4}{|c|}{$Q^2$=1.74 (GeV/c)$^2$, x=0.21} &  		      0.45 & 0.258 & 0.007 & 0.039 &			    0.18 & 1.326 & 0.004 & 0.180 &			  0.41 & 0.314 & 0.005 & 0.043   \\ \hline
                        0.12 & 2.106 & 0.006 & 0.306 &\multicolumn{4}{|c|}{$Q^2$=1.74 (GeV/c)$^2$, x=0.34} &			    0.21 & 1.115 & 0.003 & 0.151 &			  0.45 & 0.244 & 0.004 & 0.034   \\ \hline
                        0.15 & 1.717 & 0.005 & 0.244 &  		      0.18 & 0.867 & 0.018 & 0.133 &			    0.24 & 0.915 & 0.003 & 0.123 &			  0.50 & 0.213 & 0.006 & 0.030   \\ \hline
                        0.18 & 1.383 & 0.004 & 0.194 &  		      0.21 & 0.788 & 0.016 & 0.121 &			    0.27 & 0.774 & 0.003 & 0.103 &			  0.54 & 0.165 & 0.006 & 0.024   \\ \hline
                        0.21 & 1.155 & 0.004 & 0.162 &  		      0.24 & 0.640 & 0.018 & 0.099 &			    0.30 & 0.632 & 0.003 & 0.084 &\multicolumn{4}{|c|}{$Q^2$=2.02 (GeV/c)$^2$, x=0.37}   \\ \hline
                        0.24 & 0.949 & 0.004 & 0.132 &  		      0.27 & 0.569 & 0.012 & 0.088 &			    0.34 & 0.514 & 0.003 & 0.069 &			  0.15 & 0.950 & 0.010 & 0.128   \\ \hline
                        0.27 & 0.788 & 0.003 & 0.109 &  		      0.30 & 0.487 & 0.015 & 0.076 &			    0.37 & 0.410 & 0.003 & 0.055 &			  0.18 & 0.861 & 0.008 & 0.115   \\ \hline
                        0.30 & 0.644 & 0.003 & 0.089 &  		      0.34 & 0.363 & 0.010 & 0.056 &			    0.41 & 0.312 & 0.003 & 0.043 &			  0.21 & 0.784 & 0.008 & 0.105   \\ \hline
                        0.34 & 0.518 & 0.003 & 0.072 &  		      0.37 & 0.330 & 0.014 & 0.053 &			    0.45 & 0.237 & 0.003 & 0.033 &			  0.24 & 0.608 & 0.007 & 0.082   \\ \hline
                        0.27 & 0.528 & 0.005 & 0.072 &  		      0.54 & 0.105 & 0.006 & 0.016 &			    0.34 & 0.384 & 0.004 & 0.049 &			  0.15 & 1.645 & 0.008 & 0.230   \\ \hline
                        0.30 & 0.488 & 0.006 & 0.066 &\multicolumn{4}{|c|}{$Q^2$=2.37 (GeV/c)$^2$, x=0.27} &			    0.37 & 0.308 & 0.004 & 0.040 &			  0.18 & 1.345 & 0.006 & 0.186   \\ \hline
\end{tabular}
\end{center}
\end{table*}

\begin{table*}[ht]
\begin{center}
\caption{Continue from Table~\ref{table:h2_sidis}.}
\vspace{2mm}
\begin{tabular}{|c|c|c|c|c|c|c|c|c|c|c|c|c|c|c|c|} \hline
$z$ & $H_2$ & stat.err. & sys.err. & $z$ & $H_2$ & stat.err. & sys.err. & $z$ & $H_2$ & stat.err. & sys.err. & $z$ & $H_2$ & stat.err. & sys.err. \\ \hline
                        0.34 & 0.369 & 0.005 & 0.050 &  		      0.07 & 2.047 & 0.014 & 0.280 &			    0.41 & 0.286 & 0.004 & 0.037 &			  0.21 & 1.113 & 0.006 & 0.153   \\ \hline
                        0.37 & 0.320 & 0.005 & 0.044 &  		      0.12 & 2.162 & 0.006 & 0.311 &			    0.45 & 0.228 & 0.004 & 0.030 &			  0.24 & 0.926 & 0.006 & 0.127   \\ \hline
                        0.41 & 0.300 & 0.006 & 0.042 &  		      0.15 & 1.730 & 0.005 & 0.244 &			    0.50 & 0.187 & 0.005 & 0.025 &			  0.27 & 0.761 & 0.005 & 0.104   \\ \hline
                        0.45 & 0.214 & 0.005 & 0.030 &  		      0.18 & 1.437 & 0.005 & 0.200 &			    0.54 & 0.153 & 0.005 & 0.021 &			  0.30 & 0.621 & 0.005 & 0.085   \\ \hline
                        0.50 & 0.188 & 0.005 & 0.027 &  		      0.21 & 1.196 & 0.004 & 0.166 &\multicolumn{4}{|c|}{$Q^2$=2.37 (GeV/c)$^2$, x=0.41} &			  0.34 & 0.505 & 0.006 & 0.069   \\ \hline
\multicolumn{4}{|c|}{$Q^2$=2.02 (GeV/c)$^2$, x=0.41} &  		      0.24 & 0.998 & 0.004 & 0.138 &			    0.15 & 0.895 & 0.008 & 0.113 &			  0.37 & 0.388 & 0.005 & 0.054   \\ \hline
                        0.18 & 0.700 & 0.012 & 0.102 &  		      0.27 & 0.824 & 0.004 & 0.113 &			    0.18 & 0.789 & 0.006 & 0.099 &			  0.41 & 0.339 & 0.007 & 0.048   \\ \hline
                        0.21 & 0.639 & 0.008 & 0.094 &  		      0.30 & 0.675 & 0.004 & 0.093 &			    0.21 & 0.663 & 0.007 & 0.084 &			  0.45 & 0.224 & 0.005 & 0.032   \\ \hline
                        0.24 & 0.495 & 0.009 & 0.073 &  		      0.34 & 0.562 & 0.004 & 0.077 &			    0.24 & 0.580 & 0.005 & 0.073 &\multicolumn{4}{|c|}{$Q^2$=2.93 (GeV/c)$^2$, x=0.34}   \\ \hline
                        0.27 & 0.449 & 0.007 & 0.066 &  		      0.37 & 0.440 & 0.004 & 0.061 &			    0.27 & 0.477 & 0.004 & 0.061 &			  0.07 & 1.469 & 0.016 & 0.191   \\ \hline
                        0.30 & 0.420 & 0.007 & 0.063 &  		      0.41 & 0.341 & 0.004 & 0.048 &			    0.30 & 0.404 & 0.004 & 0.052 &			  0.09 & 1.926 & 0.007 & 0.268   \\ \hline
                        0.34 & 0.359 & 0.006 & 0.054 &  		      0.45 & 0.251 & 0.004 & 0.036 &			    0.34 & 0.347 & 0.005 & 0.045 &			  0.12 & 1.629 & 0.006 & 0.225   \\ \hline
                        0.37 & 0.293 & 0.005 & 0.045 &  		      0.50 & 0.188 & 0.004 & 0.028 &			    0.37 & 0.277 & 0.004 & 0.036 &			  0.15 & 1.378 & 0.005 & 0.187   \\ \hline
                        0.41 & 0.256 & 0.005 & 0.039 &  		      0.54 & 0.137 & 0.004 & 0.021 &			    0.41 & 0.257 & 0.005 & 0.034 &			  0.18 & 1.124 & 0.004 & 0.150   \\ \hline
                        0.45 & 0.216 & 0.004 & 0.033 &  		      0.59 & 0.101 & 0.004 & 0.016 &			    0.45 & 0.200 & 0.005 & 0.027 &			  0.21 & 0.943 & 0.004 & 0.125   \\ \hline
                        0.50 & 0.188 & 0.004 & 0.029 &  		      0.68 & 0.081 & 0.020 & 0.016 &			    0.50 & 0.153 & 0.003 & 0.021 &			  0.24 & 0.804 & 0.004 & 0.106   \\ \hline
                        0.54 & 0.147 & 0.004 & 0.023 &\multicolumn{4}{|c|}{$Q^2$=2.37 (GeV/c)$^2$, x=0.30} &\multicolumn{4}{|c|}{$Q^2$=2.37 (GeV/c)$^2$, x=0.45} &			  0.27 & 0.672 & 0.004 & 0.088   \\ \hline
\multicolumn{4}{|c|}{$Q^2$=2.02 (GeV/c)$^2$, x=0.45} &  		      0.15 & 1.401 & 0.005 & 0.185 &			    0.15 & 0.898 & 0.012 & 0.120 &			  0.30 & 0.560 & 0.004 & 0.073   \\ \hline
                        0.18 & 0.721 & 0.011 & 0.116 &  		      0.18 & 1.134 & 0.004 & 0.147 &			    0.18 & 0.628 & 0.009 & 0.084 &			  0.34 & 0.484 & 0.004 & 0.063   \\ \hline
                        0.21 & 0.489 & 0.009 & 0.079 &  		      0.21 & 0.989 & 0.004 & 0.128 &			    0.21 & 0.553 & 0.006 & 0.074 &			  0.37 & 0.370 & 0.004 & 0.049   \\ \hline
                        0.24 & 0.392 & 0.006 & 0.063 &  		      0.24 & 0.852 & 0.004 & 0.109 &			    0.24 & 0.469 & 0.006 & 0.063 &			  0.41 & 0.283 & 0.005 & 0.038   \\ \hline
                        0.27 & 0.368 & 0.007 & 0.060 &  		      0.27 & 0.704 & 0.004 & 0.090 &			    0.27 & 0.410 & 0.006 & 0.056 &			  0.45 & 0.221 & 0.004 & 0.030   \\ \hline
                        0.30 & 0.348 & 0.006 & 0.057 &  		      0.30 & 0.590 & 0.003 & 0.075 &			    0.30 & 0.357 & 0.005 & 0.049 &			  0.50 & 0.145 & 0.005 & 0.021   \\ \hline
                        0.34 & 0.295 & 0.006 & 0.049 &  		      0.34 & 0.511 & 0.004 & 0.065 &			    0.34 & 0.309 & 0.005 & 0.043 &			  0.54 & 0.110 & 0.004 & 0.016   \\ \hline
                        0.37 & 0.247 & 0.005 & 0.041 &  		      0.37 & 0.395 & 0.004 & 0.050 &			    0.37 & 0.246 & 0.004 & 0.035 &\multicolumn{4}{|c|}{$Q^2$=2.93 (GeV/c)$^2$, x=0.37}   \\ \hline
                        0.41 & 0.226 & 0.005 & 0.037 &  		      0.41 & 0.316 & 0.004 & 0.041 &			    0.41 & 0.234 & 0.004 & 0.033 &			  0.09 & 1.333 & 0.013 & 0.174   \\ \hline
                        0.45 & 0.188 & 0.006 & 0.031 &  		      0.45 & 0.233 & 0.004 & 0.030 &			    0.50 & 0.168 & 0.003 & 0.024 &			  0.12 & 1.282 & 0.006 & 0.168   \\ \hline
\multicolumn{4}{|c|}{$Q^2$=2.02 (GeV/c)$^2$, x=0.50} &  		      0.50 & 0.156 & 0.004 & 0.021 &\multicolumn{4}{|c|}{$Q^2$=2.37 (GeV/c)$^2$, x=0.50} &			  0.15 & 1.138 & 0.006 & 0.149   \\ \hline
                        0.21 & 0.519 & 0.011 & 0.076 &  		      0.54 & 0.152 & 0.004 & 0.021 &			    0.15 & 0.492 & 0.034 & 0.065 &			  0.18 & 0.948 & 0.005 & 0.122   \\ \hline
                        0.24 & 0.357 & 0.008 & 0.053 &  		      0.59 & 0.115 & 0.005 & 0.016 &			    0.18 & 0.590 & 0.007 & 0.078 &			  0.21 & 0.785 & 0.004 & 0.100   \\ \hline
                        0.27 & 0.289 & 0.010 & 0.043 &  		      0.63 & 0.095 & 0.006 & 0.014 &			    0.21 & 0.435 & 0.007 & 0.059 &			  0.24 & 0.711 & 0.004 & 0.090   \\ \hline
                        0.30 & 0.300 & 0.008 & 0.045 &\multicolumn{4}{|c|}{$Q^2$=2.37 (GeV/c)$^2$, x=0.34} &			    0.24 & 0.394 & 0.005 & 0.053 &			  0.27 & 0.609 & 0.004 & 0.077   \\ \hline
                        0.34 & 0.246 & 0.006 & 0.037 &  		      0.09 & 1.094 & 0.013 & 0.132 &			    0.27 & 0.350 & 0.005 & 0.048 &			  0.30 & 0.505 & 0.005 & 0.063   \\ \hline
                        0.37 & 0.197 & 0.006 & 0.030 &  		      0.12 & 1.295 & 0.006 & 0.157 &			    0.30 & 0.305 & 0.005 & 0.042 &			  0.34 & 0.427 & 0.004 & 0.054   \\ \hline
                        0.41 & 0.181 & 0.007 & 0.027 &  		      0.15 & 1.197 & 0.006 & 0.146 &			    0.34 & 0.295 & 0.005 & 0.041 &			  0.37 & 0.316 & 0.004 & 0.040   \\ \hline
\multicolumn{4}{|c|}{$Q^2$=2.02 (GeV/c)$^2$, x=0.54} &  		      0.18 & 0.990 & 0.005 & 0.120 &			    0.37 & 0.226 & 0.004 & 0.032 &			  0.41 & 0.250 & 0.004 & 0.032   \\ \hline
                        0.18 & 0.162 & 0.015 & 0.026 &  		      0.21 & 0.832 & 0.004 & 0.101 &\multicolumn{4}{|c|}{$Q^2$=2.37 (GeV/c)$^2$, x=0.54} &			  0.45 & 0.253 & 0.005 & 0.033   \\ \hline
                        0.21 & 0.207 & 0.007 & 0.034 &  		      0.24 & 0.742 & 0.004 & 0.090 &			    0.18 & 0.316 & 0.024 & 0.042 &			  0.50 & 0.171 & 0.004 & 0.023   \\ \hline
                        0.24 & 0.175 & 0.009 & 0.029 &  		      0.27 & 0.622 & 0.004 & 0.076 &			    0.21 & 0.369 & 0.006 & 0.049 &			  0.54 & 0.172 & 0.005 & 0.024   \\ \hline
                        0.27 & 0.176 & 0.006 & 0.029 &  		      0.30 & 0.544 & 0.005 & 0.067 &			    0.24 & 0.332 & 0.014 & 0.045 &			  0.59 & 0.117 & 0.005 & 0.016   \\ \hline
\multicolumn{4}{|c|}{$Q^2$=2.37 (GeV/c)$^2$, x=0.24} &  		      0.34 & 0.437 & 0.004 & 0.054 &			    0.27 & 0.297 & 0.009 & 0.041 &\multicolumn{4}{|c|}{$Q^2$=2.93 (GeV/c)$^2$, x=0.41}   \\ \hline
                        0.07 & 3.085 & 0.030 & 0.419 &  		      0.37 & 0.343 & 0.004 & 0.043 &			    0.30 & 0.246 & 0.005 & 0.033 &			  0.12 & 1.089 & 0.007 & 0.128   \\ \hline
                        0.12 & 2.502 & 0.014 & 0.354 &  		      0.41 & 0.288 & 0.004 & 0.036 &			    0.34 & 0.229 & 0.005 & 0.031 &			  0.15 & 0.961 & 0.006 & 0.114   \\ \hline
                        0.15 & 1.860 & 0.012 & 0.258 &  		      0.45 & 0.220 & 0.003 & 0.028 &			    0.37 & 0.185 & 0.005 & 0.026 &			  0.18 & 0.795 & 0.007 & 0.094   \\ \hline
                        0.18 & 1.477 & 0.010 & 0.202 &  		      0.50 & 0.182 & 0.004 & 0.024 &\multicolumn{4}{|c|}{$Q^2$=2.37 (GeV/c)$^2$, x=0.59} &			  0.21 & 0.668 & 0.005 & 0.079   \\ \hline
                        0.21 & 1.224 & 0.009 & 0.168 &  		      0.54 & 0.143 & 0.003 & 0.019 &			    0.18 & 0.175 & 0.007 & 0.024 &			  0.24 & 0.600 & 0.005 & 0.072   \\ \hline
                        0.24 & 1.000 & 0.008 & 0.136 &\multicolumn{4}{|c|}{$Q^2$=2.37 (GeV/c)$^2$, x=0.37} &			    0.21 & 0.180 & 0.004 & 0.025 &			  0.27 & 0.544 & 0.005 & 0.065   \\ \hline
                        0.27 & 0.805 & 0.008 & 0.109 &  		      0.09 & 0.755 & 0.026 & 0.093 &			    0.24 & 0.162 & 0.004 & 0.023 &			  0.30 & 0.418 & 0.004 & 0.051   \\ \hline
                        0.30 & 0.653 & 0.008 & 0.089 &  		      0.15 & 1.044 & 0.008 & 0.130 &			    0.27 & 0.142 & 0.005 & 0.020 &			  0.34 & 0.367 & 0.004 & 0.045   \\ \hline
                        0.34 & 0.509 & 0.008 & 0.070 &  		      0.18 & 0.886 & 0.007 & 0.110 &			    0.30 & 0.136 & 0.005 & 0.019 &			  0.37 & 0.277 & 0.004 & 0.034   \\ \hline
                        0.37 & 0.387 & 0.007 & 0.054 &  		      0.21 & 0.762 & 0.006 & 0.095 &			    0.34 & 0.129 & 0.003 & 0.018 &			  0.41 & 0.235 & 0.004 & 0.029   \\ \hline
                        0.41 & 0.306 & 0.006 & 0.043 &  		      0.24 & 0.653 & 0.005 & 0.082 &\multicolumn{4}{|c|}{$Q^2$=2.93 (GeV/c)$^2$, x=0.30} &			  0.45 & 0.240 & 0.006 & 0.031   \\ \hline
                        0.45 & 0.221 & 0.006 & 0.032 &  		      0.27 & 0.540 & 0.004 & 0.068 &			    0.09 & 2.623 & 0.010 & 0.377 &			  0.50 & 0.167 & 0.005 & 0.022   \\ \hline
                        0.50 & 0.160 & 0.006 & 0.024 &  		      0.30 & 0.477 & 0.005 & 0.061 &			    0.12 & 2.089 & 0.008 & 0.297 &\multicolumn{4}{|c|}{$Q^2$=2.93 (GeV/c)$^2$, x=0.45}   \\ \hline
                        0.09 & 0.522 & 0.024 & 0.063 &  		      0.24 & 0.840 & 0.018 & 0.117 &			    0.24 & 0.435 & 0.005 & 0.054 &			  0.24 & 0.566 & 0.008 & 0.076   \\ \hline
                        0.12 & 0.940 & 0.010 & 0.113 &  		      0.27 & 0.655 & 0.016 & 0.091 &			    0.27 & 0.383 & 0.004 & 0.048 &			  0.27 & 0.457 & 0.008 & 0.061   \\ \hline
                        0.15 & 0.794 & 0.007 & 0.096 &  		      0.30 & 0.545 & 0.017 & 0.076 &			    0.30 & 0.313 & 0.005 & 0.039 &			  0.30 & 0.416 & 0.009 & 0.056   \\ \hline
                        0.18 & 0.738 & 0.007 & 0.089 &  		      0.34 & 0.445 & 0.017 & 0.062 &			    0.34 & 0.259 & 0.004 & 0.032 &			  0.34 & 0.329 & 0.009 & 0.044   \\ \hline
\end{tabular}
\end{center}
\end{table*}

\begin{table*}[ht]
\begin{center}
\caption{Continue from Table~\ref{table:h2_sidis}.}
\vspace{2mm}
\begin{tabular}{|c|c|c|c|c|c|c|c|c|c|c|c|c|c|c|c|} \hline
$z$ & $H_2$ & stat.err. & sys.err. & $z$ & $H_2$ & stat.err. & sys.err. & $z$ & $H_2$ & stat.err. & sys.err. & $z$ & $H_2$ & stat.err. & sys.err. \\ \hline
                        0.21 & 0.560 & 0.007 & 0.068 &  		      0.37 & 0.335 & 0.015 & 0.047 &			    0.37 & 0.247 & 0.006 & 0.032 &			  0.37 & 0.231 & 0.011 & 0.031   \\ \hline
                        0.24 & 0.509 & 0.005 & 0.062 &  		      0.41 & 0.328 & 0.022 & 0.047 &			    0.41 & 0.184 & 0.005 & 0.024 &			  0.41 & 0.206 & 0.011 & 0.028   \\ \hline
                        0.27 & 0.461 & 0.005 & 0.057 &  		      0.45 & 0.199 & 0.020 & 0.029 &			    0.45 & 0.182 & 0.005 & 0.024 &			  0.45 & 0.138 & 0.013 & 0.019   \\ \hline
                        0.30 & 0.342 & 0.004 & 0.042 &\multicolumn{4}{|c|}{$Q^2$=3.42 (GeV/c)$^2$, x=0.37} &			    0.50 & 0.132 & 0.004 & 0.018 &\multicolumn{4}{|c|}{$Q^2$=4.10 (GeV/c)$^2$, x=0.50}   \\ \hline
                        0.34 & 0.300 & 0.004 & 0.037 &  		      0.09 & 1.847 & 0.011 & 0.261 &\multicolumn{4}{|c|}{$Q^2$=3.42 (GeV/c)$^2$, x=0.54} &			  0.09 & 0.950 & 0.010 & 0.134   \\ \hline
                        0.37 & 0.236 & 0.005 & 0.030 &  		      0.12 & 1.521 & 0.008 & 0.214 &			    0.12 & 0.775 & 0.012 & 0.091 &			  0.12 & 0.872 & 0.007 & 0.121   \\ \hline
                        0.41 & 0.215 & 0.005 & 0.027 &  		      0.15 & 1.242 & 0.008 & 0.172 &			    0.15 & 0.516 & 0.007 & 0.061 &			  0.15 & 0.747 & 0.008 & 0.102   \\ \hline
                        0.45 & 0.213 & 0.005 & 0.028 &  		      0.18 & 1.033 & 0.006 & 0.141 &			    0.18 & 0.466 & 0.006 & 0.055 &			  0.18 & 0.609 & 0.006 & 0.082   \\ \hline
                        0.50 & 0.133 & 0.003 & 0.018 &  		      0.21 & 0.880 & 0.006 & 0.119 &			    0.21 & 0.355 & 0.006 & 0.043 &			  0.21 & 0.546 & 0.007 & 0.073   \\ \hline
                        0.54 & 0.143 & 0.003 & 0.020 &  		      0.24 & 0.749 & 0.006 & 0.101 &			    0.24 & 0.321 & 0.004 & 0.039 &			  0.24 & 0.481 & 0.006 & 0.064   \\ \hline
\multicolumn{4}{|c|}{$Q^2$=2.93 (GeV/c)$^2$, x=0.50} &  		      0.27 & 0.617 & 0.006 & 0.083 &			    0.27 & 0.335 & 0.005 & 0.041 &			  0.27 & 0.393 & 0.006 & 0.052   \\ \hline
                        0.15 & 0.734 & 0.010 & 0.089 &  		      0.30 & 0.525 & 0.006 & 0.070 &			    0.30 & 0.247 & 0.004 & 0.031 &			  0.30 & 0.357 & 0.007 & 0.047   \\ \hline
                        0.18 & 0.619 & 0.007 & 0.075 &  		      0.34 & 0.442 & 0.006 & 0.059 &			    0.34 & 0.206 & 0.004 & 0.026 &			  0.34 & 0.276 & 0.006 & 0.037   \\ \hline
                        0.21 & 0.432 & 0.007 & 0.053 &  		      0.37 & 0.349 & 0.006 & 0.047 &			    0.37 & 0.221 & 0.004 & 0.028 &			  0.37 & 0.214 & 0.006 & 0.029   \\ \hline
                        0.24 & 0.405 & 0.006 & 0.050 &  		      0.41 & 0.258 & 0.007 & 0.035 &			    0.41 & 0.159 & 0.003 & 0.020 &			  0.41 & 0.180 & 0.006 & 0.024   \\ \hline
                        0.27 & 0.457 & 0.007 & 0.058 &  		      0.45 & 0.198 & 0.007 & 0.028 &			    0.45 & 0.151 & 0.004 & 0.020 &			  0.45 & 0.135 & 0.007 & 0.019   \\ \hline
                        0.30 & 0.306 & 0.005 & 0.039 &\multicolumn{4}{|c|}{$Q^2$=3.42 (GeV/c)$^2$, x=0.41} &\multicolumn{4}{|c|}{$Q^2$=3.42 (GeV/c)$^2$, x=0.59} &\multicolumn{4}{|c|}{$Q^2$=4.10 (GeV/c)$^2$, x=0.54}   \\ \hline
                        0.34 & 0.281 & 0.005 & 0.036 &  		      0.09 & 1.458 & 0.009 & 0.200 &			    0.15 & 0.406 & 0.006 & 0.049 &			  0.09 & 0.539 & 0.017 & 0.066   \\ \hline
                        0.37 & 0.219 & 0.003 & 0.028 &  		      0.12 & 1.270 & 0.007 & 0.173 &			    0.18 & 0.356 & 0.006 & 0.043 &			  0.12 & 0.629 & 0.007 & 0.077   \\ \hline
                        0.41 & 0.195 & 0.003 & 0.025 &  		      0.15 & 1.025 & 0.007 & 0.137 &			    0.21 & 0.291 & 0.007 & 0.036 &			  0.15 & 0.571 & 0.007 & 0.070   \\ \hline
                        0.50 & 0.135 & 0.003 & 0.018 &  		      0.18 & 0.866 & 0.005 & 0.114 &			    0.24 & 0.259 & 0.004 & 0.032 &			  0.18 & 0.458 & 0.006 & 0.056   \\ \hline
\multicolumn{4}{|c|}{$Q^2$=2.93 (GeV/c)$^2$, x=0.54} &  		      0.21 & 0.770 & 0.005 & 0.101 &			    0.27 & 0.237 & 0.004 & 0.030 &			  0.21 & 0.426 & 0.006 & 0.052   \\ \hline
                        0.15 & 0.605 & 0.008 & 0.074 &  		      0.24 & 0.662 & 0.005 & 0.086 &			    0.30 & 0.204 & 0.004 & 0.026 &			  0.24 & 0.364 & 0.005 & 0.045   \\ \hline
                        0.18 & 0.491 & 0.007 & 0.060 &  		      0.27 & 0.548 & 0.005 & 0.071 &			    0.34 & 0.167 & 0.003 & 0.021 &			  0.27 & 0.292 & 0.006 & 0.036   \\ \hline
                        0.21 & 0.327 & 0.006 & 0.040 &  		      0.30 & 0.474 & 0.006 & 0.061 &\multicolumn{4}{|c|}{$Q^2$=3.42 (GeV/c)$^2$, x=0.63} &			  0.30 & 0.268 & 0.006 & 0.033   \\ \hline
                        0.24 & 0.322 & 0.005 & 0.040 &  		      0.34 & 0.382 & 0.005 & 0.049 &			    0.15 & 0.193 & 0.007 & 0.024 &			  0.34 & 0.233 & 0.005 & 0.029   \\ \hline
                        0.27 & 0.329 & 0.005 & 0.042 &  		      0.37 & 0.300 & 0.005 & 0.039 &			    0.18 & 0.233 & 0.004 & 0.029 &			  0.37 & 0.204 & 0.007 & 0.026   \\ \hline
                        0.30 & 0.242 & 0.004 & 0.031 &  		      0.41 & 0.231 & 0.005 & 0.030 &			    0.21 & 0.159 & 0.006 & 0.020 &			  0.41 & 0.162 & 0.006 & 0.021   \\ \hline
                        0.34 & 0.221 & 0.004 & 0.028 &  		      0.45 & 0.200 & 0.006 & 0.026 &			    0.24 & 0.187 & 0.009 & 0.024 &			  0.45 & 0.130 & 0.006 & 0.017   \\ \hline
                        0.37 & 0.203 & 0.003 & 0.026 &  		      0.50 & 0.163 & 0.006 & 0.022 &			    0.27 & 0.181 & 0.007 & 0.023 &			  0.50 & 0.110 & 0.004 & 0.015   \\ \hline
                        0.41 & 0.136 & 0.003 & 0.018 &  		      0.54 & 0.128 & 0.006 & 0.018 &			    0.30 & 0.155 & 0.006 & 0.020 &\multicolumn{4}{|c|}{$Q^2$=4.10 (GeV/c)$^2$, x=0.59}   \\ \hline
\multicolumn{4}{|c|}{$Q^2$=2.93 (GeV/c)$^2$, x=0.59} &\multicolumn{4}{|c|}{$Q^2$=3.42 (GeV/c)$^2$, x=0.45} &			    0.34 & 0.117 & 0.004 & 0.015 &			  0.09 & 0.408 & 0.015 & 0.047   \\ \hline
                        0.18 & 0.396 & 0.006 & 0.052 &  		      0.09 & 0.982 & 0.009 & 0.118 &\multicolumn{4}{|c|}{$Q^2$=3.42 (GeV/c)$^2$, x=0.68} &			  0.12 & 0.510 & 0.007 & 0.059   \\ \hline
                        0.21 & 0.280 & 0.005 & 0.037 &  		      0.12 & 0.978 & 0.007 & 0.117 &			    0.15 & 0.083 & 0.004 & 0.010 &			  0.15 & 0.427 & 0.008 & 0.050   \\ \hline
                        0.24 & 0.248 & 0.007 & 0.033 &  		      0.15 & 0.848 & 0.007 & 0.102 &			    0.18 & 0.104 & 0.003 & 0.012 &			  0.18 & 0.337 & 0.006 & 0.039   \\ \hline
                        0.27 & 0.224 & 0.007 & 0.030 &  		      0.18 & 0.735 & 0.006 & 0.089 &			    0.24 & 0.082 & 0.003 & 0.010 &			  0.21 & 0.298 & 0.006 & 0.036   \\ \hline
                        0.30 & 0.177 & 0.004 & 0.024 &  		      0.21 & 0.623 & 0.005 & 0.075 &			    0.27 & 0.088 & 0.003 & 0.011 &			  0.24 & 0.312 & 0.007 & 0.037   \\ \hline
                        0.34 & 0.162 & 0.005 & 0.022 &  		      0.24 & 0.534 & 0.005 & 0.065 &\multicolumn{4}{|c|}{$Q^2$=4.10 (GeV/c)$^2$, x=0.41} &			  0.27 & 0.249 & 0.005 & 0.030   \\ \hline
                        0.37 & 0.149 & 0.005 & 0.020 &  		      0.27 & 0.478 & 0.005 & 0.058 &			    0.09 & 1.718 & 0.049 & 0.259 &			  0.30 & 0.229 & 0.005 & 0.028   \\ \hline
\multicolumn{4}{|c|}{$Q^2$=2.93 (GeV/c)$^2$, x=0.63} &  		      0.30 & 0.382 & 0.005 & 0.047 &			    0.12 & 1.406 & 0.039 & 0.209 &			  0.34 & 0.176 & 0.004 & 0.022   \\ \hline
                        0.15 & 0.122 & 0.006 & 0.017 &  		      0.34 & 0.308 & 0.004 & 0.038 &			    0.15 & 0.992 & 0.034 & 0.146 &			  0.37 & 0.144 & 0.004 & 0.018   \\ \hline
                        0.18 & 0.154 & 0.004 & 0.022 &  		      0.37 & 0.263 & 0.005 & 0.033 &			    0.18 & 0.918 & 0.029 & 0.133 &			  0.41 & 0.154 & 0.004 & 0.019   \\ \hline
                        0.21 & 0.138 & 0.003 & 0.020 &  		      0.41 & 0.211 & 0.004 & 0.026 &			    0.21 & 0.764 & 0.030 & 0.110 &\multicolumn{4}{|c|}{$Q^2$=4.10 (GeV/c)$^2$, x=0.63}   \\ \hline
                        0.24 & 0.125 & 0.003 & 0.018 &  		      0.45 & 0.212 & 0.007 & 0.028 &			    0.24 & 0.628 & 0.026 & 0.090 &			  0.15 & 0.333 & 0.006 & 0.041   \\ \hline
                        0.27 & 0.150 & 0.005 & 0.022 &  		      0.50 & 0.150 & 0.006 & 0.020 &			    0.30 & 0.383 & 0.024 & 0.055 &			  0.18 & 0.264 & 0.006 & 0.032   \\ \hline
                        0.34 & 0.099 & 0.002 & 0.014 &  		      0.54 & 0.137 & 0.007 & 0.019 &\multicolumn{4}{|c|}{$Q^2$=4.10 (GeV/c)$^2$, x=0.45} &			  0.21 & 0.218 & 0.005 & 0.027   \\ \hline
\multicolumn{4}{|c|}{$Q^2$=3.42 (GeV/c)$^2$, x=0.34} &\multicolumn{4}{|c|}{$Q^2$=3.42 (GeV/c)$^2$, x=0.50} &			    0.07 & 1.033 & 0.033 & 0.137 &			  0.24 & 0.207 & 0.004 & 0.026   \\ \hline
                        0.07 & 2.260 & 0.064 & 0.312 &  		      0.09 & 0.542 & 0.008 & 0.066 &			    0.09 & 1.245 & 0.013 & 0.177 &			  0.27 & 0.168 & 0.004 & 0.022   \\ \hline
                        0.12 & 1.835 & 0.024 & 0.264 &  		      0.12 & 0.827 & 0.010 & 0.101 &			    0.12 & 1.102 & 0.011 & 0.156 &			  0.30 & 0.168 & 0.004 & 0.022   \\ \hline
                        0.15 & 1.379 & 0.021 & 0.196 &  		      0.15 & 0.670 & 0.006 & 0.082 &			    0.15 & 0.896 & 0.010 & 0.124 &			  0.34 & 0.141 & 0.004 & 0.018   \\ \hline
                        0.18 & 1.133 & 0.018 & 0.159 &  		      0.18 & 0.592 & 0.007 & 0.072 &			    0.18 & 0.746 & 0.008 & 0.102 &\multicolumn{4}{|c|}{$Q^2$=4.10 (GeV/c)$^2$, x=0.68}   \\ \hline
                        0.21 & 0.977 & 0.018 & 0.137 &  		      0.21 & 0.510 & 0.006 & 0.063 &			    0.21 & 0.648 & 0.008 & 0.088 &			  0.15 & 0.147 & 0.003 & 0.018   \\ \hline
                        0.18 & 0.131 & 0.003 & 0.016 &  		      0.09 & 0.599 & 0.012 & 0.081 &			    0.15 & 0.318 & 0.008 & 0.042 &			  0.30 & 0.108 & 0.004 & 0.014   \\ \hline
                        0.21 & 0.136 & 0.006 & 0.017 &  		      0.12 & 0.492 & 0.010 & 0.064 &			    0.18 & 0.228 & 0.007 & 0.030 &			  0.34 & 0.100 & 0.004 & 0.013   \\ \hline
                        0.24 & 0.131 & 0.006 & 0.016 &  		      0.15 & 0.435 & 0.010 & 0.057 &			    0.21 & 0.216 & 0.006 & 0.029 &\multicolumn{4}{|c|}{$Q^2$=4.85 (GeV/c)$^2$, x=0.74}   \\ \hline
\multicolumn{4}{|c|}{$Q^2$=4.10 (GeV/c)$^2$, x=0.74} &  		      0.18 & 0.370 & 0.008 & 0.048 &			    0.24 & 0.189 & 0.006 & 0.025 &			  0.15 & 0.086 & 0.003 & 0.010   \\ \hline
                        0.15 & 0.076 & 0.005 & 0.009 &  		      0.21 & 0.321 & 0.008 & 0.042 &			    0.27 & 0.170 & 0.005 & 0.023 &			  0.21 & 0.082 & 0.004 & 0.010   \\ \hline
                        0.18 & 0.067 & 0.003 & 0.008 &  		      0.24 & 0.304 & 0.008 & 0.040 &			    0.30 & 0.138 & 0.006 & 0.019 &\multicolumn{4}{|c|}{$Q^2$=5.72 (GeV/c)$^2$, x=0.68}   \\ \hline
\end{tabular}
\end{center}
\end{table*}

\begin{table*}[ht]
\begin{center}
\caption{Continue from Table~\ref{table:h2_sidis}.}
\vspace{2mm}
\begin{tabular}{|c|c|c|c|c|c|c|c|c|c|c|c|c|c|c|c|} \hline
$z$ & $H_2$ & stat.err. & sys.err. & $z$ & $H_2$ & stat.err. & sys.err. & $z$ & $H_2$ & stat.err. & sys.err. & $z$ & $H_2$ & stat.err. & sys.err. \\ \hline
\multicolumn{4}{|c|}{$Q^2$=4.85 (GeV/c)$^2$, x=0.54} &  		      0.27 & 0.230 & 0.007 & 0.030 &			    0.34 & 0.144 & 0.005 & 0.020 &			  0.09 & 0.308 & 0.024 & 0.047   \\ \hline
                        0.09 & 0.817 & 0.023 & 0.124 &  		      0.30 & 0.202 & 0.009 & 0.027 &			    0.37 & 0.111 & 0.005 & 0.016 &			  0.12 & 0.265 & 0.011 & 0.041   \\ \hline
                        0.12 & 0.658 & 0.016 & 0.098 &  		      0.34 & 0.184 & 0.011 & 0.025 &\multicolumn{4}{|c|}{$Q^2$=4.85 (GeV/c)$^2$, x=0.68} &			  0.15 & 0.205 & 0.014 & 0.032   \\ \hline
                        0.15 & 0.600 & 0.018 & 0.088 &  		      0.37 & 0.127 & 0.009 & 0.017 &			    0.15 & 0.218 & 0.005 & 0.027 &			  0.18 & 0.156 & 0.013 & 0.024   \\ \hline
                        0.18 & 0.499 & 0.014 & 0.073 &  		      0.41 & 0.141 & 0.013 & 0.019 &			    0.18 & 0.227 & 0.009 & 0.028 &			  0.21 & 0.124 & 0.010 & 0.019   \\ \hline
                        0.21 & 0.424 & 0.014 & 0.061 &\multicolumn{4}{|c|}{$Q^2$=4.85 (GeV/c)$^2$, x=0.63} &			    0.21 & 0.145 & 0.006 & 0.018 &			  0.30 & 0.114 & 0.011 & 0.018   \\ \hline
                        0.27 & 0.305 & 0.015 & 0.044 &  		      0.09 & 0.345 & 0.012 & 0.045 &			    0.24 & 0.136 & 0.004 & 0.017 &\multicolumn{4}{|c|}{$Q^2$=5.72 (GeV/c)$^2$, x=0.74}   \\ \hline
\multicolumn{4}{|c|}{$Q^2$=4.85 (GeV/c)$^2$, x=0.59} &  		      0.12 & 0.350 & 0.006 & 0.046 &			    0.27 & 0.129 & 0.004 & 0.017 &			  0.15 & 0.127 & 0.008 & 0.019   \\ \hline
\end{tabular}
\end{center}
\end{table*}

\subsection{\label{sec:cos} Azimuthal Moments}
\begin{figure*}[htb]
\begin{center}
\includegraphics[bb=0cm 7cm 20cm 27cm, scale=0.9]{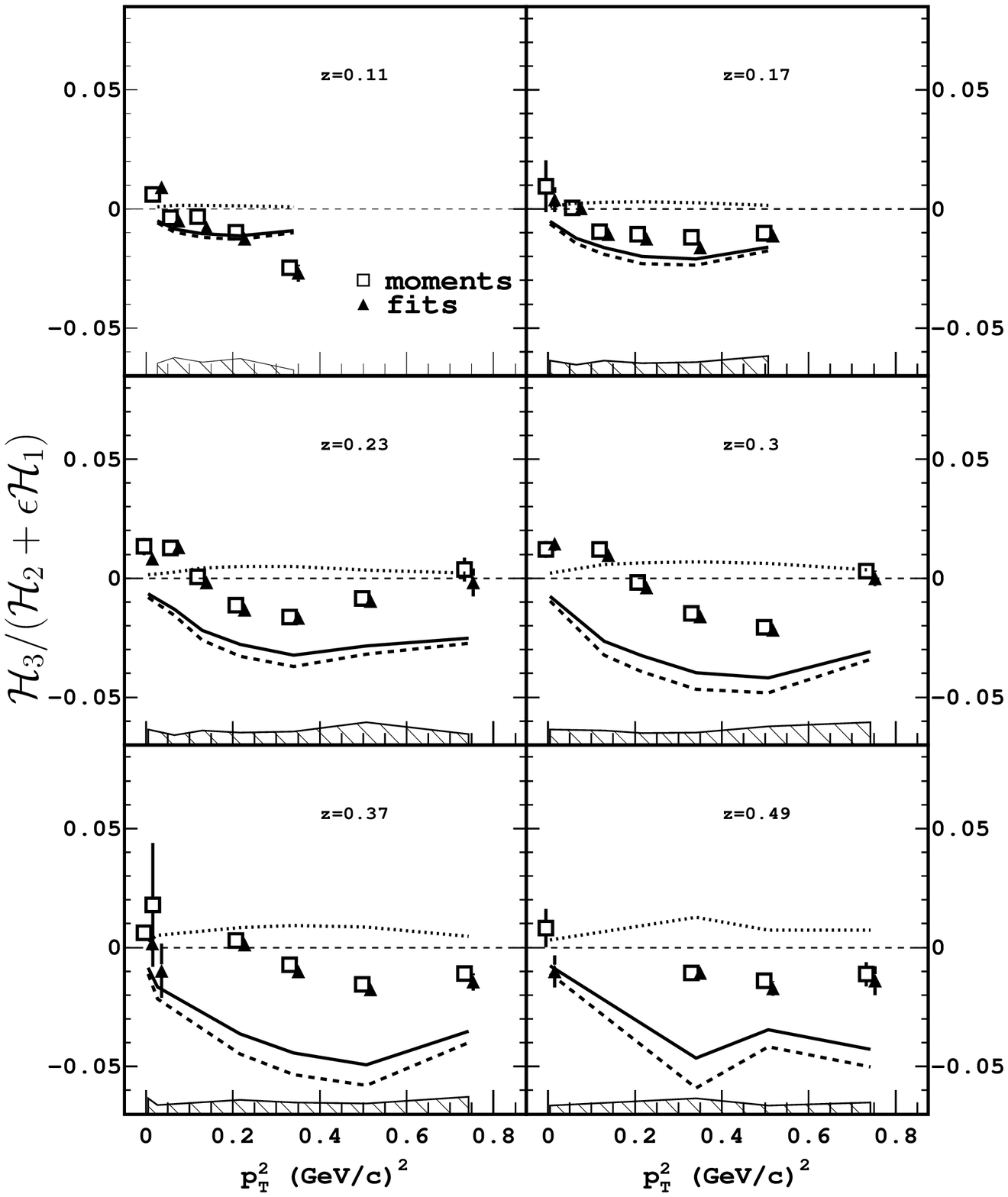}
\caption{\label{fig:cos_ptdep} The $p_T^2$-dependence of
$\mathcal{H}_3/(\mathcal{H}_2+\epsilon\mathcal{H}_1)$
(open squares - moments, full triangles - fits)
for different $z$ averaged over $x$ and $Q^2$.
The thick curves show theoretical predictions of the Cahn effect~\cite{Cahn,Ans1} (dashed),
predictions of the Berger effect~\cite{Brandenbug} using a convex pion wave function (dotted)
and their sum (solid).
The two data sets (from moments and fits extractions) are shifted equally along the $x$-axis
in opposite directions from their central values for visibility.
The hatched area show the systematic uncertainties.}
\end{center}
\end{figure*}

\begin{figure*}[htb]
\begin{center}
\includegraphics[bb=0cm 7cm 20cm 27cm, scale=0.9]{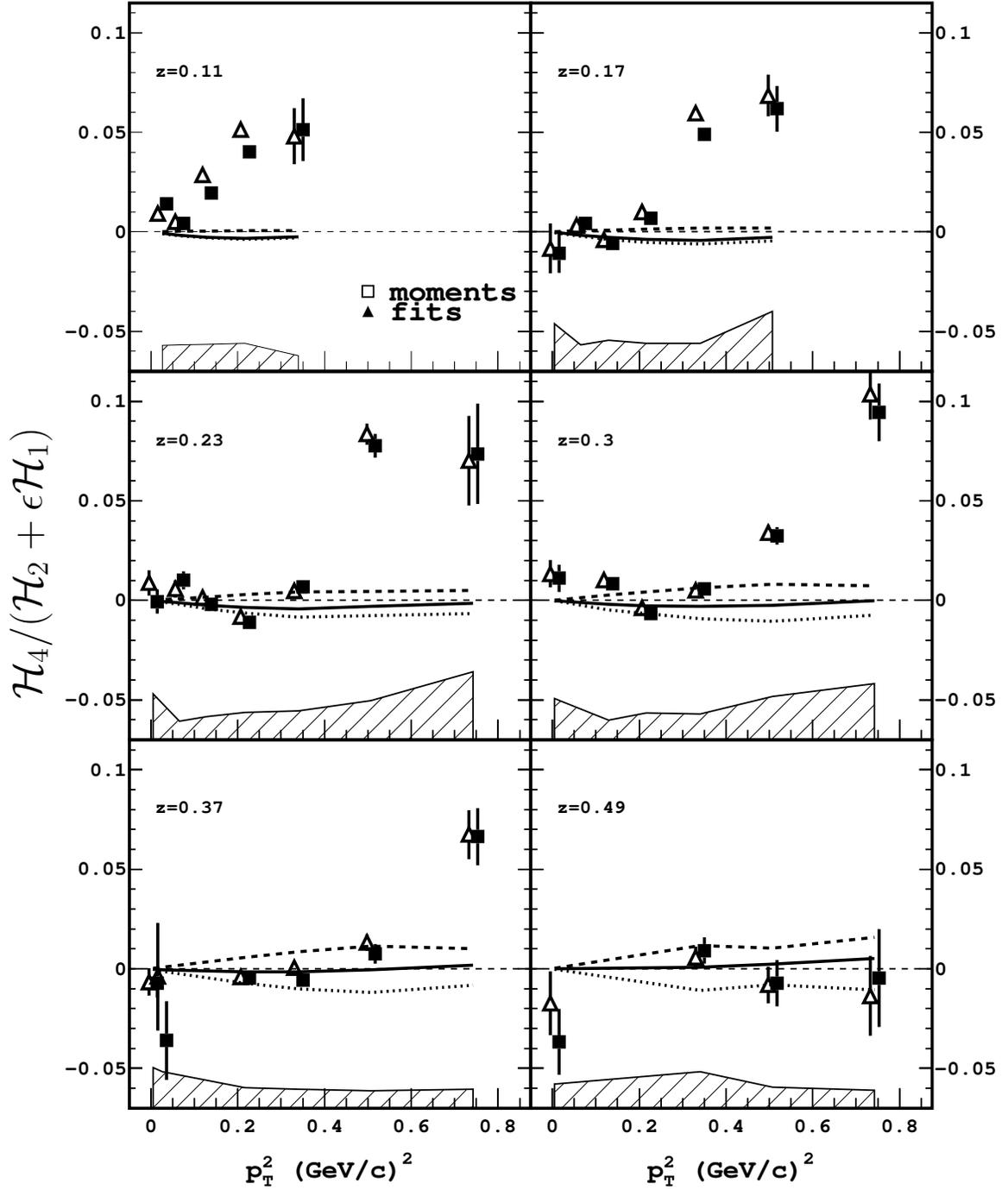}
\caption{\label{fig:cos2_ptdep} Same as Fig.~\ref{fig:cos_ptdep}
except for $\mathcal{H}_4/(\mathcal{H}_2+\epsilon\mathcal{H}_1)$
arising from the $\langle\cos{2\phi}\rangle$ moment (open triangles - moments, full squares - fits).}
\end{center}
\end{figure*}

\begin{figure*}[htb]
\begin{center}
\includegraphics[bb=0cm 7cm 20cm 27cm, scale=0.9]{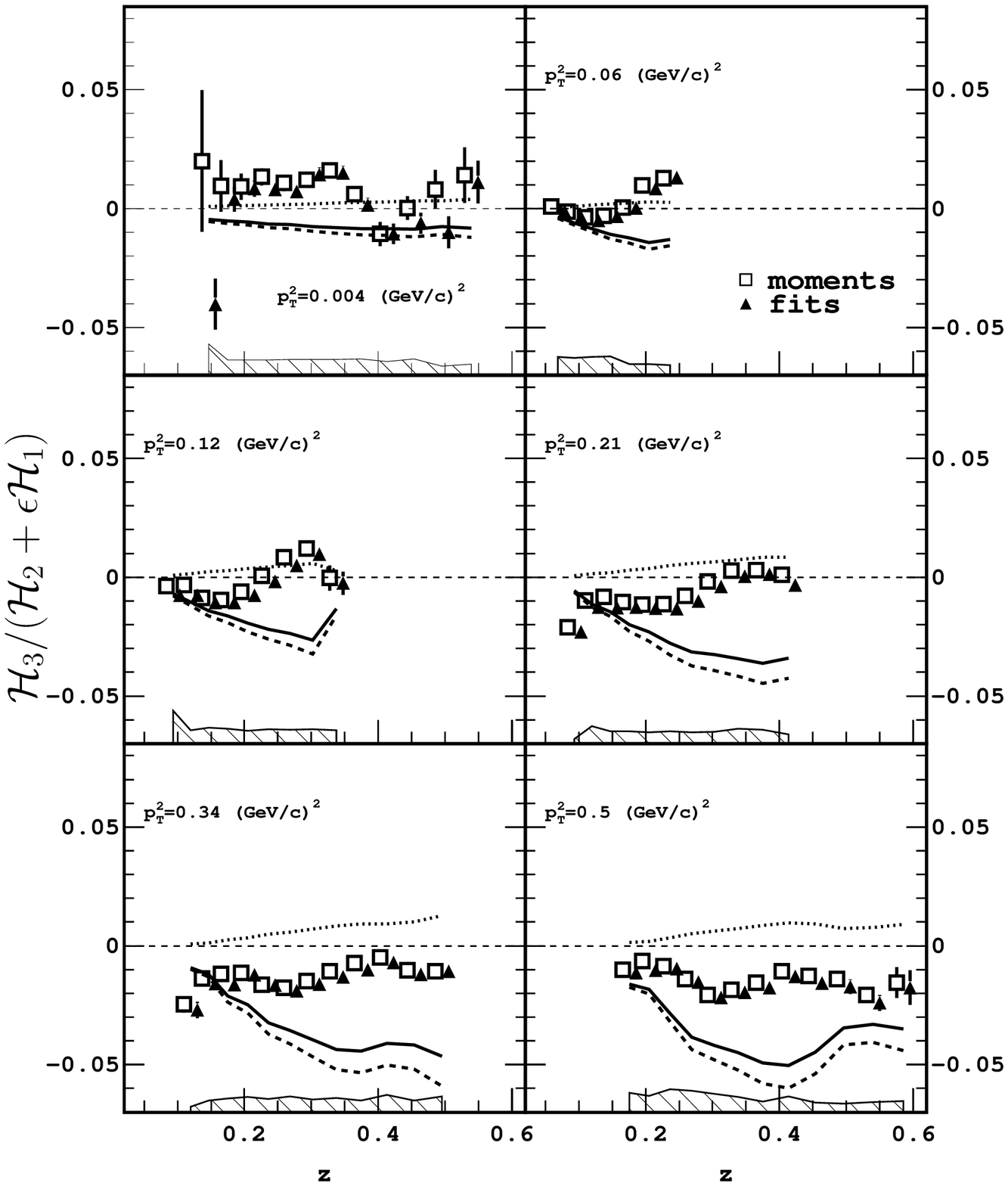}
\caption{\label{fig:cos_zdep} Same as Fig.~\ref{fig:cos_ptdep} except
with $\mathcal{H}_3/(\mathcal{H}_2+\epsilon\mathcal{H}_1)$ plotted as a
function of $z$ rather than $p_T^2$.}
\end{center}
\end{figure*}

\begin{figure*}[htb]
\begin{center}
\includegraphics[bb=0cm 7cm 20cm 27cm, scale=0.9]{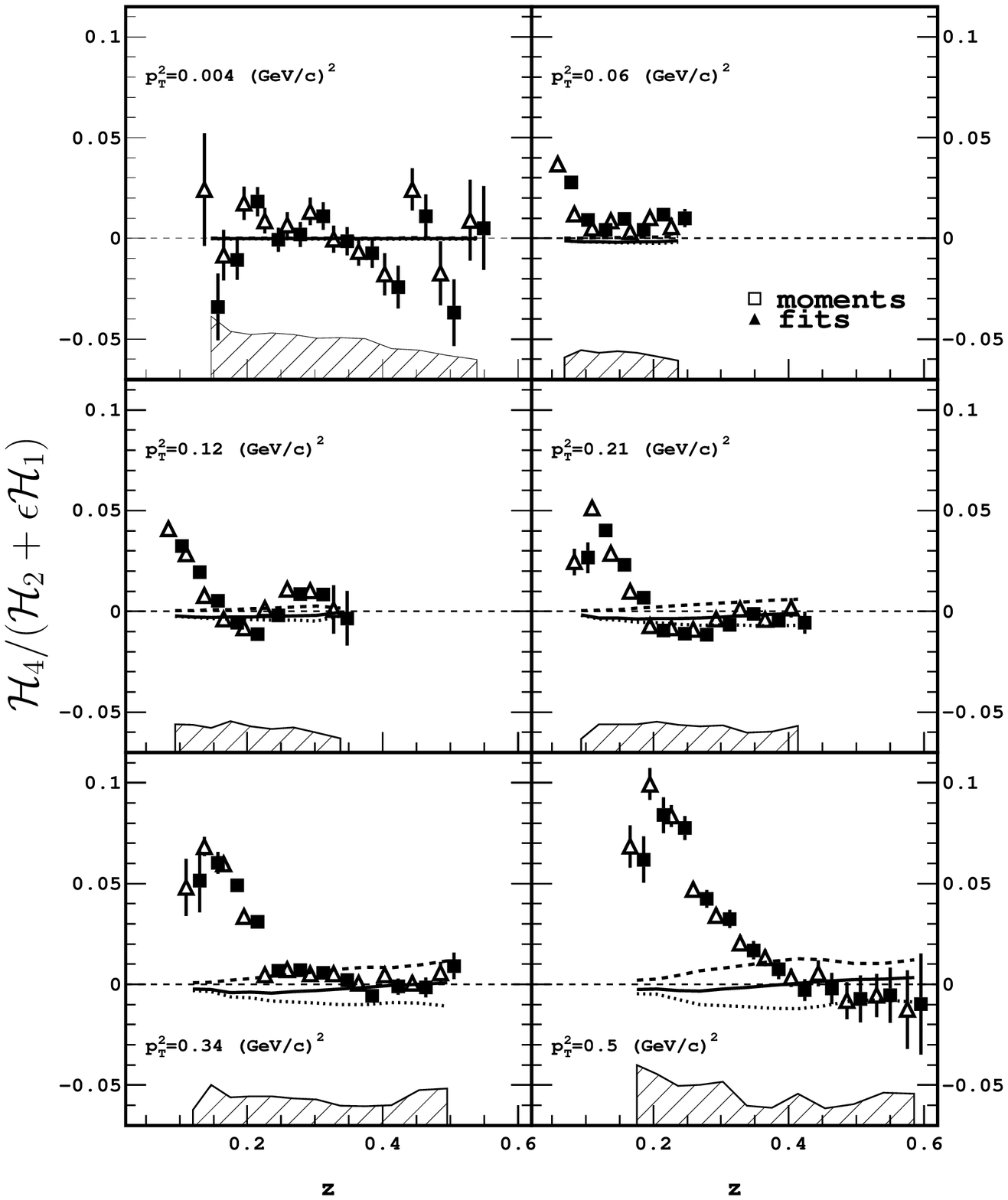}
\caption{\label{fig:cos2_zdep} Same as Fig.~\ref{fig:cos2_ptdep} except with
$\mathcal{H}_4/(\mathcal{H}_2+\epsilon\mathcal{H}_1)$ plotted as a
function of $z$ rather than $p_T^2$.}
\end{center}
\end{figure*}

Figs.~\ref{fig:cos_ptdep}, \ref{fig:cos2_ptdep}, \ref{fig:cos_zdep} and \ref{fig:cos2_zdep} show the
$p_T^2$ and $z$-dependencies of
$\mathcal{H}_3/(\mathcal{H}_2+\epsilon\mathcal{H}_1)$ and
$\mathcal{H}_4/(\mathcal{H}_2+\epsilon\mathcal{H}_1)$.
The $\phi$-dependent terms are typically less than a few percent of the $\phi$-independent part
of the semi-inclusive cross section.
The $\langle\cos{2\phi}\rangle$ moments are generally compatible with zero
within our systematic uncertainties, excluding low-$z$ and high-$p_T$ region
where they are definitely positive.
The $\langle\cos{\phi}\rangle$ moments are more significant due to smaller systematic uncertainty
and they are negative at large-$p_T$.
By exploiting the broad kinematic coverage of CLAS, we can explore
the overall trends of the data.

The $\langle\cos{\phi}\rangle$ term shown in Fig.~\ref{fig:cos_ptdep} tends to decrease
as a function of $p_T$ and eventually becomes negative. For most of the $p_T$
range the data are 2-3 systematic deviations below zero.

As one can see in Fig.~\ref{fig:cos2_ptdep}, the $\langle\cos{2\phi}\rangle$ term
is compatible with zero point-by-point, except for the low-$z$ and large $p_T$
where it is positive.

The $z$-dependence of $\langle\cos{\phi}\rangle$ shown in Fig.~\ref{fig:cos_zdep}
has a very different behavior at the lowest $p_T$ and at higher $p_T$: at the lowest $p_T$,
$\langle\cos{\phi}\rangle$ is compatible with zero,
whereas at higher $p_T$, except for the highest $p_T$ point,
$\langle\cos{\phi}\rangle$ rises from negative to positive values.

The $\langle\cos{2\phi}\rangle$ term shown in Fig.~\ref{fig:cos2_zdep} does not exhibit a clear $z$-dependence,
except for low-$z$ region where positive values decreasing with $z$ can be seen.
Above that region $\langle\cos{2\phi}\rangle$ is generally smaller than the systematic uncertainties.

Theoretical predictions in the $\langle\cos{\phi}\rangle$
are in strong disagreement with our data.
Indeed the full curve of the predictions, which has a similar dependence on $p_T$
but very different $z$-dependence, lies many
standard deviations below the measured points over much of the kinematics.
This is due to the dominant, negative Cahn effect contribution.
The positive contribution of the Berger effect slightly compensates
for the Cahn effect, but the Berger contribution is too small to bring
the sum of the two effects in agreement with the data.

Theory predicts very small $\langle\cos{2\phi}\rangle$ values partially due to
cancellation between the Cahn and Berger effect contributions.
These predictions are generally in agreement with our data.
The data points at large $p_T$ and low $z$
lie above the theoretical curves,
this difference reaches 2-3 systematic deviations.


The averaged structure function ratios $\mathcal{H}_{3,4}/(\mathcal{H}_2+\epsilon\mathcal{H}_1)$ shown in
Figs.~\ref{fig:cos_ptdep}, \ref{fig:cos2_ptdep}, \ref{fig:cos_zdep} and \ref{fig:cos2_zdep}
are listed in Tables~\ref{table:d_cos_phi} and \ref{table:d_cos_2phi}.
We notice that the use of the weighted average technique is not strictly justified
over the entire range of $z$ and $p_T$. In some points, in particular in the low-$z$ region,
the data show a clear $x$ and/or $Q^2$-dependence leading to an underestimation
of the averaged statistical uncertainties. Although, the full uncertainty on the averaged
data is in any case dominated by the systematic uncertainty.
Not averaged data obtained from the two methods are statistically compatible.

\begin{table*}[ht]
\begin{center}
\caption{The extracted data on $\mathcal{H}_3/(\mathcal{H}_2+\epsilon\mathcal{H}_1)$
averaged over $x$ and $Q^2$ with their statistical and systematic uncertainties.
The values given in brackets are obtained using the fit method.}
\label{table:d_cos_phi}
\vspace{2mm}
\begin{tabular}{|c|c|c|c|c|c|c|c|} \hline
$z$ & $p_T^2$ & $\langle y\rangle$  & $\langle Q^2\rangle$ & $\langle x\rangle$ & $\mathcal{H}_3/(\mathcal{H}_2+\epsilon\mathcal{H}_1)$ & stat. uncertainty & sys. uncertainty \\
    & (GeV/c)$^2$ & & (GeV/c)$^2$ & &  &  &  \\ \hline
0.068 & 0.026 & 0.88 (0.87) & 2.12 (2.22) & 0.23 (0.24) &  0.012 ( 0.012) & 0.0007 (0.0007) &  0.005 \\ \hline
0.068 & 0.065 & 0.88 (0.87) & 2.14 (2.19) & 0.23 (0.23) &  0.001 (-0.001) & 0.0004 (0.0005) &  0.008 \\ \hline
0.093 & 0.026 & 0.85 (0.84) & 2.14 (2.30) & 0.24 (0.26) &  0.009 ( 0.010) & 0.0005 (0.0005) &  0.006 \\ \hline
0.093 & 0.065 & 0.83 (0.83) & 2.27 (2.31) & 0.26 (0.26) & -0.001 (-0.003) & 0.0002 (0.0003) &  0.007 \\ \hline
0.093 & 0.129 & 0.87 (0.87) & 2.10 (2.13) & 0.22 (0.23) & -0.004 (-0.008) & 0.0004 (0.0004) &  0.014 \\ \hline
0.093 & 0.217 & 0.92 (0.92) & 1.85 (1.86) & 0.19 (0.19) & -0.021 (-0.023) & 0.0015 (0.0017) &  0.002 \\ \hline
0.119 & 0.026 & 0.84 (0.83) & 2.12 (2.26) & 0.24 (0.26) &  0.006 ( 0.009) & 0.0009 (0.0008) &  0.005 \\ \hline
0.119 & 0.065 & 0.81 (0.81) & 2.28 (2.35) & 0.27 (0.27) & -0.004 (-0.005) & 0.0003 (0.0003) &  0.008 \\ \hline
0.119 & 0.129 & 0.84 (0.84) & 2.23 (2.27) & 0.25 (0.25) & -0.003 (-0.008) & 0.0003 (0.0003) &  0.006 \\ \hline
0.119 & 0.217 & 0.88 (0.88) & 2.01 (2.03) & 0.21 (0.22) & -0.010 (-0.013) & 0.0006 (0.0006) &  0.007 \\ \hline
0.119 & 0.340 & 0.93 (0.93) & 1.66 (1.66) & 0.17 (0.17) & -0.025 (-0.027) & 0.0033 (0.0035) &  0.002 \\ \hline
0.147 & 0.005 & 0.61 (0.59) & 2.83 (2.65) & 0.42 (0.41) &  0.020 (-0.040) & 0.0298 (0.0107) &  0.013 \\ \hline
0.147 & 0.026 & 0.89 (0.91) & 1.89 (1.99) & 0.20 (0.20) &  0.007 ( 0.012) & 0.0035 (0.0013) &  0.003 \\ \hline
0.147 & 0.065 & 0.81 (0.80) & 2.20 (2.32) & 0.26 (0.27) & -0.003 (-0.003) & 0.0004 (0.0004) &  0.008 \\ \hline
0.147 & 0.129 & 0.82 (0.82) & 2.23 (2.28) & 0.26 (0.26) & -0.009 (-0.011) & 0.0003 (0.0004) &  0.007 \\ \hline
0.147 & 0.217 & 0.85 (0.85) & 2.14 (2.18) & 0.24 (0.24) & -0.008 (-0.013) & 0.0004 (0.0005) &  0.005 \\ \hline
0.147 & 0.340 & 0.91 (0.91) & 1.99 (2.01) & 0.20 (0.20) & -0.014 (-0.016) & 0.0011 (0.0013) &  0.005 \\ \hline
0.175 & 0.005 & 0.59 (0.60) & 2.57 (2.55) & 0.40 (0.39) &  0.009 ( 0.004) & 0.0109 (0.0052) &  0.006 \\ \hline
0.175 & 0.065 & 0.82 (0.81) & 2.19 (2.31) & 0.25 (0.27) &  0.000 ( 0.000) & 0.0005 (0.0005) &  0.004 \\ \hline
0.175 & 0.129 & 0.82 (0.82) & 2.28 (2.34) & 0.26 (0.27) & -0.010 (-0.011) & 0.0004 (0.0004) &  0.006 \\ \hline
0.175 & 0.217 & 0.84 (0.84) & 2.20 (2.24) & 0.24 (0.25) & -0.011 (-0.013) & 0.0004 (0.0005) &  0.005 \\ \hline
0.175 & 0.340 & 0.87 (0.87) & 2.02 (2.04) & 0.22 (0.22) & -0.012 (-0.016) & 0.0007 (0.0008) &  0.006 \\ \hline
0.175 & 0.507 & 0.92 (0.92) & 1.85 (1.86) & 0.19 (0.19) & -0.010 (-0.011) & 0.0024 (0.0026) &  0.008 \\ \hline
0.205 & 0.005 & 0.60 (0.62) & 2.65 (2.68) & 0.41 (0.40) &  0.009 ( 0.008) & 0.0056 (0.0035) &  0.006 \\ \hline
0.205 & 0.065 & 0.82 (0.81) & 2.12 (2.23) & 0.25 (0.26) &  0.010 ( 0.008) & 0.0008 (0.0007) &  0.004 \\ \hline
0.205 & 0.129 & 0.81 (0.81) & 2.27 (2.34) & 0.26 (0.27) & -0.006 (-0.008) & 0.0004 (0.0005) &  0.005 \\ \hline
0.205 & 0.217 & 0.84 (0.83) & 2.24 (2.28) & 0.25 (0.26) & -0.012 (-0.013) & 0.0005 (0.0005) &  0.005 \\ \hline
0.205 & 0.340 & 0.87 (0.86) & 2.14 (2.17) & 0.23 (0.23) & -0.011 (-0.012) & 0.0007 (0.0008) &  0.006 \\ \hline
0.205 & 0.507 & 0.92 (0.92) & 1.85 (1.87) & 0.19 (0.19) & -0.006 (-0.010) & 0.0018 (0.0020) &  0.007 \\ \hline
0.236 & 0.005 & 0.59 (0.61) & 2.56 (2.61) & 0.40 (0.39) &  0.013 ( 0.008) & 0.0036 (0.0027) &  0.006 \\ \hline
0.236 & 0.065 & 0.86 (0.86) & 1.86 (1.87) & 0.20 (0.20) &  0.013 ( 0.013) & 0.0024 (0.0020) &  0.004 \\ \hline
0.236 & 0.129 & 0.81 (0.81) & 2.23 (2.32) & 0.26 (0.27) &  0.001 (-0.002) & 0.0005 (0.0005) &  0.006 \\ \hline
0.236 & 0.217 & 0.82 (0.82) & 2.23 (2.28) & 0.26 (0.26) & -0.011 (-0.013) & 0.0005 (0.0005) &  0.005 \\ \hline
0.236 & 0.340 & 0.84 (0.84) & 2.15 (2.17) & 0.24 (0.24) & -0.016 (-0.017) & 0.0006 (0.0007) &  0.006 \\ \hline
0.236 & 0.507 & 0.90 (0.90) & 1.97 (1.99) & 0.20 (0.21) & -0.009 (-0.010) & 0.0014 (0.0016) &  0.010 \\ \hline
0.236 & 0.743 & 0.93 (0.93) & 1.66 (1.67) & 0.17 (0.17) &  0.004 (-0.002) & 0.0050 (0.0058) &  0.004 \\ \hline
0.269 & 0.005 & 0.61 (0.63) & 2.66 (2.67) & 0.40 (0.39) &  0.011 ( 0.007) & 0.0033 (0.0027) &  0.006 \\ \hline
0.269 & 0.129 & 0.81 (0.83) & 2.17 (2.14) & 0.25 (0.24) &  0.009 ( 0.005) & 0.0007 (0.0007) &  0.006 \\ \hline
0.269 & 0.217 & 0.81 (0.81) & 2.23 (2.28) & 0.26 (0.26) & -0.008 (-0.010) & 0.0005 (0.0006) &  0.005 \\ \hline
0.269 & 0.340 & 0.84 (0.84) & 2.15 (2.19) & 0.24 (0.24) & -0.018 (-0.019) & 0.0007 (0.0008) &  0.007 \\ \hline
0.269 & 0.507 & 0.87 (0.87) & 2.03 (2.06) & 0.22 (0.22) & -0.014 (-0.015) & 0.0011 (0.0013) &  0.009 \\ \hline
0.269 & 0.743 & 0.92 (0.92) & 1.77 (1.78) & 0.18 (0.18) &  0.005 ( 0.003) & 0.0036 (0.0041) &  0.008 \\ \hline
0.303 & 0.005 & 0.62 (0.64) & 2.48 (2.53) & 0.37 (0.37) &  0.012 ( 0.014) & 0.0035 (0.0029) &  0.007 \\ \hline
0.303 & 0.129 & 0.83 (0.82) & 2.03 (2.11) & 0.23 (0.24) &  0.012 ( 0.010) & 0.0012 (0.0012) &  0.006 \\ \hline
0.303 & 0.217 & 0.82 (0.82) & 2.21 (2.29) & 0.25 (0.26) & -0.002 (-0.004) & 0.0006 (0.0007) &  0.005 \\ \hline
0.303 & 0.340 & 0.83 (0.84) & 2.16 (2.20) & 0.24 (0.25) & -0.015 (-0.016) & 0.0007 (0.0008) &  0.005 \\ \hline
0.303 & 0.507 & 0.87 (0.87) & 2.03 (2.05) & 0.22 (0.22) & -0.021 (-0.022) & 0.0011 (0.0013) &  0.008 \\ \hline
0.303 & 0.743 & 0.92 (0.92) & 1.86 (1.86) & 0.19 (0.19) &  0.003 ( 0.000) & 0.0029 (0.0033) &  0.009 \\ \hline
0.338 & 0.005 & 0.63 (0.64) & 2.50 (2.53) & 0.37 (0.36) &  0.016 ( 0.015) & 0.0033 (0.0029) &  0.007 \\ \hline
0.338 & 0.026 & 0.63 (0.62) & 2.71 (2.68) & 0.40 (0.40) &  0.002 (-0.009) & 0.0386 (0.0138) &  0.006 \\ \hline
0.338 & 0.129 & 0.92 (0.92) & 1.74 (1.74) & 0.18 (0.18) &  0.000 (-0.003) & 0.0053 (0.0049) &  0.006 \\ \hline
0.338 & 0.217 & 0.83 (0.83) & 2.20 (2.29) & 0.25 (0.26) &  0.003 ( 0.000) & 0.0008 (0.0009) &  0.006 \\ \hline
0.338 & 0.340 & 0.83 (0.83) & 2.16 (2.20) & 0.24 (0.25) & -0.011 (-0.013) & 0.0008 (0.0009) &  0.006 \\ \hline
0.338 & 0.507 & 0.86 (0.87) & 2.03 (2.06) & 0.22 (0.22) & -0.019 (-0.020) & 0.0012 (0.0014) &  0.007 \\ \hline
0.338 & 0.743 & 0.92 (0.92) & 1.87 (1.87) & 0.19 (0.19) & -0.008 (-0.011) & 0.0030 (0.0034) &  0.010 \\ \hline
 \end{tabular}
 \end{center}
 \end{table*}
 
 \begin{table*}[ht]
 \begin{center}
 \begin{tabular}{|c|c|c|c|c|c|c|c|} \hline
 $z$ & $p_T^2$ & $\langle y\rangle$ & $\langle Q^2\rangle$ & $\langle x\rangle$ & $\mathcal{H}_3/(\mathcal{H}_2+\epsilon\mathcal{H}_1)$ & stat. uncertainty & sys. uncertainty \\
     & (GeV/c)$^2$ &  & (GeV/c)$^2$ & &  &  &  \\ \hline
0.375 & 0.005 & 0.63 (0.64) & 2.50 (2.53) & 0.37 (0.36) &  0.006 ( 0.002) & 0.0032 (0.0029) &  0.007 \\ \hline
0.375 & 0.026 & 0.67 (0.67) & 3.13 (3.11) & 0.43 (0.43) &  0.018 (-0.010) & 0.0260 (0.0113) &  0.004 \\ \hline
0.375 & 0.217 & 0.84 (0.83) & 2.10 (2.19) & 0.23 (0.25) &  0.003 ( 0.001) & 0.0012 (0.0012) &  0.006 \\ \hline
0.375 & 0.340 & 0.84 (0.84) & 2.13 (2.19) & 0.24 (0.24) & -0.007 (-0.010) & 0.0010 (0.0011) &  0.005 \\ \hline
0.375 & 0.507 & 0.86 (0.86) & 2.04 (2.07) & 0.22 (0.22) & -0.015 (-0.018) & 0.0013 (0.0015) &  0.005 \\ \hline
0.375 & 0.743 & 0.92 (0.92) & 1.84 (1.88) & 0.19 (0.19) & -0.011 (-0.014) & 0.0031 (0.0035) &  0.007 \\ \hline
0.413 & 0.005 & 0.66 (0.68) & 2.34 (2.38) & 0.33 (0.32) & -0.011 (-0.011) & 0.0051 (0.0044) &  0.006 \\ \hline
0.413 & 0.026 & 0.65 (0.65) & 2.37 (2.37) & 0.34 (0.34) & -0.003 (-0.002) & 0.1008 (0.0280) &  0.007 \\ \hline
0.413 & 0.217 & 0.87 (0.86) & 1.95 (1.99) & 0.21 (0.22) &  0.001 (-0.003) & 0.0023 (0.0023) &  0.004 \\ \hline
0.413 & 0.340 & 0.86 (0.86) & 2.17 (2.24) & 0.23 (0.24) & -0.005 (-0.007) & 0.0012 (0.0014) &  0.007 \\ \hline
0.413 & 0.507 & 0.86 (0.87) & 2.04 (2.08) & 0.22 (0.22) & -0.011 (-0.013) & 0.0015 (0.0017) &  0.007 \\ \hline
0.413 & 0.743 & 0.92 (0.92) & 1.85 (1.87) & 0.19 (0.19) & -0.016 (-0.017) & 0.0032 (0.0036) &  0.004 \\ \hline
0.454 & 0.005 & 0.66 (0.68) & 2.35 (2.37) & 0.33 (0.32) &  0.000 (-0.006) & 0.0050 (0.0045) &  0.007 \\ \hline
0.454 & 0.026 & 0.68 (0.68) & 2.61 (2.59) & 0.35 (0.35) & -0.007 ( 0.004) & 0.0401 (0.0141) &  0.004 \\ \hline
0.454 & 0.340 & 0.87 (0.87) & 2.02 (2.09) & 0.22 (0.22) & -0.010 (-0.012) & 0.0016 (0.0018) &  0.005 \\ \hline
0.454 & 0.507 & 0.89 (0.89) & 1.98 (2.02) & 0.21 (0.21) & -0.013 (-0.016) & 0.0020 (0.0023) &  0.004 \\ \hline
0.454 & 0.743 & 0.92 (0.92) & 1.85 (1.87) & 0.19 (0.19) & -0.018 (-0.024) & 0.0035 (0.0041) &  0.005 \\ \hline
0.496 & 0.005 & 0.72 (0.72) & 2.63 (2.53) & 0.34 (0.32) &  0.008 (-0.010) & 0.0080 (0.0068) &  0.004 \\ \hline
0.496 & 0.340 & 0.87 (0.86) & 1.96 (2.00) & 0.21 (0.22) & -0.011 (-0.011) & 0.0024 (0.0026) &  0.007 \\ \hline
0.496 & 0.507 & 0.92 (0.92) & 1.86 (1.88) & 0.19 (0.19) & -0.014 (-0.017) & 0.0026 (0.0030) &  0.004 \\ \hline
0.496 & 0.743 & 0.93 (0.92) & 1.67 (1.68) & 0.17 (0.17) & -0.011 (-0.014) & 0.0051 (0.0061) &  0.005 \\ \hline
0.540 & 0.005 & 0.71 (0.72) & 2.19 (2.20) & 0.28 (0.28) &  0.014 ( 0.011) & 0.0117 (0.0090) &  0.005 \\ \hline
0.540 & 0.507 & 0.92 (0.92) & 1.88 (1.91) & 0.19 (0.19) & -0.021 (-0.024) & 0.0029 (0.0034) &  0.004 \\ \hline
0.540 & 0.743 & 0.94 (0.94) & 1.49 (1.49) & 0.15 (0.15) & -0.019 (-0.042) & 0.0102 (0.0120) &  0.004 \\ \hline
0.586 & 0.507 & 0.92 (0.92) & 1.74 (1.74) & 0.18 (0.18) & -0.016 (-0.018) & 0.0065 (0.0072) &  0.005 \\ \hline
 \end{tabular}
 \end{center}
 \end{table*}

\begin{table*}[ht]
\begin{center}
\caption{The extracted data on $\mathcal{H}_4/(\mathcal{H}_2+\epsilon\mathcal{H}_1)$
averaged over $x$ and $Q^2$ with their statistical and systematic uncertainties.
The values given in brackets are obtained using the fit method.}
\label{table:d_cos_2phi}
\vspace{2mm}
\begin{tabular}{|c|c|c|c|c|c|c|c|} \hline
$z$ & $p_T^2$ & $\langle y\rangle$  & $\langle Q^2\rangle$ & $\langle x\rangle$ & $\mathcal{H}_4/(\mathcal{H}_2+\epsilon\mathcal{H}_1)$ & stat. uncertainty & sys. uncertainty \\
    & (GeV/c)$^2$ & & (GeV/c)$^2$ & &  &  &  \\ \hline
0.068 & 0.026 & 0.84 (0.84) & 2.19 (2.26) & 0.24 (0.25) &  0.001 ( 0.004) & 0.0017 (0.0021) &  0.018 \\ \hline
0.068 & 0.065 & 0.85 (0.85) & 2.15 (2.19) & 0.24 (0.24) &  0.037 ( 0.028) & 0.0013 (0.0018) &  0.011 \\ \hline
0.093 & 0.026 & 0.80 (0.79) & 2.23 (2.34) & 0.26 (0.28) &  0.006 ( 0.010) & 0.0011 (0.0012) &  0.013 \\ \hline
0.093 & 0.065 & 0.80 (0.80) & 2.29 (2.32) & 0.27 (0.27) &  0.012 ( 0.009) & 0.0007 (0.0009) &  0.015 \\ \hline
0.093 & 0.129 & 0.85 (0.85) & 2.11 (2.13) & 0.23 (0.23) &  0.041 ( 0.032) & 0.0013 (0.0016) &  0.014 \\ \hline
0.093 & 0.217 & 0.92 (0.92) & 1.86 (1.87) & 0.19 (0.19) &  0.025 ( 0.027) & 0.0066 (0.0076) &  0.007 \\ \hline
0.119 & 0.026 & 0.79 (0.78) & 2.18 (2.27) & 0.26 (0.27) &  0.009 ( 0.014) & 0.0016 (0.0017) &  0.013 \\ \hline
0.119 & 0.065 & 0.77 (0.77) & 2.31 (2.35) & 0.28 (0.29) &  0.005 ( 0.004) & 0.0008 (0.0009) &  0.013 \\ \hline
0.119 & 0.129 & 0.81 (0.81) & 2.24 (2.26) & 0.26 (0.26) &  0.028 ( 0.019) & 0.0010 (0.0012) &  0.014 \\ \hline
0.119 & 0.217 & 0.85 (0.85) & 2.01 (2.02) & 0.22 (0.22) &  0.051 ( 0.040) & 0.0020 (0.0024) &  0.014 \\ \hline
0.119 & 0.340 & 0.92 (0.92) & 1.68 (1.68) & 0.17 (0.17) &  0.048 ( 0.051) & 0.0142 (0.0157) &  0.008 \\ \hline
0.147 & 0.005 & 0.61 (0.59) & 2.75 (2.62) & 0.42 (0.41) &  0.024 (-0.034) & 0.0279 (0.0166) &  0.031 \\ \hline
0.147 & 0.026 & 0.87 (0.88) & 1.85 (1.90) & 0.20 (0.20) &  0.023 ( 0.027) & 0.0064 (0.0051) &  0.016 \\ \hline
0.147 & 0.065 & 0.75 (0.75) & 2.29 (2.34) & 0.29 (0.29) &  0.009 ( 0.010) & 0.0009 (0.0011) &  0.014 \\ \hline
0.147 & 0.129 & 0.78 (0.78) & 2.25 (2.28) & 0.27 (0.27) &  0.008 ( 0.005) & 0.0010 (0.0012) &  0.012 \\ \hline
0.147 & 0.217 & 0.82 (0.84) & 2.14 (2.43) & 0.24 (0.27) &  0.029 ( 0.023) & 0.0015 (0.0015) &  0.014 \\ \hline
0.147 & 0.340 & 0.91 (0.91) & 1.97 (1.99) & 0.20 (0.20) &  0.068 ( 0.060) & 0.0048 (0.0054) &  0.020 \\ \hline
0.175 & 0.005 & 0.59 (0.61) & 2.53 (2.60) & 0.39 (0.39) & -0.008 (-0.011) & 0.0126 (0.0100) &  0.024 \\ \hline
0.175 & 0.065 & 0.77 (0.76) & 2.26 (2.33) & 0.28 (0.29) &  0.003 ( 0.004) & 0.0011 (0.0013) &  0.013 \\ \hline
0.175 & 0.129 & 0.78 (0.78) & 2.32 (2.35) & 0.28 (0.28) & -0.004 (-0.006) & 0.0010 (0.0012) &  0.015 \\ \hline
0.175 & 0.217 & 0.81 (0.80) & 2.21 (2.23) & 0.26 (0.26) &  0.010 ( 0.007) & 0.0014 (0.0016) &  0.014 \\ \hline
0.175 & 0.340 & 0.85 (0.85) & 2.02 (2.03) & 0.22 (0.22) &  0.060 ( 0.049) & 0.0027 (0.0030) &  0.014 \\ \hline
0.175 & 0.507 & 0.92 (0.92) & 1.85 (1.86) & 0.19 (0.19) &  0.068 ( 0.062) & 0.0105 (0.0115) &  0.030 \\ \hline
0.205 & 0.005 & 0.61 (0.63) & 2.67 (2.71) & 0.40 (0.40) &  0.017 ( 0.018) & 0.0083 (0.0072) &  0.022 \\ \hline
0.205 & 0.065 & 0.77 (0.76) & 2.17 (2.25) & 0.26 (0.28) &  0.010 ( 0.012) & 0.0015 (0.0017) &  0.011 \\ \hline
0.205 & 0.129 & 0.77 (0.77) & 2.32 (2.35) & 0.28 (0.29) & -0.008 (-0.011) & 0.0011 (0.0013) &  0.013 \\ \hline
0.205 & 0.217 & 0.80 (0.80) & 2.26 (2.28) & 0.26 (0.27) & -0.007 (-0.010) & 0.0014 (0.0017) &  0.015 \\ \hline
0.205 & 0.340 & 0.84 (0.84) & 2.15 (2.17) & 0.24 (0.24) &  0.034 ( 0.031) & 0.0024 (0.0027) &  0.015 \\ \hline
0.205 & 0.507 & 0.92 (0.92) & 1.86 (1.87) & 0.19 (0.19) &  0.099 ( 0.084) & 0.0079 (0.0088) &  0.025 \\ \hline
0.236 & 0.005 & 0.61 (0.62) & 2.60 (2.62) & 0.39 (0.39) &  0.009 (-0.001) & 0.0063 (0.0060) &  0.023 \\ \hline
0.236 & 0.065 & 0.84 (0.83) & 1.87 (1.88) & 0.21 (0.21) &  0.006 ( 0.010) & 0.0044 (0.0044) &  0.009 \\ \hline
0.236 & 0.129 & 0.77 (0.77) & 2.31 (2.35) & 0.28 (0.29) &  0.001 (-0.002) & 0.0012 (0.0015) &  0.011 \\ \hline
0.236 & 0.217 & 0.78 (0.78) & 2.26 (2.28) & 0.27 (0.27) & -0.008 (-0.011) & 0.0014 (0.0017) &  0.014 \\ \hline
0.236 & 0.340 & 0.81 (0.81) & 2.15 (2.16) & 0.25 (0.25) &  0.005 ( 0.007) & 0.0021 (0.0024) &  0.014 \\ \hline
0.236 & 0.507 & 0.87 (0.87) & 1.93 (1.93) & 0.20 (0.21) &  0.084 ( 0.078) & 0.0053 (0.0060) &  0.020 \\ \hline
0.236 & 0.743 & 0.92 (0.92) & 1.67 (1.68) & 0.17 (0.17) &  0.070 ( 0.074) & 0.0226 (0.0252) &  0.034 \\ \hline
0.269 & 0.005 & 0.63 (0.64) & 2.66 (2.67) & 0.39 (0.38) &  0.007 ( 0.002) & 0.0063 (0.0062) &  0.022 \\ \hline
0.269 & 0.129 & 0.77 (0.77) & 2.23 (2.29) & 0.27 (0.28) &  0.011 ( 0.008) & 0.0015 (0.0018) &  0.012 \\ \hline
0.269 & 0.217 & 0.78 (0.78) & 2.26 (2.29) & 0.27 (0.27) & -0.009 (-0.012) & 0.0015 (0.0018) &  0.013 \\ \hline
0.269 & 0.340 & 0.81 (0.81) & 2.16 (2.17) & 0.25 (0.25) &  0.007 ( 0.007) & 0.0021 (0.0024) &  0.013 \\ \hline
0.269 & 0.507 & 0.84 (0.84) & 2.02 (2.03) & 0.22 (0.22) &  0.047 ( 0.042) & 0.0039 (0.0045) &  0.020 \\ \hline
0.269 & 0.743 & 0.92 (0.92) & 1.79 (1.80) & 0.18 (0.18) &  0.088 ( 0.072) & 0.0162 (0.0179) &  0.021 \\ \hline
0.303 & 0.005 & 0.63 (0.64) & 2.51 (2.52) & 0.36 (0.36) &  0.013 ( 0.011) & 0.0069 (0.0069) &  0.021 \\ \hline
0.303 & 0.129 & 0.79 (0.78) & 2.06 (2.11) & 0.25 (0.25) &  0.010 ( 0.008) & 0.0023 (0.0026) &  0.010 \\ \hline
0.303 & 0.217 & 0.79 (0.79) & 2.27 (2.30) & 0.27 (0.27) & -0.004 (-0.007) & 0.0017 (0.0021) &  0.013 \\ \hline
0.303 & 0.340 & 0.80 (0.80) & 2.17 (2.19) & 0.25 (0.25) &  0.005 ( 0.006) & 0.0022 (0.0025) &  0.013 \\ \hline
0.303 & 0.507 & 0.84 (0.84) & 2.02 (2.03) & 0.22 (0.22) &  0.034 ( 0.032) & 0.0039 (0.0044) &  0.022 \\ \hline
0.303 & 0.743 & 0.92 (0.92) & 1.86 (1.86) & 0.19 (0.19) &  0.103 ( 0.094) & 0.0127 (0.0144) &  0.028 \\ \hline
0.338 & 0.005 & 0.64 (0.65) & 2.50 (2.51) & 0.36 (0.36) & -0.001 (-0.001) & 0.0067 (0.0069) &  0.021 \\ \hline
0.338 & 0.026 & 0.62 (0.62) & 2.66 (2.64) & 0.39 (0.39) & -0.026 (-0.032) & 0.0351 (0.0203) &  0.022 \\ \hline
0.338 & 0.129 & 0.92 (0.92) & 1.74 (1.74) & 0.18 (0.18) &  0.001 (-0.004) & 0.0120 (0.0135) &  0.007 \\ \hline
0.338 & 0.217 & 0.79 (0.79) & 2.26 (2.31) & 0.27 (0.27) &  0.001 (-0.001) & 0.0020 (0.0024) &  0.010 \\ \hline
0.338 & 0.340 & 0.80 (0.80) & 2.17 (2.19) & 0.25 (0.26) &  0.005 ( 0.002) & 0.0023 (0.0028) &  0.010 \\ \hline
0.338 & 0.507 & 0.84 (0.84) & 2.02 (2.04) & 0.22 (0.23) &  0.020 ( 0.017) & 0.0039 (0.0046) &  0.010 \\ \hline
0.338 & 0.743 & 0.92 (0.92) & 1.86 (1.86) & 0.19 (0.19) &  0.087 ( 0.088) & 0.0124 (0.0138) &  0.015 \\ \hline
 \end{tabular}
 \end{center}
 \end{table*}
 
 \begin{table*}[ht]
 \begin{center}
 \begin{tabular}{|c|c|c|c|c|c|c|c|} \hline
 $z$ & $p_T^2$ & $\langle y\rangle$ & $\langle Q^2\rangle$ & $\langle x\rangle$ & $\mathcal{H}_4/(\mathcal{H}_2+\epsilon\mathcal{H}_1)$ & stat. uncertainty & sys. uncertainty \\
     & (GeV/c)$^2$ &  & (GeV/c)$^2$ & &  &  &  \\ \hline
0.375 & 0.005 & 0.64 (0.65) & 2.49 (2.49) & 0.36 (0.36) & -0.007 (-0.007) & 0.0069 (0.0073) &  0.020 \\ \hline
0.375 & 0.026 & 0.67 (0.67) & 3.11 (3.11) & 0.43 (0.43) & -0.004 (-0.036) & 0.0271 (0.0198) &  0.018 \\ \hline
0.375 & 0.217 & 0.80 (0.80) & 2.14 (2.20) & 0.25 (0.26) & -0.004 (-0.005) & 0.0026 (0.0031) &  0.010 \\ \hline
0.375 & 0.340 & 0.81 (0.81) & 2.14 (2.17) & 0.25 (0.25) &  0.000 (-0.006) & 0.0027 (0.0033) &  0.009 \\ \hline
0.375 & 0.507 & 0.84 (0.84) & 2.04 (2.05) & 0.23 (0.23) &  0.013 ( 0.007) & 0.0042 (0.0049) &  0.009 \\ \hline
0.375 & 0.743 & 0.92 (0.92) & 1.86 (1.87) & 0.19 (0.19) &  0.067 ( 0.066) & 0.0124 (0.0143) &  0.010 \\ \hline
0.413 & 0.005 & 0.67 (0.68) & 2.34 (2.37) & 0.32 (0.32) & -0.018 (-0.024) & 0.0105 (0.0106) &  0.015 \\ \hline
0.413 & 0.026 & 0.65 (0.65) & 2.37 (2.37) & 0.34 (0.34) & -0.020 (-0.018) & 0.0851 (0.0357) &  0.018 \\ \hline
0.413 & 0.217 & 0.84 (0.83) & 1.97 (1.99) & 0.22 (0.22) &  0.002 (-0.006) & 0.0048 (0.0054) &  0.013 \\ \hline
0.413 & 0.340 & 0.84 (0.83) & 2.20 (2.24) & 0.24 (0.25) &  0.005 (-0.001) & 0.0033 (0.0040) &  0.010 \\ \hline
0.413 & 0.507 & 0.84 (0.84) & 2.03 (2.05) & 0.22 (0.23) &  0.003 (-0.003) & 0.0046 (0.0054) &  0.016 \\ \hline
0.413 & 0.743 & 0.92 (0.92) & 1.86 (1.88) & 0.19 (0.19) &  0.037 ( 0.036) & 0.0132 (0.0153) &  0.014 \\ \hline
0.454 & 0.005 & 0.68 (0.68) & 2.36 (2.36) & 0.32 (0.32) &  0.024 ( 0.011) & 0.0104 (0.0108) &  0.014 \\ \hline
0.454 & 0.026 & 0.67 (0.68) & 2.56 (2.58) & 0.35 (0.35) & -0.053 (-0.031) & 0.0392 (0.0226) &  0.015 \\ \hline
0.454 & 0.340 & 0.84 (0.84) & 2.03 (2.06) & 0.23 (0.23) &  0.000 (-0.002) & 0.0041 (0.0050) &  0.017 \\ \hline
0.454 & 0.507 & 0.87 (0.87) & 1.95 (1.97) & 0.21 (0.21) &  0.005 (-0.002) & 0.0064 (0.0078) &  0.009 \\ \hline
0.454 & 0.743 & 0.92 (0.92) & 1.86 (1.87) & 0.19 (0.19) &  0.036 ( 0.016) & 0.0143 (0.0168) &  0.019 \\ \hline
0.496 & 0.005 & 0.72 (0.72) & 2.50 (2.49) & 0.32 (0.32) & -0.017 (-0.037) & 0.0160 (0.0165) &  0.012 \\ \hline
0.496 & 0.340 & 0.83 (0.83) & 1.98 (2.01) & 0.22 (0.22) &  0.006 ( 0.009) & 0.0055 (0.0066) &  0.018 \\ \hline
0.496 & 0.507 & 0.90 (0.91) & 1.83 (1.85) & 0.19 (0.19) & -0.008 (-0.007) & 0.0092 (0.0116) &  0.010 \\ \hline
0.496 & 0.743 & 0.92 (0.92) & 1.68 (1.68) & 0.17 (0.17) & -0.014 (-0.005) & 0.0201 (0.0245) &  0.009 \\ \hline
0.540 & 0.005 & 0.71 (0.72) & 2.18 (2.20) & 0.28 (0.28) &  0.009 ( 0.005) & 0.0201 (0.0209) &  0.010 \\ \hline
0.540 & 0.507 & 0.92 (0.92) & 1.89 (1.90) & 0.19 (0.19) & -0.006 (-0.005) & 0.0108 (0.0137) &  0.016 \\ \hline
0.540 & 0.743 & 0.94 (0.94) & 1.49 (1.49) & 0.15 (0.15) & -0.006 (-0.081) & 0.0444 (0.0569) &  0.022 \\ \hline
0.586 & 0.507 & 0.92 (0.92) & 1.74 (1.74) & 0.18 (0.18) & -0.013 (-0.010) & 0.0195 (0.0250) &  0.016 \\ \hline
 \end{tabular}
 \end{center}
 \end{table*}

\begin{figure}[htb]
\includegraphics[bb=2cm 4cm 22cm 24cm, scale=0.4]{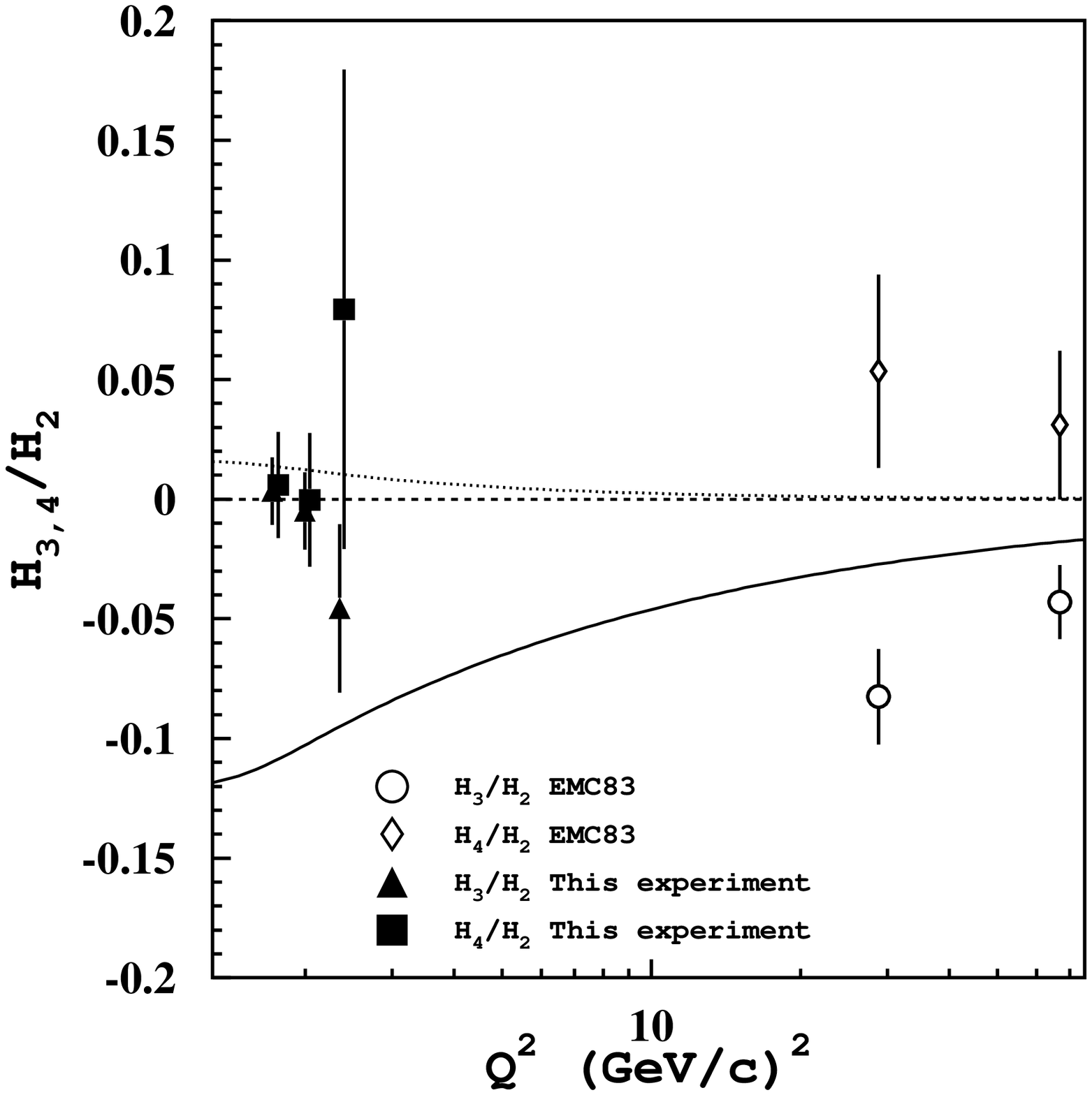}
\caption{\label{fig:cos_q2dep_emc} The $Q^2$-dependence of
$H_3/H_2$ (full triangles)
and $H_4/H_2$ (full squares) extracted using $R=0.12$
from the present data
in comparison with measurements from Ref.~\cite{EMC_cos_2phi}
(open circles - $H_3/H_2$,
diamonds - $H_4/H_2$) at $x=0.24$.
Both data sets are integrated over $z>0.2$
and $p_T>0.2$ GeV/c ($\langle z\rangle = 0.27$,
$\langle p_T^2\rangle = 0.22$ (GeV/c)$^2$).
The curves show predictions of the Cahn effect~\cite{Cahn,Ans1}
for $H_4/H_2$ (dotted)
and $H_3/H_2$ (solid),
corrected for the phase space shrinkage using Eqs.~\ref{eq:pss_4} and \ref{eq:pss_3},
respectively.
The statistical and systematic uncertainties are combined in quadrature.}
\end{figure}

The comparison with higher energy data from Ref.~\cite{EMC_cos_2phi}
shown in Fig.~\ref{fig:cos_q2dep_emc}
reveals the striking difference between the two measurements of $H_3$,
whereas both measurements of $H_4$ at large and small $Q^2$ are compatible with zero.
At large $Q^2$ the absolute values of the ratio $H_3/H_2$
reach 0.05-0.1 and seem to follow the expected $1/Q^2$ behavior. However, our data
at lower $Q^2$ do not follow this trend having values compatible with zero.
The strong suppression of $H_3$ at $Q^2\simeq 2$ (GeV/c)$^2$ with respect to
the data at $Q^2\simeq 30-60$ (GeV/c)$^2$ does not seem to be related to the threshold effect
due to the phase space shrinkage at lower energies discussed in Ref.~\cite{Kroll}.
To account for it, the ratios of the Gaussian model integrals
over the allowed kinematical region:
\begin{equation}\label{eq:pss_3}
\frac{\int_{(p_T^2)^{min}}^{(p_T^2)^{max}} p_T \exp{[-p_T^2/\langle p_T^2 \rangle]} dp_T^2}{\int_{(p_T^2)^{min}}^{(p_T^2)^{max}} \exp{[-p_T^2/\langle p_T^2 \rangle]} dp_T^2}
\end{equation}
\noindent for $H_3/H_2$ and
\begin{equation}\label{eq:pss_4}
\frac{\int_{(p_T^2)^{min}}^{(p_T^2)^{max}} p_T^2 \exp{[-p_T^2/\langle p_T^2 \rangle]} dp_T^2}{\int_{(p_T^2)^{min}}^{(p_T^2)^{max}} \exp{[-p_T^2/\langle p_T^2 \rangle]} dp_T^2}
\end{equation}
\noindent for $H_4/H_2$
are included in the Cahn effect curves in Fig.~\ref{fig:cos_q2dep_emc}.
These corrections do not affect strongly the Cahn
effect curves in the presented interval of $Q^2$.

\section{\label{sec:sum} Conclusions}
We performed a measurement of semi-inclusive $\pi^+$ electroproduction
in the $Q^2$ range from 1.4 to 5.7 (GeV/c)$^2$ with broad coverage in all other kinematic variables.
The five-fold differential cross sections allowed us to separate the contributions of
different structure functions. From these data we draw the following conclusions:
\begin{itemize}
\item The transverse momentum dependence for the $\phi$-independent term $\mathcal{H}_2+\epsilon\mathcal{H}_1$
exhibits the expected thermal Gaussian distribution.
\item At large $z$ the mean transverse momentum $\langle p_T^2 \rangle$ is found to be $x$ and $Q^2$-independent
and it rises with $z$ as expected within the naive parton model.
In the low-$z$ region, $\langle p_T^2 \rangle$ is altered by the limited phase space.
\item The comparison of the measured structure function $H_2$ to
the current fragmentation LO and NLO pQCD calculations shows that
the difference between the data and calculations reaches 20\%
at the low-$x$ limit of the $z<0.4$ region,
which is compatible with systematic uncertainties in the calculations
due to higher-order corrections and the favored fragmentation assumption.
\item The separation of the current and soft target fragmentation processes
in the CM frame shows a symmetry about $x_F=0$,
which suggests the presence of an intriguing numerical equality
between the fragmentation of the spectator diquark in the target region of SIDIS
and the fragmentation of the anti-quark in $e^+e^-$ collisions.
\item The precision of the data does not allow us to obtain information about
the contribution of the Boer-Mulders function, which is expected to be smaller than
our estimated systematic uncertainties~\cite{Gamberg,Barone2}.
\item The $\mathcal{H}_4$ structure function is compatible with zero
within our precision, except for the low-$z$ region where it is positive.
The $\mathcal{H}_3$ structure function
appears to be somewhat better determined than $\mathcal{H}_4$ and is
in strong disagreement with the predictions of the Cahn effect. Inclusion of
the Berger effect does not change significantly the disagreement in $\mathcal{H}_3$.
\item $H_3$ structure function at $Q^2\simeq 2$ (GeV/c)$^2$ found to be
strongly suppressed with respect to the data at $Q^2\simeq 30-60$ (GeV/c)$^2$.
This suppression does not seem to be related to the the phase space shrinkage
at our energies.
\end{itemize}

The data tables can be found in the CLAS physics database~\cite{CLAS_database}
and in sources of the electronic preprint~\cite{preprint_data}.


\begin{acknowledgments}
We thank P. Mulders and E. Di Salvo for helpful discussions.
We also would like to acknowledge the outstanding efforts of the staff of the 
Accelerator and the Physics Divisions at JLab that made this experiment possible.
This work was supported in part by the Istituto Nazionale di Fisica Nucleare, the 
 French Centre National de la Recherche Scientifique, 
the French Commissariat \`{a} l'Energie Atomique, the U.S. Department of Energy, the National 
Science Foundation, Emmy Noether grant from the Deutsche Forschungs gemeinschaft and
the Korean Science and Engineering Foundation.
Jefferson Science Associates (JSA) operates the 
Thomas Jefferson National Accelerator Facility for the United States 
Department of Energy under contract DE-AC05-84ER40150.
\end{acknowledgments}

\newpage 
\bibliography{sidis_pip}

\end{document}